\documentclass[finalold]{agujournal2019}
\usepackage{url} 
\usepackage{lineno}
\usepackage[inline]{trackchanges} 
\usepackage{soul}
\usepackage{lmodern}
\usepackage{subfig}
\usepackage{siunitx}
\usepackage{amsmath}
\usepackage{chngcntr}
\usepackage{graphicx}
\usepackage[normalem]{ulem}
\useunder{\uline}{\ul}{}

\draftfalse

\journalname{Space Weather}

\begin{document}


\title{Real-time prediction of two geomagnetic storms using Solar Orbiter as a far upstream solar wind monitor}


\authors{
Emma E. Davies\affil{1}, 
Eva Weiler\affil{1,2},
Christian Möstl\affil{1},
Satabdwa Majumdar\affil{1},
Hannah T. Rüdisser\affil{1,2},
Timothy S. Horbury\affil{3},
Helen O'Brien\affil{3},
Jean Morris\affil{3},
and Alastair Crabtree\affil{3}
}

\affiliation{1}{Austrian Space Weather Office, GeoSphere Austria, Reininghausstrasse 3, 8020 Graz, Austria}
\affiliation{2}{Institute of Physics, University of Graz, Universit\"atsplatz 5, 8010 Graz, Austria}
\affiliation{3}{Blackett Laboratory, Imperial College London, South Kensington Campus, London, SW7 2AZ, UK}

\correspondingauthor{Emma E. Davies}{emma.davies@geosphere.at}


\begin{keypoints}
\item We present the first real-time predictions of the geomagnetic impact of coronal mass ejections using Solar Orbiter as a far-upstream monitor
\item Predictions of the geomagnetic indices were made 33.9 and 10.3 hours in advance of the observed peak geomagnetic storm time at Earth
\item The real-time procedure developed provides critical validation for future dedicated space weather missions located further upstream than L1 
\end{keypoints}

%
%

\begin{abstract}

We present the first real-time predictions of coronal mass ejection (CME) magnetic structure and resulting geomagnetic impact at Earth for two events using far-upstream observations from Solar Orbiter during March 2024. While our approach assumes idealized conditions for CME propagation and scaling, in situ magnetic field data from upstream monitors still produced realistic predictions despite the large heliocentric distance between Solar Orbiter and L1 (0.53 and 0.60 au). Geomagnetic index predictions were made 15.3 and 4.3 hours before the CME shock arrival at L1, and 33.9 and 10.3 hours ahead of peak storm time; a large improvement over current L1-based nowcasting capabilities. We find that observationally constraining the simple drag-based models using the upstream in situ observations improved arrival time estimates for the two events in this study, although arrival time errors of several hours still remain. Our results show that good predictions of CME magnetic structure and geomagnetic indices with actionable lead-times can be made with far upstream spacecraft, even with longitudinal separations up to 10$^{\circ}$ from the Sun-Earth line, over heliocentric distance ranges where radial evolution effects dominate over longitudinal effects. Limitations include different expansion behaviors for individual CMEs and regions within. Future missions providing continuous data, including solar wind plasma parameters alongside magnetic field measurements, could account for preexisting disturbed conditions and improve geomagnetic prediction accuracy. Our findings demonstrate the substantial value of real-time upstream solar wind measurements for enhancing geomagnetic forecasting accuracy at Earth and provide critical validation for future dedicated upstream space weather missions.
\end{abstract}

\section*{Plain Language Summary}

Massive eruptions from the Sun, known as coronal mass ejections, can disrupt satellites, power grids, and communications when they reach Earth. Increasing warning times of both CME arrival and resulting geomagnetic effects is essential to protect against these natural hazards. To assess their impact, knowledge of the CME internal magnetic field structure is essential, but requires direct in situ spacecraft measurements. Currently, geomagnetic effects are predicted using spacecraft positioned just upstream of Earth, giving little time for protective action. In this study, we used Solar Orbiter, positioned far upstream between the Sun and Earth during March 2024, to directly measure these eruptions earlier. We predict the magnetic properties and potential impacts of two events in real-time, 4-15 hours before they reached near-Earth spacecraft and 10-34 hours before the resulting geomagnetic storm peaks. Even though Solar Orbiter was positioned far upstream and not directly on the Sun-Earth line, our predictions matched near-Earth observations well. We find that relatively simple models work well when used with direct upstream measurements. Although our method has limitations and arrival time uncertainties remain, the results show that placing spacecraft upstream of Earth could improve space weather forecasting abilities and provide valuable insights for future monitoring missions.

\section{Introduction} \label{sec:intro}

Coronal mass ejections (CMEs) with strong and sustained southward magnetic fields are the primary drivers of intense geomagnetic activity affecting Earth \cite{kilpua2017geoeffective}. Accurately forecasting their arrival and magnetic structure remains one of the most important challenges in space weather \cite{kilpua2019forecasting,vourlidas2019predicting}, and is essential for mitigating impacts to space and ground infrastructure \cite{eastwood2017risk,hapgood2021development}. 

CMEs are thought to comprise a magnetic flux rope structure, which is driven outwards from the Sun and propagates through the heliosphere, compressing the solar wind ahead to form a turbulent sheath region in front and further driving a forward shock if traveling faster than the local magnetosonic speed \cite<e.g., Figure 2 in>[]{zurbuchen2006situ}. However, not all CMEs are observed to have such a magnetic flux rope structure at their core when measured by spacecraft in situ \cite{gosling1990coronal}, and therefore the more general term of magnetic ejecta (ME) is often used to describe this region. Throughout this work, we use the term CME to encompass the shock (if driven), sheath region (if present) and the ME. Alterations to the structure of the ME are likely due to CMEs undergoing complex evolution during propagation \cite{manchester2017physical,scolini2022complexity,davies2022multi}, including deformations and flattening of the flux rope cross section \cite<e.g.>[]{riley2004kinematic, owens2006magnetic, liu2006curvature, savani2010observational, davies2021solo}, erosion \cite<e.g.>[]{ruffenach2012multispacecraft, ruffenach2015statistical}, rotations \cite<e.g.>[]{nieves2013inner}, and interactions with other solar wind transients \cite{lugaz2017interaction}. These processes may alter the idealized magnetic flux rope structure of a CME, as discussed in a recent review by \citeA[see Figure 10 for illustration]{alhaddad2025realistic} to a more distorted shape with overall less twist than previously assumed.

To better understand their evolution, previous studies have used in situ multi-spacecraft observations of individual CMEs close to radial alignment to track how their properties change over various heliocentric distance ranges \cite{good2016interplanetary, salman2020radial, davies2022multi, moestl2022multi}. These studies take advantage of science quality data produced by solar wind dedicated spacecraft, e.g. Wind \cite{ogilvie1997wind}, Advance Composition Explorer \cite<ACE;>[]{stone1998advanced}, Solar TErrestrial RElations Observatory \cite<STEREO;>[]{kaiser2008stereo}, Solar Orbiter \cite{mueller2020solar}, and Parker Solar Probe \cite<PSP;>[]{fox2016psp}, and by planetary mission spacecraft when outside of their respective planetary environments, e.g. The MErcury Surface, Space ENvironment, GEochemistry, and Ranging mission \cite<MESSENGER;>[]{solomon2001messenger}, Venus Express \cite{svedhem2007vex}, Juno \cite{bolton2010juno, bolton2017juno}, and BepiColombo \cite{benkhoff2010bepicolombo}. However, in terms of continuous real-time in situ measurements of CMEs, we are often limited to the suite of spacecraft located around the Sun-Earth first Lagrange point (hereafter referred to as L1), e.g., ACE and the Deep Space Climate Observatory \cite<DSCOVR;>[]{burt2012dscovr}.

To make space weather forecasts, we therefore rely on a combination of remote images and forward modeling techniques. Current real-time forecasting employs magnetohydrodynamic (MHD) models of the solar wind, comprising a coronal \cite<e.g., Wang-Sheeley-Arge (WSA)>[]{arge2000improvement, arge2004stream} and heliospheric domain \cite<e.g., Enlil>[]{odstrcil2003enlil}, to which CMEs are typically inserted at the inner heliospheric boundary \cite<e.g.>[]{pizzo2011wsaenlil}. Modeling the propagation of CMEs from the Sun to arrival at the Earth days in advance is subject to many uncertainties including those associated with modeling the ambient solar wind \cite<e.g.,>[]{reiss2023validation}, as well as those stemming from CME input parameters determined from remote images \cite<e.g.>[]{verbeke2023quantifying,kay2024llamacore}. Most operational CME predictions focus on the arrival time problem, with many forecast centers posting their predictions to the NASA Community Coordinated Modeling Center (CCMC) Scoreboard (\url{https://kauai.ccmc.gsfc.nasa.gov/CMEscoreboard/}). Thus far, most validation has therefore focused on the arrival times of CMEs at L1 \cite<e.g.,>[]{riley2018forecasting,kay2024updating}, finding that over the past 10 years, model arrival time predictions have not improved beyond a mean absolute error of approximately 10 hours.  

Knowledge of CME arrival time is an important factor, but it does not determine whether a CME will be geoeffective; other factors such as CME speed and magnetic structure are essential for increasing the value of space weather forecasts \cite{vourlidas2019predicting, owens2020value}. Although CME propagation models give estimates of arrival speed, most operational models do not model the CME internal magnetic field. For example, CMEs are commonly inserted into the WSA-Enlil model using the Cone \cite{zhao2002determination,xie2004cone} model, with the full combination of models known as WSA-Enlil+Cone \cite{mays2015ensemble,wold2018verification}. Here, CMEs are modeled as plasma cloud disturbances with uniform velocity, density, and temperature parameters. This, in combination with the complicated nature of CME evolution, means that we must rely on direct in situ spacecraft measurements to determine the exact magnetic configuration of CMEs.

The geomagnetic response, often quantified using the disturbance storm time \cite<Dst;>[]{sugiura1964,kyoto_dst}/SYM-H \cite{wanliss2006symh}, Kp \cite{bartels1957geomagnetic}, and other indices, is strongly related to the magnetic structure of CMEs, specifically the direction and duration of the north-south magnetic field component (B$_z$), amongst other factors such as speed and density \cite<e.g.,>[]{burton1975empirical}. Therefore, predictions of geomagnetic impact often rely heavily on in situ measurements from spacecraft at L1 \cite<e.g.,>[]{bailey2020prediction}. As L1 is located $\sim$0.01~au upstream of the Earth, these measurements provide a limited warning time of only 10--80 minutes (assuming previously observed CME propagation speeds at 1~au of 2500--300 km\,s$^{-1}$) before a CME impacts Earth.

To extend this warning period, research has increasingly focused on the efficacy of spacecraft positioned closer to the Sun than L1, more commonly referred to as sub-L1 monitors \cite{lugaz2024miist, lugaz2025need}, which offer the potential to increase lead times by several hours, depending on their orbital distance from the Earth and the propagation speed of the CME. Varying types of mission have been proposed that allow for different upstream separations, most promisingly those situated on distance retrograde orbits \cite<DROs;>[]{henon1969}. Such DRO missions include the European Space Agency (ESA) HENON mission \cite{cicalò2025henon} scheduled to launch with PLATO at the beginning of 2027, the ESA SHIELD mission concept, and the NASA MIIST mission concept \cite{lugaz2024miist}.

To investigate the potential of these missions, recent work has taken advantage of the Sun-Earth line crossing of STEREO-A that occurred between late 2022 and mid 2024 where STEREO-A passed $\sim$~0.05~au ahead of L1, thus well positioned to act as a sub-L1 monitor \cite{lugaz2024width, banu2025stereoa, weiler2025superstorm}. \citeA{weiler2025superstorm} demonstrated that in situ STEREO-A observations could be used to predict the geomagnetic impact of CMEs during this time, applying the Temerin~\&~Li model \cite{temerin2006magnetospheric} to time-shifted observations of the May 2024 G5 geomagnetic storm. If an operational sub-L1 monitor had existed during the event, \citeA{weiler2025superstorm} showed that the onset of the geomagnetic storm could have been predicted 2.57~hours earlier, with a predicted geomagnetic impact within 11\% of observed Dst values and 8\% of observed SYM-H values. Similarly, \citeA{liu2024pileup} derived the Dst index from both the STEREO-A and Wind data of the same event averaging the results of two empirical formulas (\citeA{burton1975empirical} and \citeA{obrien2000empirical}), finding that even a longitudinal separation of 12.6$^{\circ}$ between spacecraft observations can result in large differences in CME magnetic field strength, structure, and modeled geomagnetic storm intensity.

In terms of solar wind monitors even further upstream, previously proposed mission concepts include spacecraft on a Venus-like orbit \cite{ritter2015venus} and spacecraft powered by solar sails to stay in stationary positions along the Sun--Earth line \cite{lindsay1999dst, eastwood2015sunjammer}. Previous studies have used data from spacecraft upstream to make hindcasts of geomagnetic impact: \citeA{kubicka2016dst} used Venus Express magnetic field data whilst it was located at 0.72 au and 6$^{\circ}$ from the Sun--Earth line to constrain CME arrival time estimates at L1 and input scaled in situ data to empirical solar wind--Dst models to produce similar modeled minimum Dst values (-96 and -114~nT) in comparison to observed (-71~nT) Dst values. Similarly, \citeA{davies2021solo} used Solar Orbiter magnetic field data whilst it was located at 0.8~au and within 5$^{\circ}$ from the Sun--Earth line in combination with the PREDSTORM model \cite{bailey2020prediction} to produce a minimum Dst of -47~nT in comparison to the observed value of -60~nT. The CME event investigated by \citeA{davies2021solo} was the first CME observed after Solar Orbiter's launch in April 2020, and the spacecraft has since crossed the Sun--Earth line yearly. \citeA{laker2024upstream} took advantage of such a crossing in March 2022, using Solar Orbiter magnetic field data obtained in real time to make predictions of the CME arrival and its magnetic structure at L1 where Solar Orbiter was 0.5~au upstream of Earth. The authors found that having an upstream monitor halfway between the Sun and Earth allowed better constrained predictions to be made for CME arrival time at L1, reducing the uncertainty in arrival time from 10.4 to 2.5 hours in the case of one event, and also found a good agreement between the predicted and observed B$_z$ profiles in the case of the second event, despite a longitude separation of 10$^{\circ}$ from the Sun--Earth line.

From these studies conducted over different upstream separations, it is clear there is a trade-off between the lead time and accuracy of predictions. However, the optimal orbits for sub-L1 missions remain undetermined \cite{lugaz2025need} as there is also the issue of how aligned a sub-L1 monitor must be to the Sun--Earth line for predictions to be reliable. Multi-spacecraft observations of CMEs are generally considered radially aligned if they are within 10$^{\circ}$ of longitudinal separation \cite<e.g.,>[]{good2015radial} as spacecraft are likely to observe the same CME ME if within 30$^{\circ}$ of each other \cite{good2016interplanetary}. From recent STEREO-A--L1 observations, studies have found that the extent of the CME ME may be even smaller than previously assumed, with angular widths of $\sim$~20$^{\circ}$--30$^{\circ}$ \cite{lugaz2024width, banu2025stereoa}. However, even small longitudinal separations of less than a few degrees can lead to significant discrepancies in observations and CME properties \cite{davies2020radial, regnault2024discrepancies}, in agreement with the suggestion that CMEs are not coherent structures but may display locally apparent coherence \cite{owens2017coronal}. 

In this study, we ask whether it is possible to make accurate predictions of the CME in situ observations near-Earth and their geomagnetic effects using far upstream solar wind monitor observations in real time, despite the radial evolution undergone and possible longitudinal differences between observation at the far upstream monitor and L1. To do so, we build on the work of \citeA{laker2024upstream}, taking advantage of the yearly Sun--Earth line crossings of Solar Orbiter, going one step further to predict the geomagnetic impact based on the predicted in situ CME magnetic structure at L1. From 12 to 25 March 2024, there was an approximate two-week window in which Solar Orbiter and Earth were longitudinally separated by $\pm$15$^{\circ}$. We select this window of longitudinal separation to be consistent with previous work \cite{good2016interplanetary, lugaz2024width, banu2025stereoa}, where spacecraft are likely to observe the same CME ME. During this time, Solar Orbiter was located at heliocentric distances between 0.53--0.37~au, and close to the ecliptic plane with very small latitudinal separations between $\pm$1.1$^{\circ}$ with respect to the Earth. Gravity assist maneuvers at Venus from February 2025 have since raised the mission's inclination, making the March 2024 Sun-Earth line crossing Solar Orbiter's last whilst in the ecliptic plane. These measurements therefore provide a rare opportunity to predict the geomagnetic impact of a CME measured far upstream of the Earth without added complications due to potential differences in observations caused by large latitudinal separations.

We present the two CME events that occurred during this window and the resulting predictions of their geomagnetic impact made in real time. In Section~\ref{sec:data} we present the real-time data used throughout this study. Section~\ref{sec:realtime_procedure} presents the methods used in real time that form the pipeline from arrival prediction, to magnetic structure prediction, to forecasting the geomagnetic indices. We present the results of the real-time predictions in Section~\ref{sec:results}. We discuss the assumptions of our methodology and their effect on the results in Section~\ref{sec:discussion}, drawing conclusions and discussing the implications for future space weather missions in Section~\ref{sec:implications}. 

\section{Real-Time Data and Catalogs} \label{sec:data}

To provide the solar context, we use different passbands of the Atmospheric Imaging Assembly \cite<AIA;>[]{lemen2011atmospheric} on-board the Solar Dynamics Observatory \cite<SDO;>[]{pesnell2012sdo} to identify the source region of the CMEs. Level 1 images of SDO/AIA for the 171, 193, and 211 \AA\, channels were downloaded from the Joint Science Operations Center (JSOC). To get the magnetic field information of the source region, level 1 data of the Helioseismic and Magnetic Imager \cite<HMI;>[]{Scherrer2012SoPh} line of sight magnetograms on-board SDO was downloaded from JSOC. We also use coronagraph observations from the Large Angle Spectroscopic COronagraphs \cite<LASCO;>[]{brueckner1995lasco} C2 and C3 on-board the Solar and Heliospheric Observatory \cite<SOHO;>[]{domingo1995soho}. Level 0.5 data of LASCO C2 and C3 was reduced and calibrated to level 1 using \textit{reduce\_level\_1.pro} (included in the SolarSoft library in IDL), which involves corrections for the flat field response of the detector, radiometric sensitivity, stray light, geometric distortion, and vignetting. Then, base difference images were created by subtracting a pre-CME image from the successive images containing the CME.

To make predictions of CME arrival time, we use input parameters given by the Space Weather Database Of Notifications, Knowledge, Information (DONKI; \url{https://kauai.ccmc.gsfc.nasa.gov/DONKI/}) database provided by the Moon to Mars (M2M) Space Weather Analysis Office and hosted by the Community Coordinated Modeling Center (CCMC; \url{https://ccmc.gsfc.nasa.gov/}). The DONKI database contains estimates of parameters derived from remote sensing data, including launch time at 21.5~R$_\odot$, longitude ($lon$), latitude ($lat$), half-width ($\lambda$) and initial velocity of events ($V_0$), functioning as an operational tool for both research and forecasting purposes.

The magnetic field measurements made by the magnetometer \cite<MAG;>[]{horbury2020mag} instrument onboard Solar Orbiter \cite{mueller2013solar,mueller2020solar} were provided by the Imperial College London MAG team directly with a cadence of 1 minute. As Solar Orbiter is a science mission which does not supply a continuous stream of data to Earth, for this study, relevant data from the MAG team was sent as quickly as possible after each pass with the spacecraft had been completed. For each event, we received two sets of data: the first including the start of the CME and partially revealing its structure, and the second including the complete CME structure after it has passed the spacecraft. We note that since this work was completed, the MAG team has automated the processing and distribution of the MAG low-latency data to space weather forecasters, bringing latency down to around 1 hour during passes.

To compare predictions of the CME in situ magnetic field structure made using Solar Orbiter data to observed values at L1, we use the real-time solar wind data products released on the NOAA SWPC Services webpage (see Open Research Section), where the magnetic and plasma data are available in 1-min resolution for the previous seven days up to the current time. This data product uses either measurements taken by the magnetometer (MAG) and Faraday Cup (FC) instruments onboard DSCOVR, or the Magnetic Field Experiment \cite<MAG;>[]{smith1998ace} and Solar Wind Electron Proton Alpha Monitor \cite<SWEPAM;>[]{mccomas1998solar} instruments onboard ACE \cite{zwickl1998rtswace}. During the two-week window in which this study was conducted (12--25 March 2024, inclusive), 40.3\% of the magnetometer data was provided by DSCOVR and 59.1\% by ACE (with 0.5\% of magnetometer data missing), and 40.3\% of the plasma data was provided by DSCOVR and 49.4\% by ACE (with 10.3\% of plasma data missing).

To compare our final geomagnetic predictions (presented as a 1-min resolution Dst index), we use the NOAA real-time Dst index, similarly obtained from the NOAA SWPC Services webpage (see Open Research Section), where hourly values are given. The values provided by the NOAA data products draw from the Real-time (Quicklook) Dst index product provided by the World Data Center for Geomagnetism, Kyoto (\url{https://wdc.kugi.kyoto-u.ac.jp/dst_provisional/index.html}). 

\section{Real-Time Procedure} \label{sec:realtime_procedure}

\begin{figure}
    \centering
    \includegraphics[width=\linewidth]{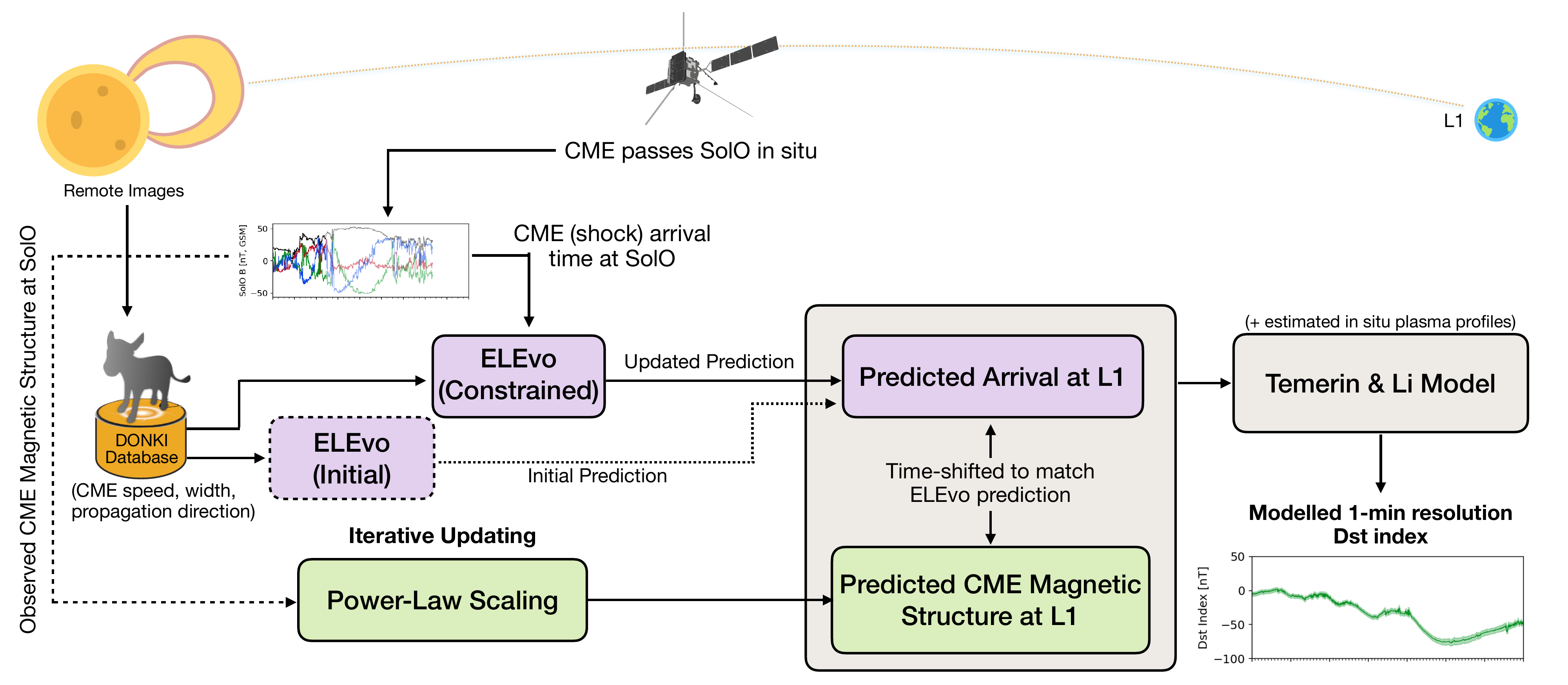}
    \caption{Schematic of the real-time procedure used to predict the CME arrival time, magnetic structure at L1, and the resulting modeled 1-min resolution Dst index predictions. The purple boxes highlight steps in the CME arrival time at L1 prediction (details given in Section~\ref{sec:pred_arrival}), the green boxes highlight steps in predicting the magnetic field structure at L1 (details given in Section~\ref{sec:pred_structure}), and the gray boxes highlight steps involved in predicting the 1-min resolution Dst index (details given in Section~\ref{sec:pred_geomag}).}
    \label{fig:procedure}
\end{figure}

Figure~\ref{fig:procedure} presents a schematic of the procedure we employ in real time, which begins once a CME erupts from the Sun and is characterized through remote imagery, with initial speed, width, and propagation direction parameters drawn from the DONKI database. For Solar Orbiter- or Earth-directed CMEs (i.e., CMEs where the half-width and longitudinal propagation direction listed by DONKI encompass the Solar Orbiter and Earth positions), these parameters feed an initial ELEvo run, providing an initial estimated time of arrival at L1 (dotted line). As the CME propagates outward and is subsequently observed in situ by Solar Orbiter, the measured shock arrival time constrains a second ELEvo run, refining the arrival prediction at L1 through an iterative updating process (updated prediction, solid line). In parallel, the magnetic structure observed in situ at Solar Orbiter is propagated to L1 using power-law scaling relations, yielding a predicted magnetic structure at L1 that is iteratively updated and time-shifted to remain consistent with the ELEvo arrival time prediction. Together, these two predicted products, arrival time and magnetic structure, are passed into the Temerin \& Li model, along with estimated in situ plasma profiles at L1, to generate the modeled 1-min resolution Dst index.

We describe the methodology used to make predictions of the CME arrival time at L1 (purple boxes) in Section~\ref{sec:pred_arrival}, predictions of the in situ CME magnetic structure at L1 (green boxes) in Section~\ref{sec:pred_structure}, and the resulting predictions of the geomagnetic impact (gray boxes) in Section~\ref{sec:pred_geomag}.

\subsection{Constraining the Arrival Time at L1} \label{sec:pred_arrival}

We make our CME arrival time predictions using the ELliptical Evolution model \cite<ELEvo;>[]{moestl2015elevo}. Generally, the model assumes an elliptical front for the CMEs and requires initial parameter estimates of the time the CME was observed at 21.5~R$_\odot$ (r$_0$), initial speed (v$_0$), and propagation direction, to propagate the CME outward using a simple drag-based model \cite{vrsnak2013propagation}, with analytical solutions as follows:

\begin{equation}
    v(t) = \frac{v_0 - w}{1 \pm \gamma (v_0 - w)t} + w,
\end{equation}

\begin{equation} \label{eq:dbm}
    r(t) = \pm \frac{1}{\gamma} \ln \left[ 1 \pm \gamma (v_0 - w)t \right] + wt + r_0,
\end{equation}

\noindent where $w$ is the ambient solar wind speed, $\gamma$ is the drag parameter, and $r$ is the heliospheric distance of the CME front. The drag-based model defines the rate at which a fast (slow) CME is decelerated (accelerated) relative to the ambient solar wind as it propagates through the heliosphere, with the analytical solutions approximating $\gamma(r)$ and $w(r)$ to be constant, i.e., these values do not vary with heliocentric distance.

The ELEvo model is used in an ensemble approach, where the ensemble is created by varying parameters of the drag-based model: the initial CME speed by $\pm$50~km\,s$^{-1}$, the ambient solar wind speed as 400~$\pm$~50~km\,s$^{-1}$, and the drag parameter as $0.1~\pm~\num{0.025e-7}$~km$^{-1}$, where the uncertainties correspond to the 1$\sigma$ uncertainty of the normal distribution from which the ensemble members are randomly drawn 100,000 times for each time step \cite<see also>[for this type of methodology]{calogovic2021dbem}. We note that these parameters are an initial starting point, and later updated after the CME is observed in situ at Solar Orbiter. 

The launch time, CME propagation direction and aspect ratio of the (assumed) elliptical front (set to 1.43) are kept fixed in this study, using values derived from forecaster-analyzed DONKI entries. Although these parameters are subject to significant uncertainty \cite{kay2024llamacore}, explicitly varying them in a real-time ensemble would likely lead to excessive ensemble spread in arrival times. The associated uncertainty for these parameters is therefore not explicitly quantified and is instead assumed to be implicitly accounted for by the ensemble-derived arrival time distribution obtained from varying the dominant propagation parameters, as well as the choice of using a simplified model. In general, the ensemble ranges adopted here are not intended to represent a statistically validated set of input uncertainties. The ELEvo setup used here should therefore be regarded as an experimental, real-time prototype rather than a fully validated operational forecast system. A systematic assessment of input parameter uncertainties and their relative importance has been carried out in recent work by \citeA{ruedisser2026towards}, which proposes globally optimized parameter ranges for ELEvo (not available at the time this study was conducted and are therefore not adopted here). However, as the parameters in our approach are subsequently constrained using upstream in situ observations on an event-by-event basis, the results presented in this work are only weakly sensitive to the choice of initial parameter ranges.

For this study, ELEvo is set up to run continuously, downloading new CME input parameters every hour from the CCMC DONKI database. If multiple DONKI entries are available for a given CME (often denoted as shock, SH, or leading edge, LE), we use the average of the reported parameters. Given the simplified nature of the model and the already substantial uncertainties associated with input parameters (which are accounted for through the ensemble approach), the additional uncertainty introduced by averaging multiple entries is expected to be of comparable magnitude to other sources of uncertainty and therefore not expected to dominate the resulting arrival time prediction. With these initial parameters from DONKI, we can estimate the arrival of the CME at both Solar Orbiter and L1 (if the CME is Solar Orbiter- or Earth-directed), with the ensemble spread providing a measure of the arrival time uncertainty.

After the CME shock has been observed to arrive at Solar Orbiter in situ, we then make a second `updated' CME arrival time at L1. To do this, we search for an updated set of parameters for the drag-based model, i.e., the ideal combination of initial CME speed, ambient solar wind speed and drag parameter, that best reproduce the observed arrival time at Solar Orbiter (keeping initial time, propagation direction, and aspect ratio of the CME fixed as before). The optimal parameters are determined using the \texttt{scipy.optimize.minimize} function to minimize the root mean squared error (RMSE) between the modeled and observed arrival at Solar Orbiter. We run the ELEvo model with the new ensemble, where the mean values for the normal distribution are the updated initial CME speed, ambient solar wind speed and drag parameter from the minimization, to produce an updated arrival time prediction at L1. 

\subsection{Predicting the In Situ CME Magnetic Structure at L1} \label{sec:pred_structure}

To quantitatively predict the geomagnetic impact of the CME, we first must predict the magnetic structure of the CME at the Sun--Earth L1 point. To do this, we first transform the data from radial-tangential-normal (RTN) coordinates to geocentric solar magnetospheric (GSM) coordinates, so that the x-axis points toward the Sun (origin is Earth center) and the z-axis is the projection of the Earth's magnetic dipole axis to the plane perpendicular to the x-axis. We then time-shift the transformed magnetic data observed by Solar Orbiter so that the CME shock front aligns with the predicted L1 arrival time given by the ELEvo model. We note that ultimately, we time-shift the predicted structure to the updated ELEvo predicted arrival time, however, we may first time-shift to the initial predicted arrival time if the updated prediction has not yet been made.

As CMEs propagate away from the Sun, they expand due to their high internal magnetic pressure until they reach equilibrium with the solar wind background through which they propagate. In situ, such CME expansion often results in both larger radial widths of CMEs and lower measured magnetic field magnitudes. To account for this and achieve more realistic CME magnetic field observations at L1, we therefore scale the Solar Orbiter magnetic field data in terms of magnitude and stretch the data temporally.

The rate at which their magnetic field strength decreases can be fit with a power law, $B \propto r^{\alpha}$, where $r$ is the heliocentric distance and $\alpha$ is the constant to be determined. Previous statistical studies of CME properties measured in situ have found a range of $\alpha$ values, with steeper rates of expansion over heliocentric distance ranges closer to the Sun \cite{gulisano2010global, winslow2015interplanetary, good2019self}, and slower rates of expansion beyond 1~au \cite{richardson2014identification, davies2021catalogue}. In this study, we use an $\alpha$ value of -1.64 to scale the magnetic field values measured at Solar Orbiter, as found by \citeA{leitner2007consequences}. We scale the magnitude and each magnetic field component using the same $\alpha$ value, in line with previous studies which found that different components decrease at similar rates with increasing heliocentric distance \cite{lugaz2020inconsistencies, davies2021solo}, contrary to more theoretical models of CME expansion \cite{farrugia1993study}. We set the upper and lower uncertainty range to values of -1.2 and -2, respectively, reflecting the slowest rate of CME expansion found \cite{richardson2014identification, davies2021catalogue}, and the theoretical steepest rate of expansion for a self-similarly expanding cylindrical force-free magnetic flux rope \cite{farrugia1993study,kumar1996magnetic}. For the preceding solar wind data, we also scale the magnetic field values in terms of magnitude (but not temporally) applying the same $\alpha$ value and wide uncertainty range, capturing the expected radial dependence of the interplanetary magnetic field magnitude over the heliocentric distance range in this study \cite{parker1958dynamics}.

After scaling the magnitude of the magnetic field values, we then consider the temporal expansion of the CME. As solar wind plasma data from Solar Orbiter was not available in real time during this study, and therefore we did not have in situ speed information with which to assess the CME local expansion directly, we instead use the dimensionless expansion parameter \cite{demoulin2009causes, gulisano2010global}. This parameter provides a measure of local expansion, typically denoted as $\zeta$, which is independent of radial distance. This parameter can therefore be related to the global expansion of CMEs: where the magnetic field strength would be expected to decrease as $r^{-2\zeta}$ ($\alpha = 2\zeta$) and the radial size of the CME expected to increase as $r^\zeta$. $\zeta$ has been found to be $\approx$~0.8 for non-perturbed magnetic clouds \cite{gulisano2010global}. We note that the radial width of a CME is calculated as $D \times v_c$, where D is the duration and $v_c$ is the cruise velocity i.e., the speed at the mid-point of the in situ CME speed profile. However, as we do not have in situ speed information at Solar Orbiter in real-time, here, we have to make the assumption that duration scales similarly to radial width (we discuss the implications of this assumption in Section~\ref{sec:discussion}). We know the heliocentric distance of Solar Orbiter (r$_1$) and the duration of the CME (D$_1$), so we can therefore apply the power law with $\zeta$ = 0.8 to calculate the estimated duration (D$_2$) at L1 (r$_2$):

\begin{equation} \label{eq:temp_scaling}
    D_2 = D_1 \times (\dfrac{r_2}{r_1})^{0.8}. 
\end{equation}

Assuming a constant expansion within the CME, we multiply the time cadence of the Solar Orbiter data by the ratio of the predicted L1 duration and observed Solar Orbiter duration (D$_2$/D$_1$), effectively stretching the data timestamps linearly to fit the calculated larger duration (D$_2$). In practice, this means that the Solar Orbiter magnetic field data with a cadence of 1 minute gets stretched to a cadence of $1 \times D_2/D_1$ minutes within the CME. 

We note that for the events in this study, we apply the same expansion laws over the whole CME interval i.e., for the sheath and magnetic ejecta regions alike. We choose this method to simplify the complicated nature of sheath evolution: of course, in reality, we do not expect the sheath to expand in the same manner as the ME, but we also have to consider its growth as solar wind is accumulated and compressed ahead of the CME. In the case of this study, we also have the added complications of possible CME interactions, possibly leading to the compression of CMEs ahead within the sheath region. We discuss this in more detail in Section~\ref{sec:discussion}.

It is also important to note that we expect the predicted observations at L1 to look different from those observed in situ due to differences in longitudinal separation, potential interactions with other CMEs, solar wind transients, the background solar wind, and other processes such as rotations, deformations, stream interactions within the CME, and magnetic erosion, that may occur as the CME events propagate between Solar Orbiter and L1. As previous studies have found, individual CMEs vary in their evolution from statistical trends, however, by using a wide uncertainty range in our scaling we hope to account for this global variation, even though we cannot account for the likely smaller scale changes to the ICME structure. In this study, we therefore aim to test whether we can still make worthwhile geomagnetic predictions with the extra in situ magnetic field information gained by the far upstream solar wind measurements of Solar Orbiter.

\subsection{Prediction of the Geomagnetic Indices} \label{sec:pred_geomag}

The Dst index quantifies axially symmetric disturbances in Earth's equatorial magnetic field, with negative excursions primarily driven by the magnetospheric ring current and positive variations resulting from magnetospheric compression due to enhanced solar wind pressure \cite{sugiura1964,mayaud1980geomagnetic}. Previous studies have found empirical relationships between measured magnetic field and plasma parameters and the resulting geomagnetic indices \cite{burton1975empirical}. One such model, the Temerin~\&~Li model \cite{temerin2002newmodel,temerin2006magnetospheric}, is a semi-empirical model optimized using eight years of solar wind data from 1995--2002, where parameters in the model were found empirically via minimizing the RMSE between the modeled Dst and observed Dst. Equation~\ref{eq:temerin_and_li} presents the six terms the model uses to calculate the Dst index:

\begin{equation} \label{eq:temerin_and_li}
    D_{ST} = dst1 + dst2 + dst3 + (pressure~term) + (direct~B_z~term) + (offset~terms). 
\end{equation}

The dst1, dst2, and dst3 terms can be understood as the ring current, the partial ring current, and the magnetotail current, respectively, where the interpretation of dst3 is less well understood \cite{temerin2006magnetospheric}. All of these magnetospheric current systems contribute to the Dst index, with the ring current being considered the most important contributor, as described above. The dst1, dst2, and dst3 terms each consist of driver and decay terms and are calculated using current values of the interplanetary magnetic field, speed, and density as well as past values of the interplanetary magnetic field, dst1, and dst2 values. The driver terms dictate the geomagnetic response to the solar wind and depend generally more on the solar wind velocity ($v^2$) than on the density ($\sqrt{density}$). The pressure term and direct B$_z$ term are calculated directly using measured in situ solar wind as input.

As we investigate a time period after 2002, we include a simple linear correction that accounts for a linear drift of the modeled Dst away from the real one by -5.3nT/year caused by a time-dependent variable in the model \cite{temerin2015underestimates, bailey2020prediction}. Since the model also depends on local time, it is essential to ensure that the input data, in this case the in situ Solar Orbiter data, is time-shifted to match the expected arrival time of the CME at L1.

We input the predicted magnetic field structure for L1 to the model to produce a predicted 1-minute resolution Dst index, similarly to the procedure used by \citeA{weiler2025superstorm}. As the Temerin~\&~Li model also depends on solar wind plasma parameters which we did not have in real-time from Solar Orbiter, we make best estimates of the speed and density profiles prior to and during the event at L1. As the Temerin~\&~Li model, and especially the driver terms, are more responsive to speed values than density, we focus on producing a more realistic speed profile. To do so, we also use speed estimates from ELEvo, however, no intrinsic quantities such as density. We therefore set the density to a constant 5~cm$^{-3}$ based on ambient solar wind conditions, prior to and during the event as we have no information on how the density will vary during the sheath and ME region. To create the speed profile, we set the speed prior to the event based on the average solar wind speed at L1 over the 12 hours prior to when the prediction is made (we note that, realistically, this is likely to change in the time after prediction). For slower events, where the predicted L1 arrival speeds given by the ELEvo model are close to the constant background speed of the model, we continue to use the same observed average solar wind speed to avoid creating a stepped profile that may introduce an artificially strong geomagnetic storm onset output by the Temerin \& Li model. For faster events where the modeled speeds given by ELEvo are much greater than the background value, we aim to create a stepped speed profile: we take the mean of the initial and updated ELEvo model estimates of the arrival speed at L1 and set the input speed from the predicted arrival time onward to this mean value. We assess how these assumptions compare to the observed values and their effect on the modeled 1-min resolution Dst index in Section~\ref{sec:discussion_geomag}.

To estimate the uncertainty when calculating the 1-min resolution Dst index, we randomly vary the scaled magnetic field components (within the lower bound obtained using $\alpha=-2$, and an upper bound using $\alpha=-1.2$, as described in Section~\ref{sec:pred_structure}), our estimates for the initial dst1, dst2, and dst3 values, the speed ($\pm$ 50~km\,s$^{-1}$), and the density ($\pm$ 30\%), all along a normal distribution using 10,000 ensemble members. We also randomly vary the time-shift obtained from the ELEvo model within the arrival time error. We note that these uncertainty ranges used are more appropriate for the in situ data inputs the Temerin \& Li model usually takes, however, it is likely that these ranges will not capture the larger uncertainties between the modeled input values derived from remote images (i.e., speed) and those measured in situ. Therefore, we emphasize that the resulting 1-min resolution Dst index calculated by the Temerin \& Li model is an estimated prediction rather than a fully physically constrained forecast.

The complete procedure described in Section~\ref{sec:realtime_procedure} takes less than 5 minutes to run: optimizing the initial parameters for the ELEvo model takes less than one second, both the initial and updated ELEvo model runs take a total of 16~s, performing the in situ data transforms and power law scaling takes less than 1~minute, and finally, modeling the 1-min resolution Dst index with 10,000 ensemble members takes 42~s. The complete pipeline is therefore well suited for real-time applications. 

\section{Results} \label{sec:results}

During March 12--25, there were two opportunities to test our real-time procedure: the first with two CMEs launched in close succession on March 17, and the second also with two CMEs launched in close succession on March 23, as listed in the DONKI CCMC catalog. We present the results of applying the methodology described in Section~\ref{sec:realtime_procedure} to Event 1 (Section~\ref{sec:event1}) and Event 2 (Section~\ref{sec:event2}), close to the chronological order in which the steps were performed, as listed in Tables~\ref{tab:timeline_event_1} and \ref{tab:timeline_eevent_2}, respectively. The resulting predictions for each event are compared to the real-time observations in Table~\ref{tab:elevo_parameters} (arrival times) and Table~\ref{tab:event_predictions} (CME properties and geomagnetic indices).

\subsection{Event 1: March 17, 2024} \label{sec:event1}

\begin{table}[]
\caption{Timeline of actions during the real-time procedure applied to Event 1. The first column provides a description of the action with the corresponding methodology step (as given in Section \ref{sec:realtime_procedure}) if applicable, the second column provides the time at which this happened in UT, the third column provides the time difference between the corresponding action and the observed arrival of the CME shock at L1 in hours, and the fourth column provides the time difference between the corresponding action and the observed minimum Dst index in hours.}
\label{tab:timeline_event_1}
\resizebox{\textwidth}{!}{%
\begin{tabular}{llll}
{\textbf{Action}} & {\textbf{Time}} & {\textbf{$\Delta$ t$_{L1}$}} & {\textbf{$\Delta$ t$_{Dst}$}} \\
 & \textbf{[UT]} & {\textbf{[hrs]}} & {\textbf{[hrs]}} \\
\hline
CME launch observed & 2024-03-17 03:36 & 94.8 & 113.4 \\
\hline
\begin{tabular}[c]{@{}l@{}}Initial ELEvo arrival time predictions \\ for Solar Orbiter and L1 using DONKI input parameters (2)\end{tabular} & 2024-03-18 09:29 & 64.9 & 83.5 \\
\hline
Received first MAG data from Solar Orbiter & 2024-03-19 14:58 & 35.4 & 54.0 \\
\hline
Made first CME magnetic field structure prediction at L1 (4) & 2024-03-19 16:56 & 33.5 & 52.1 \\
\hline
\begin{tabular}[c]{@{}l@{}}Updated ELEvo arrival time prediction at L1 using \\ Solar Orbiter arrival time constraint (3)\end{tabular} & 2024-03-19 19:35 & 30.8 & 49.4 \\
\hline
\begin{tabular}[c]{@{}l@{}}Received full Solar Orbiter MAG data and \\ updated CME magnetic field structure prediction (4)\end{tabular} & 2024-03-19 22:28 & 27.9 & 46.5 \\
\hline
Used Temerin \& Li model to produce 1-min Dst index prediction (5) & 2024-03-20 11:06 & 15.3 & 33.9 \\
\hline
Time CME (shock) arrived in situ at L1 & 2024-03-21 02:24 & - & 18.6 \\
\hline
Time observed minimum Dst index & 2024-03-21 21:00	 & -18.6 & - \\
\hline
\end{tabular}%
}
\end{table} 

\begin{figure}
    \centering
    {\includegraphics[width=\textwidth]{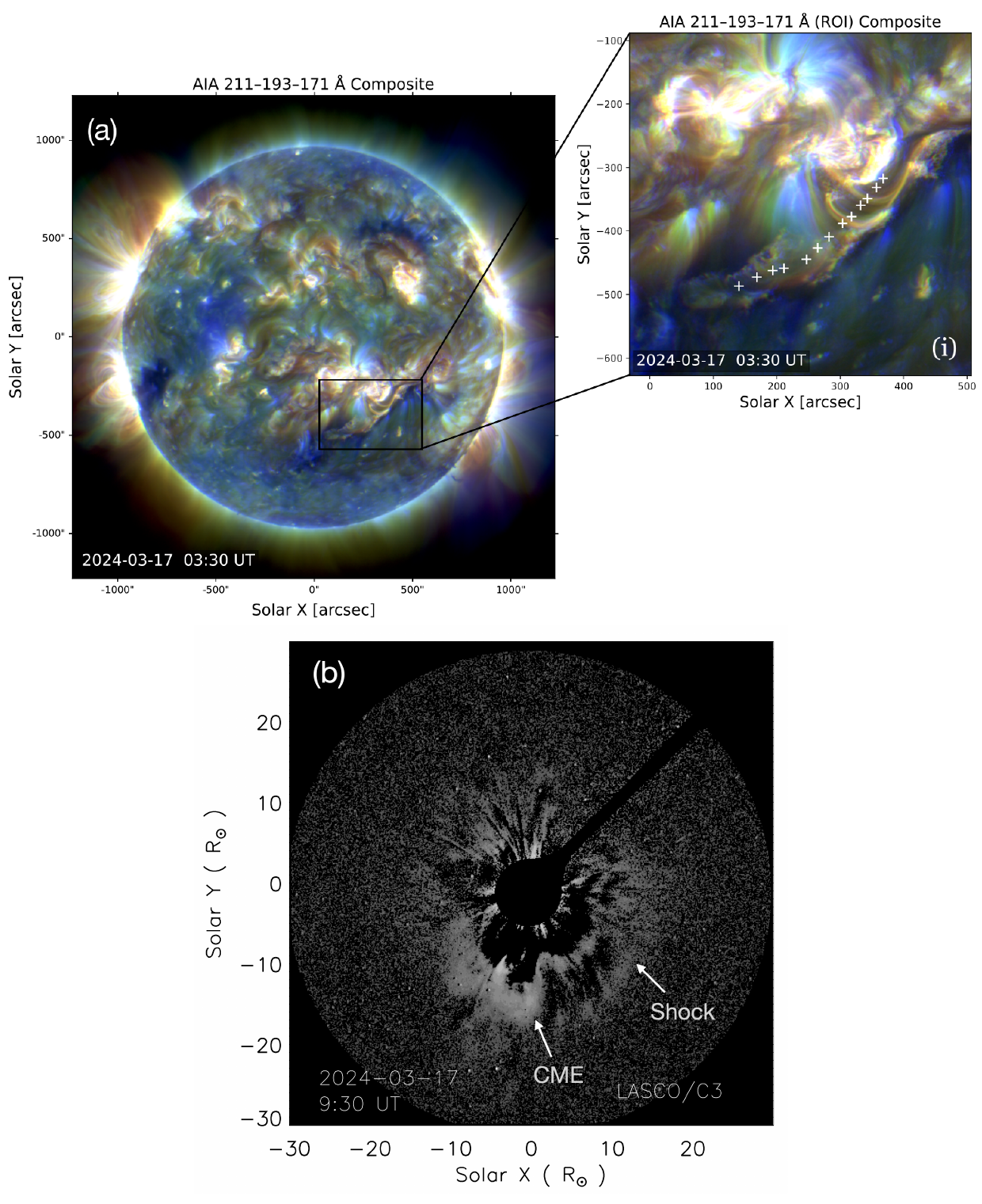} }
    \caption{a) SDO/AIA 211, 193, and 171~Å composite image showing the CME source region (outlined by the black rectangle) at 2024-03-17 03:30~UT. Panel (i) presents a zoomed view of the source region, highlighting the skewed post-eruption arcades, with white ‘+’ symbols marking the filament footpoints. b) SOHO LASCO C3 image at 2024-03-17 9:30~UT of the CME directed towards the South with arrows marking the shock and CME.}
    \label{fig:sdo_17march}
\end{figure}

Two CMEs were successively launched from the same source region located between S45E05 and S15W40 on 2024 March 17 at 03:12~UT and 03:36~UT, respectively. The source region of the CMEs was identified using the JHelioviewer software \cite<>[]{mueller2017jhelioviewer} that enables back-projecting the CMEs onto the solar disk. This is done by combining coronagraph observations (SOHO/LASCO C2) with disk imaging observations (different extreme passbands of SDO/AIA). CME speeds reported in DONKI were used to define a temporal search window, while the central position angle and angular width provided a spatial constraint for associating the CME with its source region. Base-difference images from AIA and LASCO C2 were then inspected for outward-propagating features within these windows, following the methodology described in \citeA{majumdar2023source}. Figure~\ref{fig:sdo_17march}(a) shows a SDO/AIA 211, 193, and 171 Å composite image taken at 03:30~UT, with the source region enclosed inside a black rectangle. A zoomed-in view of the region (panel (i)) highlights the skewed post-eruption arcades, and the white crosses mark the traced filament axis corresponding to the polarity inversion line. Taking into account some proxies on the Sun that indicate the handedness of the flux rope \citeA<e.g.,>{palmerio2017determining,palmerio2018coronal}, we would expect a right-handed flux rope in situ. Since the source region of this CME is located in the southern hemisphere, this is also consistent with the hemisphere rule \cite<e.g.,>[]{rust1994spawning,pevtsov2003helicity,zhou2020relationship}. Panel (b) presents the associated CME observed in the LASCO C3 field of view, with arrows marking the leading edge and the shock front.

Both CMEs were launched southward, as shown by the SOHO/LASCO C3 image in Figure~\ref{fig:sdo_17march}(b). The CCMC DONKI catalog lists the CME propagation directions and speeds determined using the SWPC CME Analysis Tool (CAT). For the first CME, parameters were determined using SOHO/LASCO C2 and STEREO-A/SECCHI COR2 remote image data, for both the faint shock observed in the early image frames and the leading edge based on the bulk portion of the CME, with the estimated propagation directions (lat/lon) listed as -31.0$^{\circ}$/31.0$^{\circ}$ and -35.0$^{\circ}$/10.0$^{\circ}$ for the shock and leading edge, respectively. For the second CME, parameters were determined using SOHO/LASCO C3 and STEREO-A/SECCHI COR2 remote image data, again for both the faint shock and the leading edge, with estimated propagation directions of  -53.0$^{\circ}$/-24.0$^{\circ}$ and -48.0$^{\circ}$/-24.0$^{\circ}$, respectively. The later CME was faster than the first, with estimated propagation speeds of 558~km\,s$^{-1}$ for the shock and 499~km\,s$^{-1}$ for the leading edge, compared with a shock speed of 502~km\,s$^{-1}$ and a leading edge speed of 325~km\,s$^{-1}$ for the first CME. Performing a simple analysis assuming constant speeds, the shock of the faster second CME would catch the leading edge of the first slower CME within approximately 33 minutes after the second CME was launched. This is also observed remotely in Figure~\ref{fig:sdo_17march}(b), where the two CME structures are already difficult to distinguish from one another in the LASCO C3 field of view. It is therefore likely that the CME signatures observed in situ, at both Solar Orbiter and L1, would also be a combination of both CMEs.

We use the average of the shock and leading edge properties listed in the DONKI catalog to produce initial in situ arrival time predictions for each of the two CMEs using the ELEvo model (as described in Section~\ref{sec:pred_arrival}). Both CME arrival time predictions at L1 were then posted to the CCMC Scoreboard, \url{https://kauai.ccmc.gsfc.nasa.gov/CMEscoreboard/PreviousPredictions/2024}. The ELEvo model predicts that the later and faster CME would arrive at both Solar Orbiter and L1 earlier than the CME launched slightly earlier, and therefore we use the parameters of this CME to continue our real-time procedure. We note that this procedure excludes any interactions that would have occurred between the CMEs as the ELEvo model is not capable of modeling such CME-CME interactions, however, as the interaction took place close to the Sun, we continue our real-time procedure assuming a merged single structure driven by the faster CME.

\begin{table}[]
\caption{Input parameters to the ELEvo model and resulting arrival time and speed predictions for each event. Fixed ELEvo model parameters include the CME propagation longitude (in HEEQ coordinates), half-width, time at 21.5~R$_s$, and the ellipse aspect ratio. The initial ELEvo run is performed with the free parameters of CME propagation speed, ambient solar wind speed and drag parameter, and the resulting arrival time and speed predictions are listed for both Solar Orbiter and L1. The updated ELEvo run parameters, found by constraining the free model parameters to best match the observed CME shock arrival time at Solar Orbiter, are given again with the resulting updated arrival time and speed prediction at L1. These predictions can then be compared to the observed CME arrival time and speeds at Solar Orbiter and L1.}
\label{tab:elevo_parameters}
\resizebox{\textwidth}{!}{%
\begin{tabular}{lll}
 & \textbf{Event 1} & \textbf{Event 2} \\ \hline
\multicolumn{1}{l|}{{\ul Fixed ELEvo parameters:}} & \multicolumn{1}{l|}{} &  \\
\multicolumn{1}{l|}{Longitude [$^{\circ}$, HEEQ]} & \multicolumn{1}{l|}{-24} & 2 \\
\multicolumn{1}{l|}{Half-width [$^{\circ}$]} & \multicolumn{1}{l|}{33} & 41 \\
\multicolumn{1}{l|}{Time at 21.5~R$_s$} & \multicolumn{1}{l|}{2024-03-17 10:36} & 2024-03-23 03:19 \\
\multicolumn{1}{l|}{Ellipse aspect ratio} & \multicolumn{1}{l|}{1.43} & 1.43 \\ \hline
\multicolumn{1}{l|}{{\ul Initial ELEvo prediction:}} & \multicolumn{1}{l|}{} &  \\
\multicolumn{1}{l|}{CME prop. speed [kms$^{-1}$]} & \multicolumn{1}{l|}{528.5 $\pm$ 50} & 1613 $\pm$ 50 \\
\multicolumn{1}{l|}{Ambient SW speed [kms$^{-1}$]} & \multicolumn{1}{l|}{400 $\pm$ 50} & 400 $\pm$ 50 \\
\multicolumn{1}{l|}{Drag parameter [km${-1}$]} & \multicolumn{1}{l|}{0.1 $\pm$ 0.025 $\times 10^{-7}$} & 0.1 $\pm$ 0.025 $\times 10^{-7}$ \\
\multicolumn{1}{l|}{Solar Orbiter pred. arrival   time} & \multicolumn{1}{l|}{2024-03-18 16:34 $\pm$ 5.3 hrs} & 2024-03-23 11:39 $\pm$ 0.7 hrs \\
\multicolumn{1}{l|}{Solar Orbiter pred. arrival   speed  [kms$^{-1}$]} & \multicolumn{1}{l|}{512 $\pm$ 75} & 1307 $\pm$ 182 \\
\multicolumn{1}{l|}{L1 pred. arrival time} & \multicolumn{1}{l|}{2024-03-20 18:26 $\pm$ 13.3 hrs} & 2024-03-24 11:19 $\pm$ 5.4 hrs \\
\multicolumn{1}{l|}{L1 pred. arrival speed  [kms$^{-1}$]} & \multicolumn{1}{l|}{493 $\pm$ 75} & 963 $\pm$ 182 \\ \hline
\multicolumn{1}{l|}{{\ul Updated ELEvo prediction:}} & \multicolumn{1}{l|}{} &  \\
\multicolumn{1}{l|}{CME prop. speed  [kms$^{-1}$]} & \multicolumn{1}{l|}{519 $\pm$ 50} & 1313 $\pm$ 50 \\
\multicolumn{1}{l|}{Ambient SW speed  [kms$^{-1}$]} & \multicolumn{1}{l|}{398 $\pm$ 50} & 420 $\pm$ 50 \\
\multicolumn{1}{l|}{Drag parameter [km${-1}$]} & \multicolumn{1}{l|}{0.104 $\pm$ 0.025 $\times 10^{-7}$} & 0.114 $\pm$ 0.025 $\times 10^{-7}$ \\
\multicolumn{1}{l|}{L1 pred. arrival time} & \multicolumn{1}{l|}{2024-03-20 19:06 $\pm$ 13.8 hrs} & 2024-03-24 16:49 $\pm$ 5.4 hrs \\
\multicolumn{1}{l|}{L1 pred. arrival speed  [kms$^{-1}$]} & \multicolumn{1}{l|}{485 $\pm$ 75} & 829 $\pm$ 132 \\ \hline
\multicolumn{1}{l|}{{\ul Observed:}} & \multicolumn{1}{l|}{} &  \\
\multicolumn{1}{l|}{Solar Orbiter arrival time} & \multicolumn{1}{l|}{2024-03-18 16:55} & 2024-03-23 13:30 \\
\multicolumn{1}{l|}{Solar Orbiter arrival   speed  [kms$^{-1}$]} & \multicolumn{1}{l|}{-} & - \\
\multicolumn{1}{l|}{L1 arrival time} & \multicolumn{1}{l|}{2024-03-21 02:24} & 2024-03-24 14:11 \\
\multicolumn{1}{l|}{L1 arrival speed  [kms$^{-1}$]} & \multicolumn{1}{l|}{356} & 853
\end{tabular}%
}
\end{table}

\begin{figure}
\noindent\includegraphics[width=\textwidth]{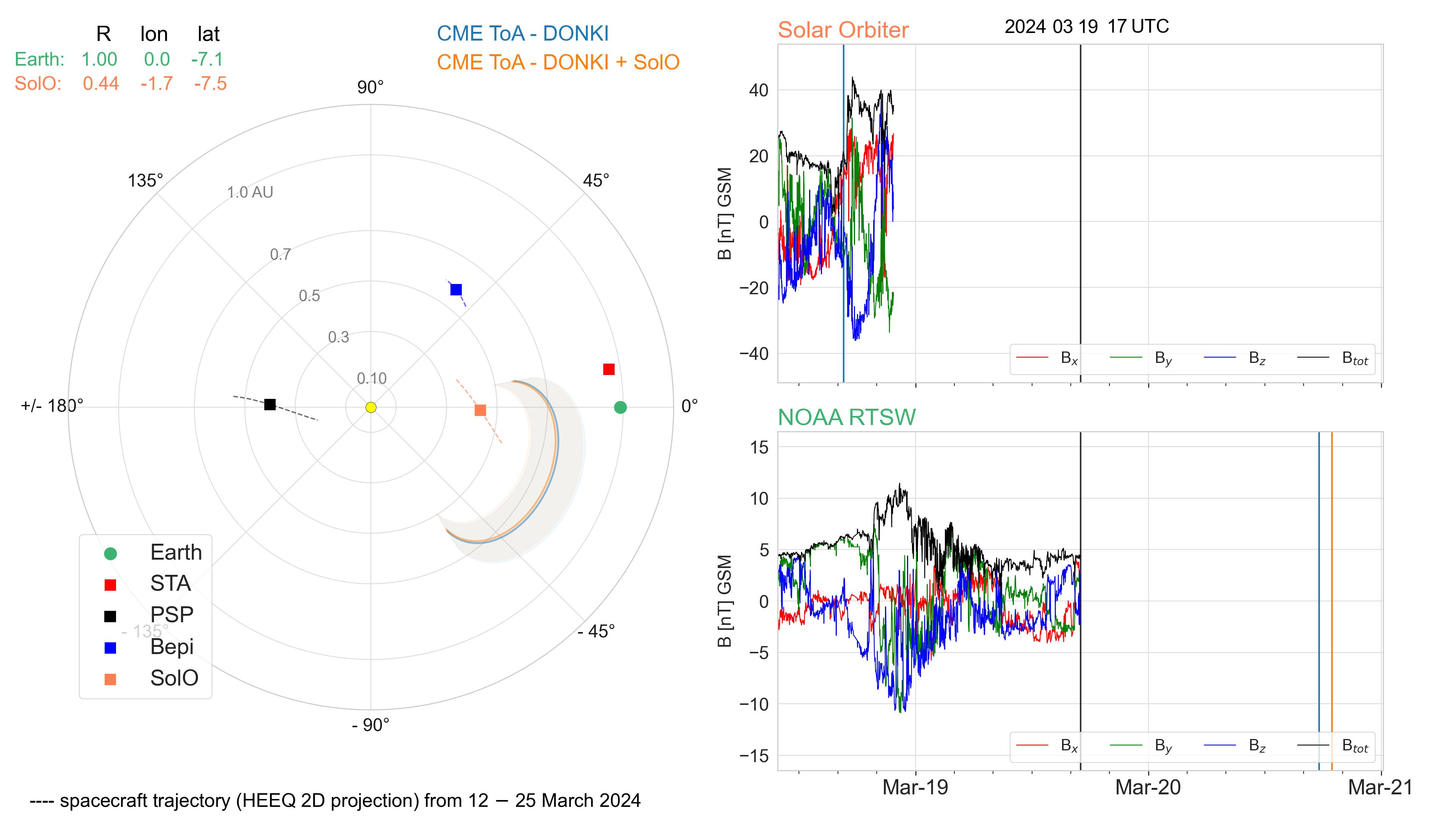}
\caption{Snapshot of the ELEvo model visualization at 2024 March 19 17:00~UT. Left: A top-down view of the solar equatorial plane with the spacecraft locations of Solar Orbiter (orange), BepiColombo (blue), Parker Solar Probe (black), STEREO-A (red) and Earth (green) shown in Heliocentric Earth Equatorial (HEEQ) coordinates. Their trajectories over the two-week window of this study (March 12--25) are indicated by the dashed lines in the corresponding color to the spacecraft marker. The heliocentric distance (R), longitude (lon), and latitude (lat) of Solar Orbiter and Earth are listed, showing that Solar Orbiter is well aligned with the Sun-Earth line during measurement of the CME and located $\sim~$0.56~au upstream of Earth. The propagation of the CME launched 2024 March 17 03:36~UT is represented by the elliptical fronts, where the shaded areas indicate the $\pm 1 \sigma$ uncertainties of the arrival time. The two fronts displayed correspond to two different arrival time predictions: The blue elliptical CME front represents the propagation of the CME using only DONKI kinematics producing predicted arrival times at both Solar Orbiter and L1. The orange front represents the second arrival time prediction, where the observed arrival time at Solar Orbiter is used to constrain the ELEvo model ensemble, producing an updated arrival time prediction at L1. Right: In situ magnetic field data in Geocentric Solar Magnetospheric (GSM) coordinates at Solar Orbiter (top panel) and the real-time solar wind (RTSW) data produced by NOAA at L1, available at the time the snapshot was taken (black vertical line). The predicted CME arrival times corresponding to the modeled fronts are represented by the blue (initial prediction) and orange (updated prediction) vertical lines.}
\label{fig:elevo_17march}
\end{figure}

The exact input parameters used to make the initial CME arrival time predictions at Solar Orbiter and L1 are given in Table~\ref{tab:elevo_parameters}, where the propagation longitude, half-width, time at 21.5~R$_\odot$, and ellipse aspect ratio remain fixed, and the CME propagation speed, ambient solar wind speed, and drag parameter are varied in the ensemble. Figure~\ref{fig:elevo_17march} presents a snapshot of the ELEvo model visualization on 17 March 2024 21:00~UT. The spacecraft positions (left) presented in Heliocentric Earth Equatorial (HEEQ) coordinates show that Solar Orbiter (orange marker) is well aligned with Earth (green marker) during the measurement of the CMEs, longitudinally separated by only -1.7 degrees at this time, while it was located at a heliocentric distance of 0.44~au. The blue elliptical CME front represents the propagation of the CME using the initial ELEvo input parameters described. The real-time magnetic field data available at both Solar Orbiter and L1 at the time of the snapshot (represented by the black vertical line) is presented on the right of Figure~\ref{fig:elevo_17march}. The resulting predicted arrival times from this initial run are represented by the blue vertical lines plotted over the real-time magnetic field data, with a predicted arrival time of 2024-03-18 16:34~UT $\pm$~5.3~h at Solar Orbiter and 2024-03-20 18:26~UT $\pm$~13.3~h at L1. The predicted arrival time at L1 was submitted to the CCMC Scoreboard on 2024-03-18 09:25~UT, before the CME arrived at Solar Orbiter. 

\begin{figure}
\noindent\includegraphics[width=0.9\textwidth]{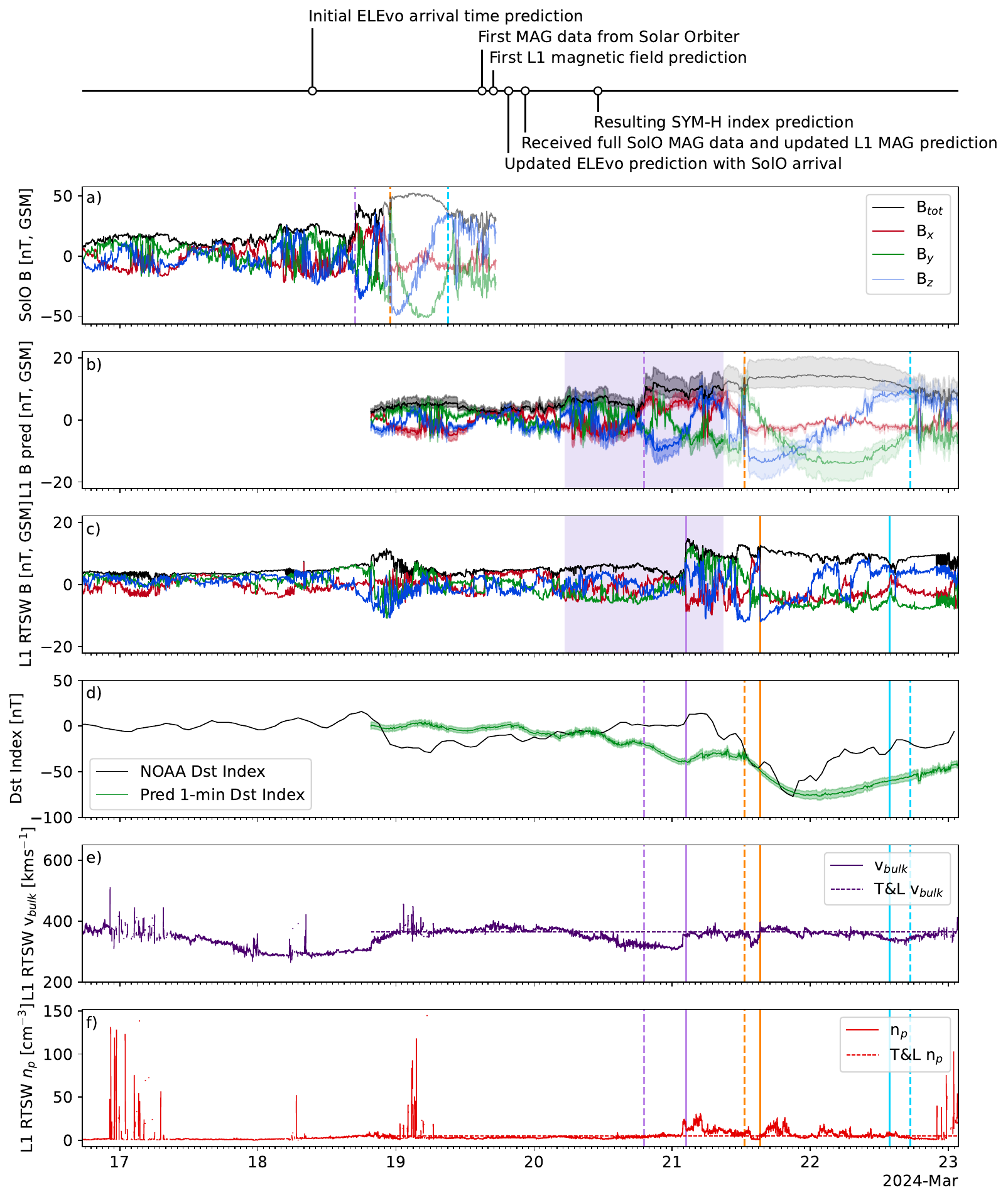}
\caption{\footnotesize{In situ CME and geomagnetic indices observations and predictions. Top: a timeline indicating when the steps of the real-time procedure were performed (times are also listed in Table~\ref{tab:timeline_event_1}). a) the observed Solar Orbiter magnetometer data, displayed in Geocentric Solar Magnetospheric (GSM) coordinates with the x, y, and z components in red, green and blue, respectively. The initial Solar Orbiter data is presented by the full color lines, with the data received later represented by fainter color lines. Purple, orange and light blue vertical dashed lines delineate the CME shock front, leading and trailing edge of the CME, respectively. b) the predicted magnetic field at L1 using scaled ($\alpha$ = -1.64) Solar Orbiter MAG observations as described in Section~\ref{sec:pred_structure}, in the same format as panel a. The associated shaded regions correspond to the uncertainty range, calculated by scaling the data using $\alpha$ = -2 for the lower bound and $\alpha$ = -1.2 for the upper bound. The same dashed vertical lines as panel a indicate where the predicted CME shock front, leading and trailing edge of the CME will occur at L1. These dashed vertical lines are carried through other panels to indicate the predicted CME boundaries where appropriate. The purple shaded region around the estimated time of arrival represents the model uncertainty. c) the observed real-time magnetic field data provided by the NOAA RTSW data product. Purple, orange and light blue vertical solid lines delineate the observed CME shock front, leading and trailing edge of the CME, respectively. These vertical lines are carried through other panels to indicate the observed boundaries where appropriate. d) the predicted 1-min Dst index produced by the Temerin~\&~Li model is shown by the green line, where the shaded green region represents the uncertainty of the model. The real-time Dst index is shown in black. e) the observed real-time solar wind speed and f) the real-time proton density provided by the NOAA RTSW data product, with dashed lines corresponding to the plasma profiles input to the Temerin~\&~Li model.}}
\label{fig:results_17march}
\end{figure}

Figure~\ref{fig:results_17march} presents the in situ CME and geomagnetic indices observations and predictions. We received the first Solar Orbiter magnetic field data on 2024-03-19 14:58~UT, presented in panel a of Figure~\ref{fig:results_17march}, having been converted from RTN to GSM coordinates. Here, the black line represents the total magnetic field strength, and the x, y, z components in red, green, and blue, respectively. The initial datafile provided contains magnetic field data between 2024-03-16 17:28~UT and 2024-03-18 21:44~UT, represented by the full color lines. From this initial data, we observe a shock arrival time of 2024-03-18 16:55~UT, delineated by the dashed vertical purple line, just 21 minutes later than the modeled estimate and well within our modeled uncertainty of 5.3~h. The remaining data reveals sheath-like magnetic field behavior, though some strong rotation of the transverse and normal components, indicating that this may also include the compressed earlier CME.

\begin{table}[]
\caption{Comparison between predictions and observations of CME parameters at L1 and geomagnetic indices at Earth. From top to bottom: the radial, latitudinal, and longitudinal spacecraft separations between Solar Orbiter and L1, the duration of the CME sheath region, the duration of the ME, the mean magnetic field magnitude in the sheath, the mean magnetic field magnitude in the ME, the minimum B$_z$ component value in the sheath, the minimum B$_z$ component value in the ME, and the minimum predicted 1-min Dst index vs the minimum Dst index.}
\label{tab:event_predictions}
\resizebox{\textwidth}{!}{%
\begin{tabular}{lll|ll}
 & \multicolumn{2}{c|}{\textbf{Event 1}} & \multicolumn{2}{c}{\textbf{Event 2}} \\
 & {\ul Predicted} & {\ul Observed} & {\ul Predicted} & {\ul Observed} \\
\multicolumn{1}{l|}{SC $\Delta$r [au], $\Delta$lat [$^{\circ}$], $\Delta$lon [$^{\circ}$]} &  & 0.53, 0.3, 6.3 &  & 0.60, 1.6, 8.5 \\
\multicolumn{1}{l|}{Sheath duration [hrs]} & 17.5 & 12.9 & 4.8 & 4.2 \\
\multicolumn{1}{l|}{ME duration [hrs]} & 28.8 & 22.5 & 14.8 & 15.6 \\
\multicolumn{1}{l|}{Mean sheath $|B|$ [nT]} & $10.0^{+4.1}_{-2.4}$  & 10.3 & $17.1^{+8.6}_{-4.8}$ & 20.6 \\
\multicolumn{1}{l|}{Mean ME $|B|$ [nT]} & $13.3^{+5.5}_{-3.3}$ & 9.2 & $23.2^{+11.8}_{-6.6}$ & 21.0 \\
\multicolumn{1}{l|}{Min. sheath B$_z$ [nT]} & -9.9 & -12.0 & -18.1 & -26.5 \\
\multicolumn{1}{l|}{Min. ME B$_z$ [nT]} & -13.3 & -12.0 & - & - \\
\multicolumn{1}{l|}{Min. Dst index [nT]} & -77 $\pm$ 4 & -77 & -88 $\pm$ 6 & -130
\end{tabular}%
}
\end{table}

We applied the magnetic field and temporal scaling laws as described in Section~\ref{sec:pred_structure} at 16:56~UT, less than two hours after receiving the data. Using the observed in situ CME shock arrival time at Solar Orbiter, we update the ELEvo parameters (as described in Section \ref{sec:pred_arrival} and listed in Table~\ref{tab:elevo_parameters}). We produce an updated CME arrival time at L1 of 2024-03-20 19:06~UT $\pm$ 13.8~h, 40 minutes later than the initial L1 arrival time prediction (2024-03-20 18:26~UT $\pm$~13.3~h). This updated ELEvo model run is represented by the orange elliptical front and vertical line in Figure~\ref{fig:elevo_17march}. The predicted in situ magnetic field structure is then shifted to align with the updated arrival time given by the ELEvo model, as shown by the full color lines in panel b of Figure~\ref{fig:results_17march}. 

We received the second Solar Orbiter datafile on 2024-03-19 22:28~UT, still well in advance of the updated predicted CME arrival time at L1. This data reveals the magnetic ejecta of the CME, as shown by the faded lines in panel a of Figure~\ref{fig:results_17march}, bounded by the orange and light blue dashed lines. The magnetic ejecta comprises a clear flux rope structure rotating from South to North, while the transverse component points West. We therefore discern that the flux rope type is SWN, and thus right-handed and low inclination, consistent with observations of the source region.

With the full CME data at Solar Orbiter, we updated our in situ magnetic structure prediction at L1, shown by the fainter lines in panel b of Figure~\ref{fig:results_17march}. This prediction was then input to the Temerin~\&~Li model, as described in Section~\ref{sec:pred_geomag}, to produce the predicted 1-min  Dst index shown by the green line in panel d of Figure~\ref{fig:results_17march}, where the uncertainties are represented by the shaded region. This prediction was made at 2024-03-20 11:06~UT, 8 hours in advance of the predicted CME arrival time. 

The real-time magnetic field, solar wind speed ($v_{bulk}$), and proton density data observed at L1 are presented in panel c, e, and f of Figure~\ref{fig:results_17march}, respectively. Table~\ref{tab:event_predictions} presents a comparison of the predictions and observations of CME parameters at L1 and geomagnetic indices at Earth. Directly comparing the predicted magnetic field structure to that observed at L1, we observe the true CME arrival time at L1 to be 2024-03-21 02:24~UT, 7.3~hours later than the updated arrival time prediction (and almost 8 hours later than the initial ELEvo arrival time prediction). The CME arrived with an observed speed of 356~km\,s$^{-1}$, slower than that produced by the updated ELEvo model run of 485~$\pm$~75~km\,s$^{-1}$. We observe that the temporal expansion was overestimated for our predicted structure in this case, with longer sheath (17.5~h) and magnetic ejecta (28.8~h) durations in comparison to those observed (12.9 and 22.5~h, respectively), as shown in Table~\ref{tab:event_predictions}. However, in terms of the global expansion, we slightly overestimate the mean magnetic field strength, with a predicted value of 13.3~nT compared with the observed 9.2~nT. Due to differences in sheath temporal expansion, the observed arrival time of the leading edge of the flux rope is very similar to that of the predicted structure: 2024-03-21 15:17~UT in comparison to 2024-03-21 12:36~UT. The overall structure of the magnetic ejecta observed at L1 displays a rotation similar to that of the predicted magnetic structure (also a SWN flux rope type), albeit less coherent, likely due to the interactions and other processes affecting the CME as it propagated between Solar Orbiter and L1.

The real-time Dst index obtained via NOAA SWPC is presented in black in panel d of Figure~\ref{fig:results_17march}. Here, we can see that the Dst index had returned to initial background levels following a disturbance late on March 18, prior to the arrival of the CME. After CME arrival, the Dst index decreases in three discrete steps, corresponding to the three periods of strong southwards magnetic field: two short periods in the sheath region, and the first part of the flux rope. Similarly, the Temerin~\&~Li model responds to the two main periods of southwards magnetic field in the predicted magnetic field structure: the first period in the sheath region, and the second at the leading edge of the flux rope, as shown by the two prolonged depressions in the predicted 1-min Dst index. Due to the predicted and observed leading edges occurring at similar times, with similar minimum southward magnetic field strengths (-13.3~nT and -12.0~nT, respectively), the time at which we observe the minimum Dst index aligns well with the minimum predicted 1-min Dst index. We observe a minimum real-time Dst index of -77~nT, within the uncertainties of the minimum predicted 1-min Dst index of -77$\pm$4~nT, showing that our predictions captured the strength of the geomagnetic storm very well. In addition, we compare the predicted Dst index profile with the real-time observed Dst index over the duration of the geomagnetic storm. To do so, we resample our predictions to 1-hour resolution (shown in Figure \ref{fig:l1_dst}) and calculate an RMSE of 31~nT. To put this result into context, we also retrospectively calculate the RMSE between the model output calculated using (1) the observed real-time values at L1 and the observed Dst index, and (2) science-quality Wind data and observed Dst indices, finding values of 14~nT and 16~nT, respectively. The minimum Dst index was observed at 2024-03-21 21:00~UT, 33.9 hours after the 1-min Dst predictions were made, showing that in this case, accurate predictions of the geomagnetic impact could be made using a far upstream monitor with a lead time much greater than current nowcasting capabilities at L1.

\subsection{Event 2: March 23, 2024} \label{sec:event2}

\begin{table}[]
\caption{Timeline of actions during the real-time procedure applied to Event 2 in the same format as Table \ref{tab:timeline_event_1}.}
\label{tab:timeline_eevent_2}
\resizebox{\textwidth}{!}{%
\begin{tabular}{llll}
\textbf{Action} & \textbf{Time} & \textbf{$\Delta$ t$_{L1}$} & \textbf{$\Delta$ t$_{Dst}$} \\
 & \textbf{UT} & \textbf{hrs} & \textbf{hrs} \\ \hline
CME launch observed & 2024-03-23 01:25 & 36.8 & 42.6 \\ \hline
\begin{tabular}[c]{@{}l@{}}Initial ELEvo arrival time predictions \\ for Solar Orbiter and L1 using DONKI input parameters (2)\end{tabular} & 2024-03-23 16:51 & 21.3 & 27.2 \\ \hline
Received first MAG data from Solar Orbiter & 2024-03-23 17:59 & 20.2 & 26.0 \\ \hline
Made first CME magnetic field structure prediction at L1 (4) & 2024-03-23 18:15 & 19.9 & 25.8 \\ \hline
\begin{tabular}[c]{@{}l@{}}Updated ELEvo arrival time prediction at L1 using \\ Solar Orbiter arrival time constraint (3)\end{tabular} & 2024-03-23 22:26 & 15.8 & 21.6 \\ \hline
Used Temerin \& Li model to produce 1-min Dst index prediction (5) & 2024-03-24 09:42 & 4.5 & 10.3 \\ \hline
Time CME (shock) arrived in situ at L1 & 2024-03-24 14:11 & - & 5.8 \\ \hline
Time observed minimum Dst index & 2024-03-24 20:00 & -5.8 & - \\ \hline
Received full Solar Orbiter MAG data & 2024-03-25 10:04 & -19.9 & -14.1 \\ \hline
\end{tabular}%
}
\end{table} 

\begin{figure}
    \centering
    {\includegraphics[width=\textwidth]{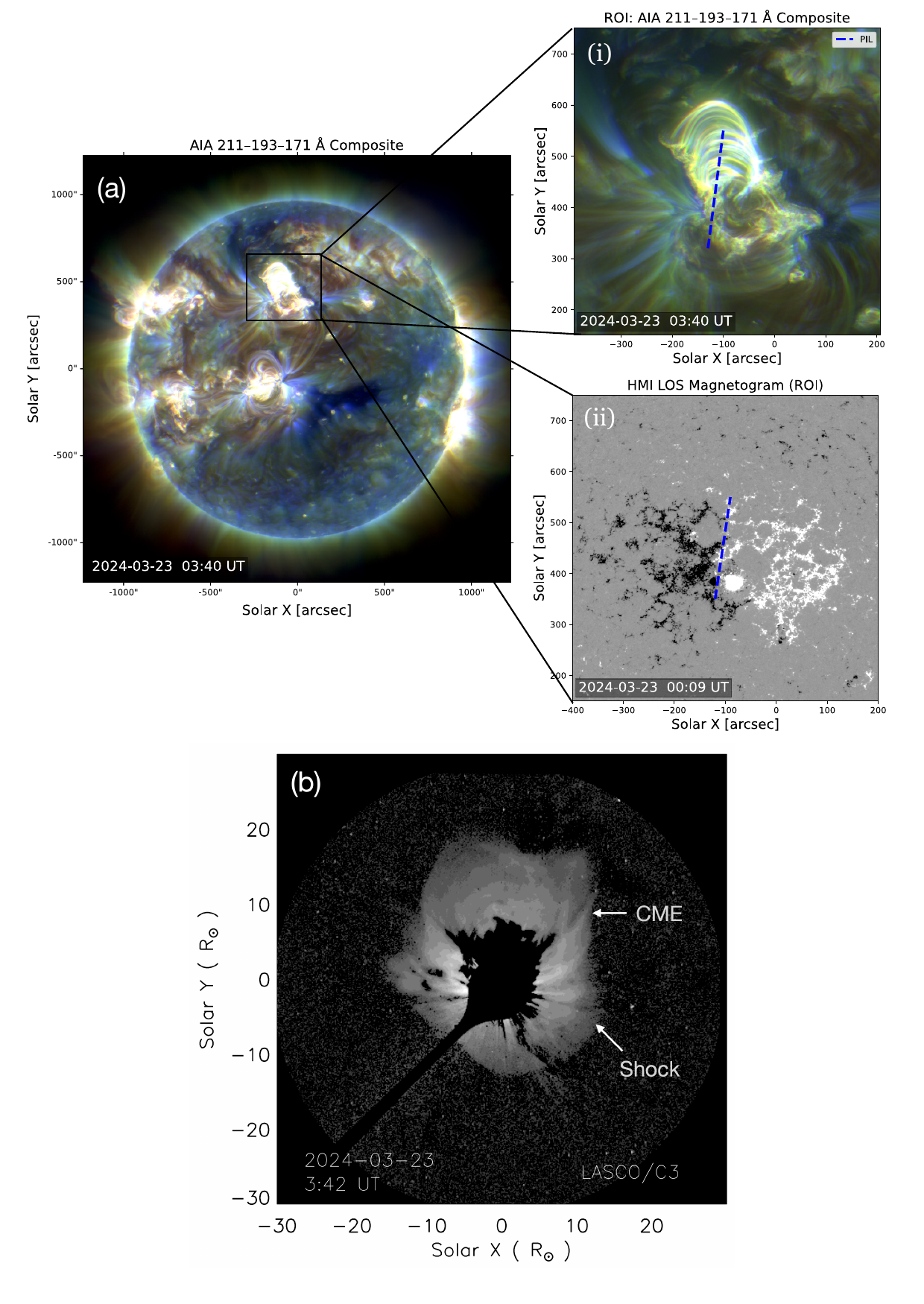} }
    \caption{a) Composite SDO/AIA 211, 193, and 171~Å image at 03:40~UT showing the source regions of the CMEs, where the black rectangle outlines the AR~3614 located around N25E07 (the source of the first and faster CME). Panel (i) presents a zoomed view of AR~3614 highlighting the skewed post-eruption loop system at 03:40~UT. Panel (ii) presents the SDO/HMI LOS magnetogram at 00:09~UT and the corresponding pre-eruption photospheric magnetic field context, with the polarity inversion line (PIL) traced in blue. The same PIL is overlaid in panel (i). The  b)SOHO LASCO C3 image at 2024-03-23 03:42~UT of the halo CME eruption where arrows indicate the shock and CME.}
    \label{fig:sdo_23march}
\end{figure}

As with Event 1, Event 2 on March 23 also comprised two CMEs launched in close succession. In contrast to Event 1, which consisted of slow CMEs, these CMEs were much faster, with a high potential for a strong geomagnetic storm to develop. Here, the first CME was launched on 2024-03-23 01:25~UT, associated with an X1.1 class solar flare from Active Region (AR) 3614, around N25E07. The second CME was launched 23~minutes later, at 01:48~UT. It was initially unclear whether the CME shared the same source region as the first CME, but was later attributed by the DONKI CCMC catalog to the concurrent flare from AR 3615 (S14E15). Both source regions are clearly visible in Figure~\ref{fig:sdo_23march}(a), which shows a composite SDO/AIA 211, 193, and 171~Å image at 2024-03-23 03:40~UT. To examine the first CME source region at N25E07 in greater detail, panel (i) presents a zoomed-in view highlighting the skewed post-eruption loop system. Panel (ii) shows the corresponding pre-eruption photospheric magnetic field context from the SDO/HMI LOS magnetogram at 0:09 UT, with the polarity inversion line (PIL) traced in blue; the same PIL is overlaid in panel (i) for reference.

The morphology of the post-eruption loops in panel (i), together with flare ribbon signatures observed in AIA 1600 Å (not shown), indicates the presence of a left-handed flux rope, consistent with the established hemispheric helicity rule. Given the strongly inclined PIL (approximately 90° relative to the solar equator) shown in panel (ii), the expected magnetic cloud configuration at 1~au would correspond to an ENW-type flux rope \cite{bothmer1998structure,palmerio2018coronal}. We also find that the source regions (especially the one at S14E15) is located adjacent to a sub-equatorial coronal hole in the southern hemisphere. It is worth noting that low-latitude coronal holes routinely generate high-speed solar wind streams \cite<>[]{krieger1973coronal,majumdar2025solwind}, which can modify CME kinematics. In particular, the enhanced background wind can alter the drag force acting on a CME \cite<>[]{gopalswamy2000interplanetary} or contribute to early-phase deflection of the CME trajectory \cite<>[]{kay2015deflect,majumdar2020kinem}. In this study however, we have not examined these effects in detail. In Figure~\ref{fig:sdo_23march}(b), a view of the CME in the LASCO C3 field of view is presented, with the shock and the leading edge of the CME indicated with arrows.

As in Section~\ref{sec:event1}, we obtain the CME kinematic parameters from the CCMC DONKI catalog. For the first CME, parameters were determined using only SOHO/LASCO C3 remote image data as there was a data gap in STEREO-A COR2 imagery. For the second CME, DONKI lists the parameters for both the fainter shock front using SOHO/LASCO C2, C3, and STEREO-A/SECCHI COR2 images, and the leading edge using SOHO/LASCO C3 and STEREO-A/SECCHI COR2 images. Both CMEs had very fast initial propagation speeds, with the leading edge of the first CME estimated to be 1613~km\,s$^{-1}$, and the shock front and leading edge of the second CME to be 1571~km\,s$^{-1}$ and 1572~km\,s$^{-1}$, respectively. The CMEs were launched almost directly towards the Earth, as shown by the halo CME in Figure~\ref{fig:sdo_23march}b, with estimated propagation directions of the leading edges (lat/lon) of 22$^{\circ}$/2$^{\circ}$ and 5$^{\circ}$/5$^{\circ}$, respectively. With such similar estimated propagation speeds and directions, it is again likely that the signatures observed in situ may be a combination of both CMEs. 

We use the ELEvo model with the estimated DONKI parameters to make initial arrival time predictions at L1 for both the CMEs: the first CME predicted to arrive at 2024-03-24 11:19~UT $\pm$ 5.4~h (posted to the CCMC Scoreboard at 2024-03-23 17:18~UT), and the second to arrive slightly later at 2024-03-24 12:30~UT $\pm$ 5.6~h (posted at 2024-03-23 17:21~UT). We therefore focus on the earlier and faster CME for our overall predictions. 

\begin{figure}
\noindent\includegraphics[width=\textwidth]{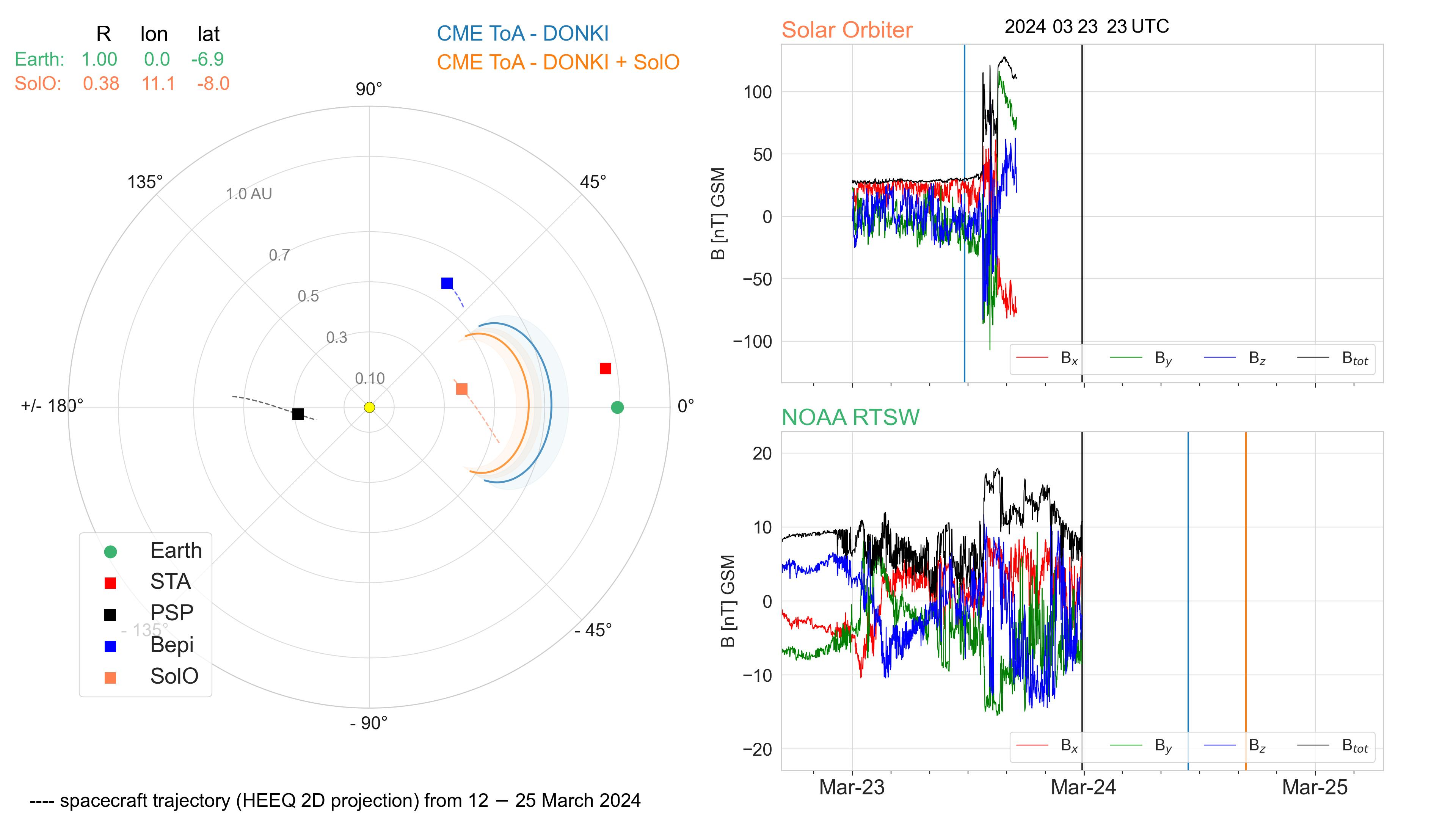}
\caption{Snapshot of the ELEvo model visualization at 2024 March 23 23:00~UT presented in the same format as Figure~\ref{fig:elevo_17march}. The spacecraft positions show that Solar Orbiter is located at a heliocentric distance of 0.38~au with a longitudinal separation of 11.1$^{\circ}$ with respect to the Sun-Earth line at the time of the snapshot.}
\label{fig:elevo_23march}
\end{figure}

Figure~\ref{fig:elevo_23march} presents the visualization of the ELEvo model on 23 March 2024 23:00~UT, with the input parameters listed in Table~\ref{tab:elevo_parameters}. As there is only one entry for the leading edge in the DONKI catalog for this CME, we note that these initial parameters are the same values as listed in the catalog. Solar Orbiter (orange marker) is fairly well aligned with Earth (green marker) during the measurement of the CMEs, longitudinally separated by -11.1$^{\circ}$, while it was located at a heliocentric distance of 0.38~au. The right side of Figure~\ref{fig:elevo_23march} presents the initial ELEvo arrival time predictions as vertical blue lines overlaying the in situ data at Solar Orbiter and L1 available at the time of the snapshot (marked by the vertical black line). We see that the initial ELEvo arrival time prediction at Solar Orbiter is 2024-03-23 11:39~UT $\pm$ 0.7~h, approximately 2 hours earlier than the observed arrival time of 2024-03-23 13:30~UT. 

\begin{figure}
\noindent\includegraphics[width=\textwidth]{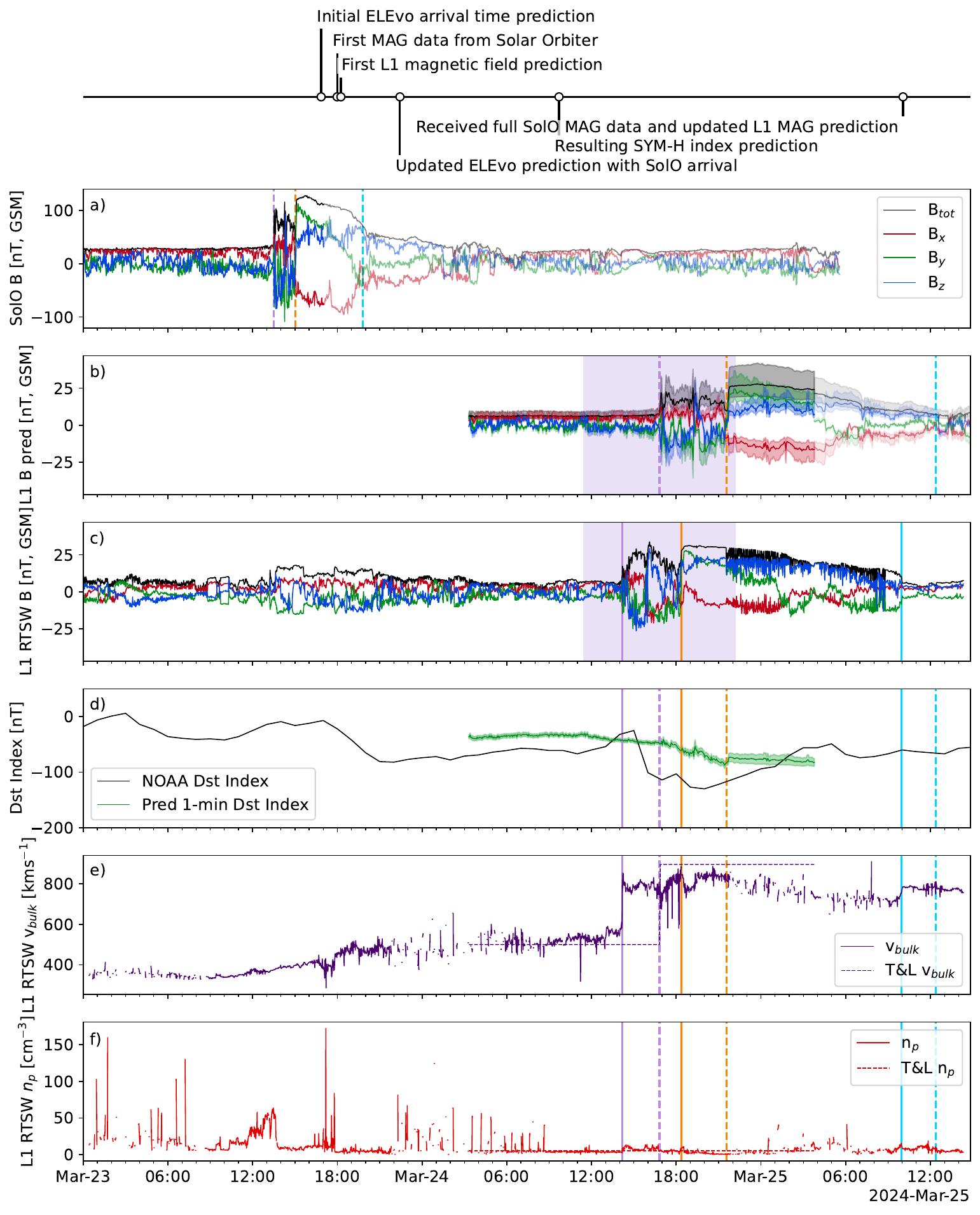}
\caption{In situ CME observations and predictions in the same format as Figure~\ref{fig:results_17march}.}
\label{fig:results_23march}
\end{figure}

Similarly to Event 1, Figure~\ref{fig:results_23march} presents the in situ CME and geomagnetic indices observations and predictions. We received the initial Solar Orbiter MAG datafiles at 2024-03-23 17:59~UT, obtaining an early glimpse of the in situ magnetic structure of the CME up to 23 March 17:02~UT, as shown by the solid color lines in panel a of Figure~\ref{fig:results_23march}. The shock of the CME arrives on 23 March 13:30~UT at Solar Orbiter, approximately 27.5~hours earlier than the predicted arrival time at L1, giving a possible lead time of over one day. The shock is followed by the turbulent magnetic field of the sheath region which includes a relatively strong southward component, before an increase in magnetic field strength at 15:02~UT indicates the arrival of the smoothly rotating magnetic flux rope. The initial structure of the flux rope, with a steadily northward pointing magnetic field, and a dominant, eastward pointing, transverse magnetic field component that has started to rotate, indicates a highly inclined flux rope with respect to the ecliptic plane (ENW flux rope type, matching the deduction from the solar source region), from which we expect a low geomagnetic impact from the flux rope alone. From these observations, we can deduce that much of the geomagnetic impact will be a result of the CME sheath region.

We make our first prediction of the CME magnetic structure at 2024-03-23 18:15~UT, 16 minutes after receiving the Solar Orbiter MAG files, using the same scaling laws described in Section~\ref{sec:pred_structure}. As before, we then constrain the ELEvo model parameters using the observed in situ arrival time at Solar Orbiter (where the updated ELEvo parameter set is given in Table~\ref{tab:elevo_parameters}), to produce an updated L1 arrival time prediction of 2024-03-24 16:49~UT $\pm$ 5.4~h (orange elliptical front of Figure~\ref{fig:elevo_23march}), 5.5~hours later than the initial ELEvo arrival time prediction at L1. The predicted in situ CME magnetic field structure (using the data available from the Solar Orbiter MAG files received at this point) is shown by the full color lines in panel b of Figure~\ref{fig:results_23march}, shifted to the updated L1 arrival time given by the ELEvo model. 

As in Figure~\ref{fig:results_17march}, the lower four panels of Figure~\ref{fig:results_23march} present the real-time data given by NOAA. From these measurements, we observe that the Earth was already under the influence of higher speed solar wind ($\sim$~500~km\,s$^{-1}$ compared to usual ambient solar wind speeds of 300--350 km\,s$^{-1}$) associated with the coronal hole observed in remote images, which had resulted in a geomagnetic response with the Dst index decreasing to a minimum value of -82~nT, recovering very slowly. Based on this slow rate of recovery, we projected that the Dst index would recover to $\sim$~-40~nT around the time of predicted CME arrival at L1. We therefore take this into account when making our geomagnetic index predictions: as we input only the predicted magnetic structure based on Solar Orbiter magnetic field observations (panel b) and the idealized solar wind speed and density profiles to the Temerin~\&~Li (represented by the dashed lines in panels e and f, respectively), the model does not have enough information about the prior conditions to automatically take these into account. We therefore use the projected recovery value of -40~nT as a baseline instead of the assumed 0~nT, manually shifting the resulting model output to attempt to capture a more realistic geomagnetic impact due to the ongoing disturbed conditions.

The resulting 1-min resolution Dst index prediction is shown by the green line in panel d of Figure~\ref{fig:results_23march}. This prediction was made at 2024-03-24 09:42~UT, 7.1~hours in advance of the predicted CME arrival time. However, the shock of the CME was measured in situ at L1 on 24 March 14:08 UT (Figure~\ref{fig:results_23march}, panel c), 2.6~hours earlier than the updated ELEvo model prediction (but 2.9~hours later than the initial arrival time prediction), reducing the lead time of the geomagnetic index prediction. The true time difference between making the 1-min Dst prediction and observing the CME shock at L1 is therefore 4.5 hours, as listed in Table~\ref{tab:timeline_eevent_2}. The observed CME arrival time lies between the initial and updated ELEvo arrival time predictions, arriving with an observed speed of 853~km\,s$^{-1}$, between the two model estimations of 963~$\pm$~182~km\,s$^{-1}$ (inital ELEvo parameters) and 829~$\pm$~132~km\,s$^{-1}$ (updated ELEvo parameters), as listed in Table~\ref{tab:elevo_parameters}. 

The real-time magnetic field measurements at L1 presented in Figure~\ref{fig:results_23march} reveal an in situ CME structure remarkably similar to that predicted using the upstream Solar Orbiter data. Table~\ref{tab:event_predictions} lists the predicted and observed in situ CME parameters at L1 and the geomagnetic indices for Event 2. The observed CME sheath had a duration of 4.2 hours in comparison to 4.8 hours for the predicted structure. The sheath of both the observed and predicted structure have two periods of southward magnetic field, leading to a double dip in the observed Dst index, for which the same profile was replicated well by the Temerin~\&~Li model. However, the first region of southward pointing magnetic field had a stronger minimum B$_z$ value than the predicted magnetic structure (observed: -26.5~nT, predicted: -18.1~nT), driving a stronger initial geomagnetic response than the model prediction, in addition to the already disturbed geomagnetic conditions. Although there was less disparity between minimum B$_z$ values for the second period of southward field in the sheath (observed: -12.2~nT, predicted: -10.0~nT), from the L1 observations we see that this period is coincident with another increase in speed and density, not accounted for by the Temerin~\&~Li model inputs, again driving a second stronger geomagnetic response. The minimum observed Dst index was -130~nT, outside of the predicted 1-min Dst index range of -88$\pm$6~nT. In this case, the 1-min Dst prediction was made 10.3 hours in advance of the minimum observed Dst index, again increasing warning time in comparison to nowcasting capabilities based on L1 in situ data, but perhaps too far in advance to fully capture the already ongoing geomagnetic activity at Earth. However, if we compare the modeled Dst profiles for both the predictions using Solar Orbiter data and those using the observed L1 RTSW data with the real-time observed Dst index (Figure~\ref{fig:l1_dst}), we find RMSE values of 40~nT and 38~nT, respectively, showing that inputting the observed L1 RTSW data did not lead to significant modeled improvements in this case. When comparing Dst indices modeled from science-quality Wind data to observed indices, the RMSE amounts to 19~nT, which is significantly better than when calculating the Dst from L1 RTSW data and comparable to the corresponding RMSE of 16~nT for Event 1 (see Section~\ref{sec:event1}). This suggests that the model itself introduces an error of approximately 15–20~nT to the predicted Dst profile.

We received the full Solar Orbiter magnetic field observations at 2024-03-25 10:04~UT, revealing the full structure of the CME at Solar Orbiter (Figure~\ref{fig:results_23march}, fainter color lines in panel a): an ENW type flux rope, as predicted. After the full CME had passed over L1 (the end of the magnetic flux rope was observed at 2024-03-25 09:57~UT), a comparison between the real-time observations at L1 and a hindcast of the full predicted CME structure could be made (fainter color lines of panel b, Figure~\ref{fig:results_23march}). The hindcasted and observed magnetic structures present the same pattern of magnetic field rotation, similar magnetic flux rope durations (observed: 15.6~h, predicted: 14.8~h), and similar mean magnetic field strengths (observed: 21.0~nT, predicted: 23.2~nT). In this case, we therefore show that it is possible to make good predictions of the in situ magnetic field structure, with actionable lead times, using far upstream spacecraft, even with longitudinal separations up to 10$^{\circ}$ from the Sun-Earth line. However, despite the similarity of the magnetic field predictions to observations, the Temerin~\&~Li model failed to capture the magnitude of the geomagnetic storm, suggesting that real-time plasma data may be essential for better forecasting the geomagnetic response; we discuss this further in Section~\ref{sec:discussion_geomag}. 

\section{Discussion and Recommendations} \label{sec:discussion}

In this study, our aim is to investigate whether it is possible to make accurate predictions of CME in situ observations near-Earth and their geomagnetic effects using far upstream solar wind monitor observations in real time, and by doing so, extend the lead time of such predictions compared to current L1-based capabilities. To do so, our methodology makes a series of assumptions, with each step building on the next: from arrival time prediction with the ELEvo model, to statistical relationships used to scale the observed magnetic field data between Solar Orbiter and L1, and then the inputs needed to produce a final 1-min resolutin Dst index prediction with the Temerin~\&~Li model.

\subsection{Solar Context and CME Arrival Time} \label{sec:discussion_solar_time}

As previously described in Section~\ref{sec:pred_arrival}, the ELEvo model used to predict CME arrival times in this study is a simple drag based model and therefore does not take into account a perhaps more realistic solar wind background through which the CMEs propagate. In the case of the two events in this study, it is clear from the in situ data observed at L1 that higher speed streams were present in the days before the CME events arrived at L1 (Panel e of Figures~\ref{fig:results_17march} and \ref{fig:results_23march}) that could have affected CME arrival times (and further, the geomagnetic impact of the CME events having already disturbed geomagnetic conditions). The solar context for the high speed streams observed in situ is most clear for Event 2, where a sub-equatorial coronal hole is clearly observed in the SDO AIA composite images (Figure~\ref{fig:sdo_23march}a). Both events are further complicated by the fact that they comprise two CMEs launched in close succession, and therefore, CME-CME interactions may also affect arrival time predictions. For these cases, would using a model with a solar wind background have improved predicted arrival times at L1? We investigate this by comparing the L1 arrival time predictions made by ELEvo with those posted to the CCMC Scoreboard by the NASA M2M Office using the WSA-Enlil+Cone model: for Event 1, the M2M Office posted two predictions for the first CME launched at 2024-03-17~03:12~UT (2024-03-21 07:11~UT and 2024-03-21 05:53~UT) and two predictions for the second CME launched at 2024-03-17~03:36~UT (2024-03-21~01:00~UT and 2024-03-21~05:53~UT), with differences to the observed time of +4.8, +3.5, -1.4 and +3.5 hours, respectively. The initial predicted arrival time produced by the ELEvo model in this study was 2024-03-20 18:26 UT, with a difference of -8.0 hours with respect to the observed in situ L1 arrival time. In this case, the WSA-Enlil+Cone model predictions were closer to the observed L1 arrival time than that produced by the ELEvo model. For Event 2, the M2M office posted one arrival time prediction for each of the two CMEs launched 2024-03-23 01:25~UT and 2024-03-23 01:48~UT, of 2024-03-24 17:55~UT (the same arrival time prediction attached to the two different CMEs), with a difference of +3.7 hours with respect to the observed arrival time in situ at L1. For this event, the initial arrival time prediction at L1 produced by the ELEvo model is 2024-03-24 11:19, -2.8 hours to the observed arrival time. In this case, the ELEvo model initial prediction was closer to the observed L1 arrival time than that produced by the WSA-Enlil+Cone model.

These arrival time prediction comparisons discussed above do not account for potential CME-CME interactions as the CMEs were modeled as individual CME pulses/fronts. Models of CME propagation based on kinematics derived from the remote images allow estimates of when the CME interactions may occur: In the case of Event 1, the interaction between the two CMEs likely occurs quickly after the second faster CME was launched, and in the case of Event 2, the faster CME is launched first with the slower CME following so interactions may be less of an issue with regard to affecting arrival time. In this study, in situ Solar Orbiter observations allow us to confirm not only whether the CME-CME interactions have already occurred, but also observe their effect on the magnetic structure. Due to the early interaction in the case of Event 1, they also present a clear advantage when it comes to constraining the ELEvo input parameters, and allow us to confidently model the CME as one front in the updated ELEvo prediction run due to the interaction having already taken place forming one complex structure. If we compare the updated arrival time predictions at L1 for Event 1 (2024-03-20 19:06~UT) and Event 2 (2024-03-24 16:49), we find a time difference to the observed CME arrival time at L1 for Event 1 of -7.3 hours and Event 2 of +2.6 hours. For both events in this study, this presents an improvement over the initial predictions made with only inputs based on the remote images obtained from the DONKI catalog (-7.3 vs. -8.0, and +2.6 vs. -2.8 hours). 

As noted above, the ELEvo model is not able to model CME-CME interactions, which in this case is mitigated by the far upstream measurements provided by Solar Orbiter. It is, however, possible to run the WSA-Enlil+Cone model with multiple CMEs as input. For the case of Event 1, the CMEs were modeled individually by the NASA M2M Office with the WSA-Enlil+Cone model. However, for Event 2, the NASA M2M Office later posted an ensemble prediction modeling both CME pulses, producing a predicted arrival time at L1 of 2024-03-24~22:32~UT. Here, the difference between predicted and observed arrival time is +8.4 hours, compared to the difference of +2.6 hours produced by the updated ELEvo model prediction for Event 2 (and -2.8 hours produced by the initial ELEvo model prediction). In this case, despite modeling the CME-CME interaction, the WSA-Enlil+Cone model prediction is further from the observed arrival time than that of the ELEvo model. We conclude that the large uncertainties involved in reconstructing the CMEs and the background through which they propagate may have a greater effect on arrival time error than any model-driven systematic errors, similarly to the conclusions of \citeA{kay2024updating}. These results suggest that one major advantage of upstream monitors is that even simpler models can be observationally constrained to produce improved arrival time estimates.

\subsection{In Situ CME Structure} \label{sec:discussion_structure}

The next assumptions made in our real-time procedure relate to the in situ magnetic field structure of the CME events, and how this may evolve between Solar Orbiter and L1. Due the complex nature of CME evolution, we expect the predicted magnetic field structures at L1 to look different from those observed in situ. We therefore aim to assess to what extent general assumptions hold and whether these can still lead to good predictions of the in situ magnetic field structure at L1. The main assumption made to scale the magnetic field data is that the CME comprises a force-free self-similarly expanding flux rope. This assumption is held by most forward-models of CME propagation \cite<e.g.,>[]{weiss2021analysis,ruedisser2024understanding} and found to explain in situ observations of CME expansion between spacecraft close to radial alignment \cite{good2018correlation}. However, previous multi-spacecraft studies in which in situ observations of CMEs are made by spacecraft over wider longitudinal and radial separations are hard to reconcile with the assumption of self-similar expansion \cite{weiss2021triple,davies2024fluxrope}. Thereby focusing on the two-week window within which Solar Orbiter was longitudinally separated by $\pm$15$^{\circ}$ upstream of Earth (a smaller separation in comparison to \citeA{davies2024fluxrope}), we trust self-similar expansion to be a reasonable assumption in this study. 

As mentioned in Section~\ref{sec:pred_structure}, the theoretical steepest rate of expansion ($\alpha$) of the axial magnetic field strength for a self-similarly expanding cylindrical force-free magnetic flux rope would follow $B \propto r^{-2}$ \cite{farrugia1993study}. Statistical studies of CME properties observed in situ have found a range of $\alpha$ values, from steeper values in the inner heliosphere of $\sim$-1.9 \cite{gulisano2010global, winslow2015interplanetary} to $\sim$-1.2 beyond 1~au \cite{richardson2014identification, davies2021catalogue}. As shown by previous studies, individual CMEs vary in their evolution from statistical trends \cite{salman2020radial, davies2022multi}, with $\alpha$ values sometimes greater than -2 due to magnetic erosion of the CME \cite{salman2020radial}, or even positive due to longitudinal separation differences \cite{davies2022multi}. By using a wide uncertainty range in our scaling of the magnetic field values in this study, we attempted to account for most of the global variation expected for a non-perturbed magnetic flux rope. Retrospective investigation of the power laws for each event shows that for Event 1, $B \propto r^{-2.11}$, and for Event 2, $B \propto r^{-1.75}$. The $\alpha$ value for Event 2 is close to the employed value of -1.64 \cite{leitner2007consequences}, and is within the uncertainty range used in this study. However, for Event 1, the $\alpha$ value of -2.11 is outside of our uncertainty range and, as mentioned, may be steeper than expected due to magnetic erosion between the interacting CMEs in this case. Therefore, for cases of CME-CME interaction, a wider uncertainty range of $\alpha$ values may be necessary to more accurately capture the additional variability introduced and the resulting uncertainty of the predicted magnetic field magnitude of CMEs at L1 when scaling data from upstream solar wind monitors.

To scale the magnetic field data temporally, we also assume an idealized relationship between the global expansion ($\alpha$) and the local expansion (dimensionless expansion parameter, $\zeta$), and use a value of $\zeta = 0.8$ assuming the CMEs are non-perturbed magnetic clouds \cite{gulisano2010global}. We emphasize that we used the ideal non-perturbed magnetic cloud $\zeta$ value to temporally scale the data in this study due to the availability of only magnetic field data at Solar Orbiter. However, if solar wind plasma data is available, the dimensionless expansion parameter can be calculated using Equation~\ref{eq:expansion},

\begin{equation} \label{eq:expansion}
    \zeta = \dfrac{\Delta V}{\Delta t} \dfrac{d}{V_c^2},
\end{equation}

\noindent where $V_c$ is the speed at the mid-point of the ME (also known as the cruise velocity), $\Delta V/\Delta t$ is the rate of expansion, and $d$ is the heliocentric distance at which the mid-point of the ME is measured. The value of $\zeta$ used in this study fits well with the $\alpha$ value of -1.64 used to scale the magnetic field values as, ideally, $\alpha \approx 2\zeta$. However, previous multi-spacecraft studies have found a weak relationship between the local and global expansion \cite{lugaz2020inconsistencies, davies2022multi}. Similarly as above, using Equation~\ref{eq:expansion} we retrospectively calculate the values of $\zeta$, using both the real-time L1 data and the science-quality Solar Orbiter solar wind plasma data \cite{owen2020swa} now available, to investigate whether this idealized relationship holds for the events of this study. Considering a calculated $\alpha$ value of -2.11 for Event 1, one would expect a local expansion value of $\zeta = 1.06$ (positive half of $\alpha$); instead, we calculate a value of $\zeta = 0.72$ at Solar Orbiter and $\zeta = 0.56$ at L1. Similarly, for Event 2, one would expect a local expansion value of $\zeta = 0.88$ ($\alpha = -1.75$), whereas we calculate a value of $\zeta = 0.46$ at Solar Orbiter and $\zeta = 0.36$ at L1. These results show that the local expansion is not consistent at each spacecraft for the same event, and therefore, it is perhaps unsurprising that the values do not follow the idealized relationship with global expansion. This result also has the implication that even if solar wind plasma data was available at Solar Orbiter to calculate and use this value of $\zeta$ instead of the ideal non-perturbed magnetic cloud $\zeta$ value of 0.8, the local expansion of the CME at L1 may still be different from when it was measured at Solar Orbiter.

Similarly, we can also use the solar wind plasma data now available to retrospectively investigate the impact of using Equation~\ref{eq:temp_scaling} to estimate the duration of the ME, rather than its radial width for which it was designed. For Event 1, the radial width of the ME at Solar Orbiter is 0.089~au and 0.195~au at L1, giving a $\zeta$ of 1.0 (instead of the $\zeta$ = 0.8 assumed in Equation~\ref{eq:temp_scaling}). This value is higher than both the calculated dimensionless expansion parameters ($\zeta = 0.72$ at Solar Orbiter and $\zeta = 0.56$ at L1) but very similar to half of the global expansion power ($\alpha$ = -2.11). If we consider the observed durations for Event 1, we find a power of 1.03. This is very similar to the calculated power based on the radial widths, therefore in the case of Event 1, using the duration rather than the radial width did not affect the results significantly. For Event 2, we calculate a ME radial width of 0.105~au at Solar Orbiter and 0.289~au at L1, giving a $\zeta$ of 1.08, larger than half the global expansion value ($\alpha = -1.75$), and the calculated local expansion values ($\zeta = 0.46$ at Solar Orbiter and $\zeta = 0.36$ at L1). Similarly, if we consider the observed durations for Event 2, we find a power of 1.27 which is slightly larger than that found considering the radial widths. This difference for Event 2 compared to the very similar values for Event 1 is likely due to its faster propagation speed, demonstrating the value of having real-time in situ speed measurements, especially for faster CMEs. 

The expansion scaling laws used in this study are derived from previous statistical relationships that focus on the ME, and are therefore not necessarily appropriate to scale the magnetic field values within the sheath region. The evolution of the sheath region and the ME are different, with the sheath region also growing due to accumulation of the solar wind in front of the CME as the shock propagates. The sheath evolution is further complicated in this study due to the CME-CME interactions for both events, especially in the case of Event 1 where the preceding CME was likely compressed into the sheath region preceding the second faster CME. Calculating the observed global expansion powers, we find $\alpha = -1.60$ for Event 1, and $\alpha = -1.43$ for Event 2. These values are not outside of the wide uncertainty range used in the scaling (-1.2 to -2) of the magnetic field data despite being based on power laws derived from ME observations. The calculated Event 1 $\alpha$ is very close to that used in this study of -1.64, hence the similarity between the observed and predicted mean magnetic field strength within the sheath for this event, whereas the higher calculated Event 2 $\alpha$ is why the mean magnetic field strength was underestimated in this case. 

Focusing on the temporal expansion, we also investigate the durations of the sheaths finding that the sheath of Event 1 increases with heliocentric distances as $D \propto r^{0.97}$, and Event 2 increases as $D \propto r^{1.08}$ between Solar Orbiter and L1. Instead, if we calculate the radial width using the mean speed measured in situ within the sheath, we find the radial width for Event 1 also increases with heliocentric distances as $r^{0.97}$ and for Event 2 as $r^{0.89}$ between Solar Orbiter and L1. As for the ME, the difference between powers for the radial width and duration is greater for the faster Event 2, whereas there is almost no difference for the slower Event 1. Comparing the sheath values to those derived for the ME, the values for Event 1 are very similar, whereas for Event 2, the sheath expands at a slower rate than the ME, as can be seen by the overestimation of the predicted sheath duration and the underestimation of the predicted ME duration for Event 2. 

A recent study conducted by \citeA{larrodera2024sheaths} studied the evolution of both the radial size of ME regions and their sheaths over varying heliocentric distance ranges between 0.25 and 5.42~au. For ME regions (that had associated sheath regions), they found the size of the ME scales as $r^{1.15\pm0.33}$ between 0.25--5.42~au and $r^{2.42\pm0.78}$ between 0.25--0.99~au. The values found in this study fall within the uncertainty range of the 0.25--5.42~au relationship, but are far lower than the relationship found between 0.25--0.99~au. For the sheath region, \citeA{larrodera2024sheaths} found the radial width to increase as $r^{0.80\pm0.37}$ between 0.25--5.42~au and $r^{1.72\pm0.74}$ between 0.25--0.99~au. Again, the values calculated for the sheath expansion in this study fall within the uncertainty range for the relationship over the wider heliocentric distance range of 0.25--5.42~au, but are at the lower limit of the uncertainty range for the 0.25--0.99~au relationship. As with the global expansion, it is clear there are also large variations of local and temporal expansions for individual events (and for different regions, i.e., sheath and ME, within the CME), compounding uncertainties of arrival time with duration of the sheath and ME, increasing the uncertainties around the predicted observations including key parameters (e.g., time of minimum B$_z$). 

\subsection{Geomagnetic Indices} \label{sec:discussion_geomag}

The final set of assumptions made in our real-time procedure relate to the Temerin~\&~Li model and its inputs: With respect to the magnetic field strengths and durations of the sheath and ME regions of Event 1, both the durations were overestimated, however, the minimum B$_z$ was not as low as observed within the sheath region and lower than observed within the ME, perhaps balancing out when it came to the final 1-min Dst prediction produced by the model which matched observations well. In the case of Event 2, the negative magnetic field within the sheath region provided the main contribution to the geomagnetic storm, where despite estimating the duration of the sheath well, the minimum B$_z$ was not as low as observed, leading to underestimating the 1-min Dst index in this case (in combination with the already disturbed geomagnetic conditions). However, the Temerin~\&~Li model also takes the density and speed profiles across the event as input, for which real-time solar wind plasma data was unavailable at the time of this study. 

As described in Section~\ref{sec:pred_geomag}, we take a constant estimated density of 5~cm$^{-3}$ prior to and during the events in lieu of in situ measurements, as without these, it is not possible to estimate realistic sheath or ME densities. We note that if we did have real-time plasma density data available at Solar Orbiter, theoretical scaling laws $ \propto r^{-3}$ could have been used to make estimates of the density at L1 \cite<for magnetic clouds $<$~1~au;>[]{kumar1996magnetic}, however, those based on observed in situ data follow closer to $r^{-2}$ over large heliocentric distance separations \cite{liu2005statistical, richardson2014identification}. For Event 1, prior to CME arrival at L1, the mean in situ proton density was measured as 5.4~cm$^{-3}$, increasing to a mean of 10.7~cm$^{-3}$ within the sheath and 7.7~cm$^{-3}$ within the ME, showing that for this event, inputting a stepped density profile may have been more realistic. However, for Event 2, the mean observed in situ values prior to CME arrival are 6.1~cm$^{-3}$, 6.9~cm$^{-3}$ within the sheath, and 5.5~cm$^{-3}$ within the ME, therefore remaining fairly constant. For the speed profiles input into the Temerin~\&~Li model, we created a stepped profile only for faster events, where modeled L1 arrival speeds given by ELEvo were much greater than the constant background speed of the model. This was a sensible assumption in the case of Event 1, considered a slower event, so a constant speed of 365~km~s$^{-1}$ was used. The observed mean values for Event 1 were fairly constant throughout the event: prior to CME arrival, mean speeds were 321~km~s$^{-1}$, and the mean observed speeds within the sheath region and ME were 354.4~km~s$^{-1}$ and 361~km~s$^{-1}$, respectively. Event 2 was predicted to be much faster, so a stepped profile was employed from the observed 500~km~s$^{-1}$ prior to the prediction up to the average of the two estimated L1 arrival speeds given by ELEvo of 896~km~s$^{-1}$. The true observed mean speed prior to CME arrival at L1 was 524~km~s$^{-1}$, with mean speeds of 788~km~s$^{-1}$ and 774~km~s$^{-1}$ throughout the sheath and ME regions, respectively. 

Based on these observed mean values for the density and speed we create stepped profiles to investigate how the Temerin~\&~Li model output is affected and whether using the mean measured values with our predicted in situ magnetic field and durations would have improved the estimated minimum 1-min Dst index. We find that for Event 1, the model estimates a minimum 1-min Dst index of -69~nT, not as low as our original predicted value of -77~nT; despite the density increase that was not used in our initial inputs, the lower observed speeds produce a higher minimum 1-min Dst index value, further from the minimum observed Dst index. For Event 2, the model estimates a minimum 1-min Dst index of -85~nT, not as low as the original predicted value of -88~nT but within its uncertainty range. Similarly to Event 1, the slight density change seems to have little effect, and the lesser step in speed values drives less of a response in the model. We note that the speed and density profiles input are still idealized and that the magnetic field strength and orientation seems to be the dominant input factor for the Temerin~\&~Li model. Perhaps using more realistic profiles based on the in situ data may produce more realistic predictions, however, a more detailed investigation of how the in situ CME profiles of these parameters vary over sub-L1 monitor distances is necessary.

\subsection{Considerations for Future Upstream Monitors}

To achieve the aim of extending the lead time of geomagnetic index predictions for events observed within the time window of our study, our event analysis focuses on Solar Orbiter in situ data as the spacecraft was positioned far upstream of Earth around 0.4~au. However, as seen in Figures~\ref{fig:elevo_17march} and \ref{fig:elevo_23march}, STEREO-A is also situated close to the Earth; during observation of the two events, STEREO-A is located 0.02~au upstream of L1, with longitudinal separations of 9.1$^{\circ}$ and 9.3$^{\circ}$, respectively. Over such longitudinal separations (in combination with such small radial separations), STEREO-A and the L1 spacecraft take different paths through the CME, which could lead to different in situ observations of the same CME. Interestingly, for Event 1, we observe the CME shock in situ at STEREO-A at 2024 March 21 04:16 UT, 1 hour and 52 minutes after L1 arrival (02:24 UT). Similarly, for Event 2, the shock arrives at STEREO-A on 2024 March 24 at 14:23~UT, 12 minutes after the shock is observed at L1 (14:11 UT). In the case of the events in this study, STEREO-A would not have served as an early warning monitor, nor could we have used it to further constrain predicted arrival times. This is an important consideration for future upstream solar wind monitors, for which it is necessary to consider their location with respect to Earth so that radial evolution effects are likely to dominate over longitudinal effects. 

In fact, the positioning of an upstream solar wind monitor is vital: the further upstream it is placed, the more the CME may evolve since in situ observation; however, the closer it is placed, the less the warning time gained to mitigate potentially adverse effects. For the cases in this study, upstream observations were particularly useful in determining the effect of the CME-CME interactions on the in situ parameters. However, similar CME-CME interactions have occurred between initial upstream spacecraft measurement and arrival at L1, leading to different observed in situ profiles \cite{laker2024upstream, zhuang2024combining}. \citeA{lugaz2020inconsistencies} suggested that the expansion of the ME between $\sim$0.8~au, Earth and beyond, is more likely controlled by the change in solar wind pressure and thus is less dependent on the initial magnetic field strength of the ME. A sub-L1 monitor placed around $\sim$0.8~au would therefore capture most of the complex CME evolution undergone, but lessen the uncertainty around further large changes to the in situ parameters. 

While our approach introduces errors by assuming idealized conditions for both CME propagation and scaling, the simple scaling laws offer the advantage of being agnostic to the complexity of solar wind structures and interactions, making them broadly applicable even in cases where more complex models might struggle with highly intricate or interacting structures. In situ observations capture the true variability of CME parameters, often smoothed out by more idealized CME flux rope models, providing a more realistic basis from which to make further predictions. However, it is also clear that a realistic solar wind background is also a necessary input to capture already disturbed conditions and accurately propagate CME observations. A continuous data input to models such as  Temerin~\&~Li where prior conditions would be taken into account, as well as assimilation of future in situ solar wind background observations and sub-L1 monitor data, may improve the accuracy of geomagnetic predictions. 

We also note that many of the delays associated with performing the steps of the procedure are human delays e.g., receiving the data, running scripts, etc. Chaining together and automating many of the steps would increase gain in lead-time even further than the time differences listed in Tables~\ref{tab:timeline_event_1} and \ref{tab:timeline_eevent_2}, reducing delays after receiving data to only the short performance times of the models themselves. For future sub-L1 monitors providing real-time data streams, this means that actionable lead times can still be achieved, even for those not as far upstream as Solar Orbiter in this study.

\section{Conclusions and Implications for Future Space Weather Prediction Missions} \label{sec:implications}

In this study, we present the first real-time predictions of CME magnetic structure and geomagnetic impact using far-upstream Solar Orbiter observations at $\sim$0.4~au. During the two-week window in which Solar Orbiter crossed the Sun-Earth line, two CME events were observed in situ. We used these observations to assess our real-time procedure, demonstrating how measurements from future dedicated upstream solar wind monitors could be used to improve space weather forecasting.  

Our results demonstrate very similar predictions of the CME magnetic structure at Earth in comparison to observations for both CME events despite the early CME-CME interactions that occurred for both events. Given the large heliocentric distance separation between Solar Orbiter ($\sim$0.4~au) and L1 ($\sim$1~au), it is perhaps surprising how effective the application of general statistical power laws to Solar Orbiter magnetometer data was at capturing in situ observations at L1, considering the numerous physical processes that can alter the structure of the CME during propagation. Over the radial distance separations in this study, we suggest that radial expansion effects likely dominated over longitudinal effects.

A limitation identified in our approach concerns the different expansion behaviors of CME sheath regions compared to the magnetic ejecta of CMEs. The scaling laws applied in this study, derived from CME magnetic ejecta and magnetic flux rope observations, do not accurately capture the true compression and accumulation of solar wind ahead of the CME as it propagates. This is likely reflected by the longer predicted sheath durations in comparison to the observed values.

While our in situ CME magnetic structure predictions were successful, arrival time predictions remain challenging, with errors of several hours persisting even with updated constraints from Solar Orbiter observations. For Event 1, the CME arrived 7.3 hours later than predicted, while Event 2 arrived 2.6 hours earlier than the updated prediction. While the updated arrival times constrained by the observed CME arrival time at Solar Orbiter are slight improvements over the initial model estimates using the DONKI input parameters, observations so far upstream still leave plenty of margin for error. ELEvo is a simple drag based CME propagation model, so improvements to the arrival time predictions likely require better characterization of the ambient solar wind environment that account for dynamic solar wind conditions. However, we note that the arrival time errors in this study are within the ~10--13 hour uncertainty margins in Sun-to-L1 propagation found across all types of CME propagation models which represent a fundamental limitation that has not significantly improved over time \cite{riley2018forecasting,kay2024updating}. Our results support the value of upstream in situ monitors for constraining CME arrival time models, showing that even just one upstream monitor can lead to improvements. In the future, an optimal solution would likely include a combination of upstream monitors at different heliocentric distances and longitudinal separations to better constrain CME propagation parameters and improve arrival time forecasts.

Our study also highlights the essential role of plasma parameters in accurate geomagnetic impact modeling. The Temerin~\&~Li model requires inputs of the solar wind speed and density for which we made assumptions of constant density (5 cm$^{-3}$) and simplified speed profiles for the CME events. Realistic speed and density profiles, particularly the enhanced densities typically observed in CME sheath regions and the decreasing speed profiles observed within CME magnetic ejecta, produce more accurate geomagnetic storm onset values and stronger overall responses in the Temerin~\&~Li model. The input of more realistic plasma parameter profiles would likely improve the geomagnetic index predictions for this study, and are therefore essential measurements that should be provided by future upstream monitoring missions.

We also show that geomagnetic impact predictions can be complicated by pre-existing disturbed geomagnetic conditions, particularly evident for Event 2 where there was enhanced geomagnetic activity prior to CME arrival. It is likely that these complications would be lessened with continuous data input to the Temerin~\&~Li model, as would be the case with future space weather missions, where prior conditions would be taken into account. The assimilation of in situ solar wind measurements (such as those made by the future Vigil mission at L5) and upstream sub-L1 monitor data provide a promising avenue for more realistic predictions of CME geomagnetic impact in the future.

Using our real-time procedure, we demonstrate that even with relatively simple scaling laws, models and limited plasma data, accurate predictions of the CME magnetic structure and geomagnetic impact can be made, with actionable warning times of 4--15 hours before CME arrival at L1 and 10--34 hours before the resulting geomagnetic storm peak time. Our findings directly support proposed and future dedicated upstream space weather missions \cite<e.g.,>[]{lugaz2024miist,cicalò2025henon} and the ESA SHIELD concept, demonstrating that useful predictions can be made with longitudinal separations up to 10$^{\circ}$ over heliocentric distance ranges where radial evolution effects dominate.

%
%

\section*{Open Research Section} \label{sec:open_research}

The python scripts and notebooks used to predict the CME magnetic field structure and geomagnetic indices are available as GitHub projects accessed via \url{https://github.com/ee-davies/solar_orbiter_real_time_predictions} \cite{davies_2026_github}, including the python and package versions used. These packages include NumPy \cite{harris2020numpy}, pandas \cite{mckinney2010pandas, reback2020pandas}, SpacePy \cite{niehof2022spacepy}, SciPy \cite{scipy2020}, AstroPy \cite{astropy2013, astropy2018, astropy2022}, Plotly \cite{plotly2026} and Matplotlib \cite{hunter2007matplotlib}.

A repository of the material posted as the work was conducted in real time can be found at \url{https://figshare.com/projects/Real-Time_ICME_Predictions_using_Solar_Orbiter_as_a_Far_Upstream_Solar_Wind_Monitor/250433}. This figshare project includes screenshots of the tweets made by the Austrian Space Weather Office \cite{davies_2025_17march_posts, davies_2025_23march_posts} and submissions to the CCMC scoreboard \cite{davies_2025_ccmc}.

The NOAA L1 real-time solar wind data is obtained from \url{https://services.swpc.noaa.gov/products/solar-wind/} and the NOAA real-time Dst data from
\url{https://services.swpc.noaa.gov/products/kyoto-dst.json}. Solar Orbiter magnetometer data is publicly available from the ESA Solar Orbiter Archive, \url{https://soar.esac.esa.int/soar/}. The processed low-latency files received directly from the Imperial College London Solar Orbiter MAG team and NOAA real-time solar wind files are available at \citeA{davies_2025_data}. The solar images were obtained using the JHelioviewer software \cite<>[]{mueller2017jhelioviewer}. 


\acknowledgments
E.~E.~D., E.~W., C~M., and H.~T.~R. are funded by the European Union (ERC, HELIO4CAST, 101042188). Views and opinions expressed are however those of the author(s) only and do not necessarily reflect those of the European Union or the European Research Council Executive Agency. Neither the European Union nor the granting authority can be held responsible for them. S.~M. is supported by the Austrian Science Fund (FWF) [10.55776/P34437]. T.~H. is supported by STFC grant ST/W001071/1. Solar Orbiter magnetometer operations are funded by the UK Space Agency (grant UKRI943). 

This study has benefited from the availability of NOAA RTSW (ACE and DSCOVR), Wind, SDO, SOHO, and Solar Orbiter data, and thus we would like to thank the mission and instrument teams, and the NOAA SWPC, JSOC, SPDF CDAWeb, M2M Space Weather Analysis Office and CCMC for their distribution of data. We are grateful to the reviewers for the insightful and constructive comments that helped improve the manuscript.

\section*{Conflict of Interest Statement}

The authors have no conflicts of interest to disclose.

\section*{Appendix A: Dst Profile Comparison}

\renewcommand{\thefigure}{A\arabic{figure}}
\setcounter{figure}{0}

\begin{figure}
    \centering
    \includegraphics[width=\linewidth]{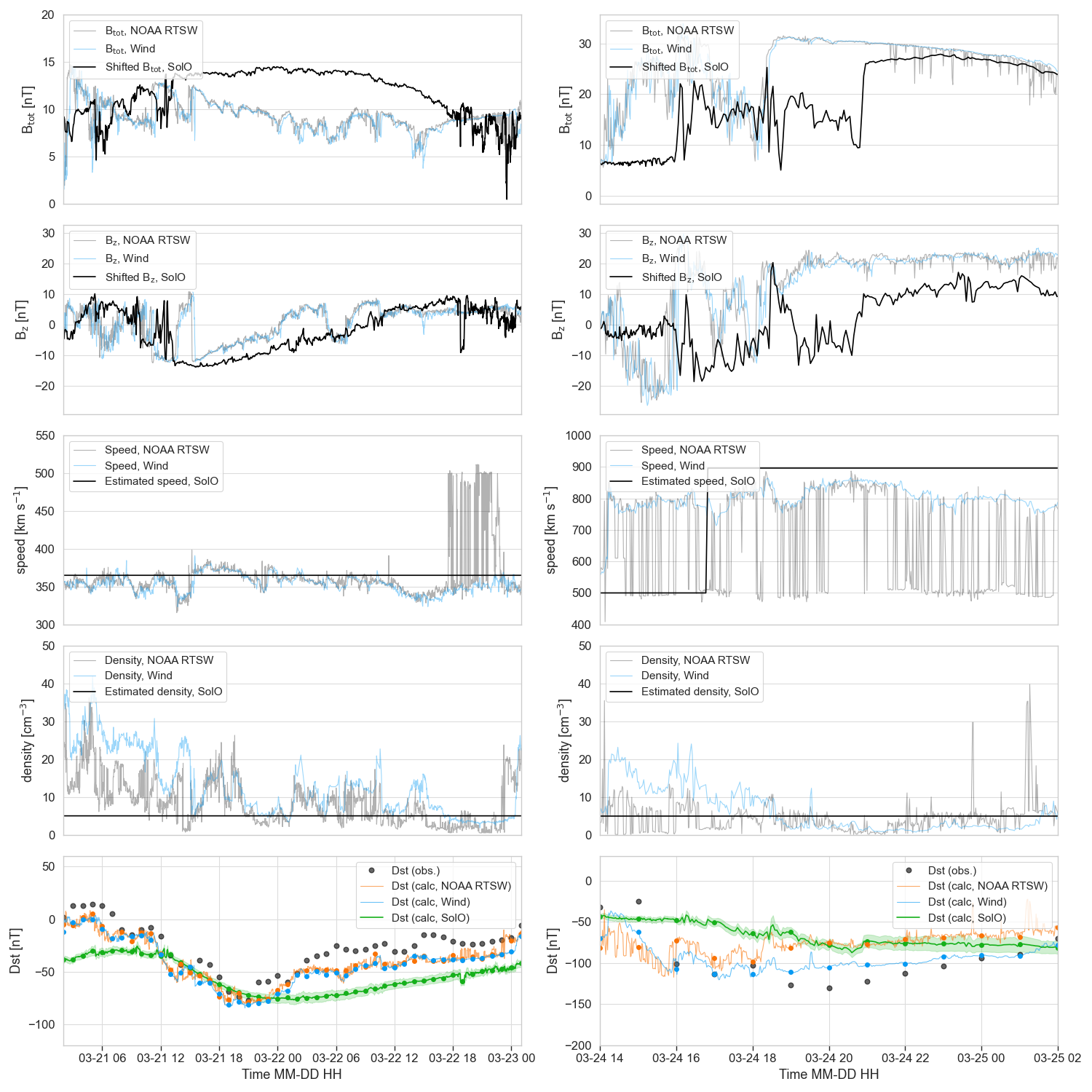}
    \caption{Comparison of in situ L1 RTSW data (gray), Wind science data (blue) and Solar Orbiter real-time procedure predicted values (black) input into the Temerin \& Li model and the resulting Dst indices (observed, black; calculated using Solar Orbiter predicted values, green; calculated using observed L1 RTSW values, orange; calculated using Wind data, blue) over observed geomagnetic storm duration at Earth for Event 1 (left) and Event 2 (right). From top to bottom: the total magnetic field, B$_z$ component, speed, density, and Dst profiles. The calculated Dst indices are also presented in 1-hour resolution (markers) for direct comparison with the observed real-time Dst index.}
    \label{fig:l1_dst}
\end{figure}

Figure~\ref{fig:l1_dst} presents the in situ data input to the Temerin \& Li model and the resulting calculated Dst indices for Event 1 (left) and Event 2 (right) over the observed geomagnetic storm duration. We compare the input values predicted using our real-time procedure (similarly to Figures \ref{fig:results_17march} and \ref{fig:results_23march}) with those of the L1 RTSW data product and science-quality Wind data. Both sets of input values are used to calculate the 1-min resolution Dst index (predicted real-time procedure results in green, L1 RTSW results in orange and Wind science data in blue), also resampled to an hourly cadence for comparison with the observed real-time Dst index (black).

\bibliography{bibliography}

\begin{thebibliography}{}

\bibitem [\protect \citeauthoryear {%
{Al-Haddad}%
\ \BBA {} {Lugaz}%
}{%
{Al-Haddad}%
\ \BBA {} {Lugaz}%
}{%
{\protect \APACyear {2025}}%
}]{%
alhaddad2025realistic}
\APACinsertmetastar {%
alhaddad2025realistic}%
\begin{APACrefauthors}%
{Al-Haddad}, N.%
\BCBT {}\ \BBA {} {Lugaz}, N.%
\end{APACrefauthors}%
\unskip\
\newblock
\APACrefYearMonthDay{2025}{{\APACmonth{02}}}{}.
\newblock
{\BBOQ}\APACrefatitle {{The Magnetic Field Structure of Coronal Mass Ejections: A More Realistic Representation}} {{The Magnetic Field Structure of Coronal Mass Ejections: A More Realistic Representation}}.{\BBCQ}
\newblock
\APACjournalVolNumPages{\ssr}{221}{1}{12}.
\newblock
\begin{APACrefDOI} \doi{10.1007/s11214-025-01138-w} \end{APACrefDOI}
\PrintBackRefs{\CurrentBib}

\bibitem [\protect \citeauthoryear {%
{Arge}%
, {Luhmann}%
, {Odstrcil}%
, {Schrijver}%
\BCBL {}\ \BBA {} {Li}%
}{%
{Arge}%
\ \protect \BOthers {.}}{%
{\protect \APACyear {2004}}%
}]{%
arge2004stream}
\APACinsertmetastar {%
arge2004stream}%
\begin{APACrefauthors}%
{Arge}, C\BPBI N.%
, {Luhmann}, J\BPBI G.%
, {Odstrcil}, D.%
, {Schrijver}, C\BPBI J.%
\BCBL {}\ \BBA {} {Li}, Y.%
\end{APACrefauthors}%
\unskip\
\newblock
\APACrefYearMonthDay{2004}{{\APACmonth{10}}}{}.
\newblock
{\BBOQ}\APACrefatitle {{Stream structure and coronal sources of the solar wind during the May 12th, 1997 CME}} {{Stream structure and coronal sources of the solar wind during the May 12th, 1997 CME}}.{\BBCQ}
\newblock
\APACjournalVolNumPages{Journal of Atmospheric and Solar-Terrestrial Physics}{66}{15-16}{1295-1309}.
\newblock
\begin{APACrefDOI} \doi{10.1016/j.jastp.2004.03.018} \end{APACrefDOI}
\PrintBackRefs{\CurrentBib}

\bibitem [\protect \citeauthoryear {%
{Arge}%
\ \BBA {} {Pizzo}%
}{%
{Arge}%
\ \BBA {} {Pizzo}%
}{%
{\protect \APACyear {2000}}%
}]{%
arge2000improvement}
\APACinsertmetastar {%
arge2000improvement}%
\begin{APACrefauthors}%
{Arge}, C\BPBI N.%
\BCBT {}\ \BBA {} {Pizzo}, V\BPBI J.%
\end{APACrefauthors}%
\unskip\
\newblock
\APACrefYearMonthDay{2000}{{\APACmonth{05}}}{}.
\newblock
{\BBOQ}\APACrefatitle {{Improvement in the prediction of solar wind conditions using near-real time solar magnetic field updates}} {{Improvement in the prediction of solar wind conditions using near-real time solar magnetic field updates}}.{\BBCQ}
\newblock
\APACjournalVolNumPages{\jgr}{105}{A5}{10465-10480}.
\newblock
\begin{APACrefDOI} \doi{10.1029/1999JA000262} \end{APACrefDOI}
\PrintBackRefs{\CurrentBib}

\bibitem [\protect \citeauthoryear {%
{Astropy Collaboration}%
\ \protect \BOthers {.}}{%
{Astropy Collaboration}%
\ \protect \BOthers {.}}{%
{\protect \APACyear {2022}}%
}]{%
astropy2022}
\APACinsertmetastar {%
astropy2022}%
\begin{APACrefauthors}%
{Astropy Collaboration}%
, {Price-Whelan}, A\BPBI M.%
, {Lim}, P\BPBI L.%
, {Earl}, N.%
, {Starkman}, N.%
, {Bradley}, L.%
\BDBL {}{Astropy Project Contributors}%
\end{APACrefauthors}%
\unskip\
\newblock
\APACrefYearMonthDay{2022}{{\APACmonth{08}}}{}.
\newblock
{\BBOQ}\APACrefatitle {{The Astropy Project: Sustaining and Growing a Community-oriented Open-source Project and the Latest Major Release (v5.0) of the Core Package}} {{The Astropy Project: Sustaining and Growing a Community-oriented Open-source Project and the Latest Major Release (v5.0) of the Core Package}}.{\BBCQ}
\newblock
\APACjournalVolNumPages{\apj}{935}{2}{167}.
\newblock
\begin{APACrefDOI} \doi{10.3847/1538-4357/ac7c74} \end{APACrefDOI}
\PrintBackRefs{\CurrentBib}

\bibitem [\protect \citeauthoryear {%
{Astropy Collaboration}%
\ \protect \BOthers {.}}{%
{Astropy Collaboration}%
\ \protect \BOthers {.}}{%
{\protect \APACyear {2018}}%
}]{%
astropy2018}
\APACinsertmetastar {%
astropy2018}%
\begin{APACrefauthors}%
{Astropy Collaboration}%
, {Price-Whelan}, A\BPBI M.%
, {Sip{\H{o}}cz}, B\BPBI M.%
, {G{\"u}nther}, H\BPBI M.%
, {Lim}, P\BPBI L.%
, {Crawford}, S\BPBI M.%
\BDBL {}{Astropy Contributors}%
\end{APACrefauthors}%
\unskip\
\newblock
\APACrefYearMonthDay{2018}{{\APACmonth{09}}}{}.
\newblock
{\BBOQ}\APACrefatitle {{The Astropy Project: Building an Open-science Project and Status of the v2.0 Core Package}} {{The Astropy Project: Building an Open-science Project and Status of the v2.0 Core Package}}.{\BBCQ}
\newblock
\APACjournalVolNumPages{\aj}{156}{3}{123}.
\newblock
\begin{APACrefDOI} \doi{10.3847/1538-3881/aabc4f} \end{APACrefDOI}
\PrintBackRefs{\CurrentBib}

\bibitem [\protect \citeauthoryear {%
{Astropy Collaboration}%
\ \protect \BOthers {.}}{%
{Astropy Collaboration}%
\ \protect \BOthers {.}}{%
{\protect \APACyear {2013}}%
}]{%
astropy2013}
\APACinsertmetastar {%
astropy2013}%
\begin{APACrefauthors}%
{Astropy Collaboration}%
, {Robitaille}, T\BPBI P.%
, {Tollerud}, E\BPBI J.%
, {Greenfield}, P.%
, {Droettboom}, M.%
, {Bray}, E.%
\BDBL {}{Streicher}, O.%
\end{APACrefauthors}%
\unskip\
\newblock
\APACrefYearMonthDay{2013}{{\APACmonth{10}}}{}.
\newblock
{\BBOQ}\APACrefatitle {{Astropy: A community Python package for astronomy}} {{Astropy: A community Python package for astronomy}}.{\BBCQ}
\newblock
\APACjournalVolNumPages{\aap}{558}{}{A33}.
\newblock
\begin{APACrefDOI} \doi{10.1051/0004-6361/201322068} \end{APACrefDOI}
\PrintBackRefs{\CurrentBib}

\bibitem [\protect \citeauthoryear {%
{Bailey}%
\ \protect \BOthers {.}}{%
{Bailey}%
\ \protect \BOthers {.}}{%
{\protect \APACyear {2020}}%
}]{%
bailey2020prediction}
\APACinsertmetastar {%
bailey2020prediction}%
\begin{APACrefauthors}%
{Bailey}, R\BPBI L.%
, {M{\"o}stl}, C.%
, {Reiss}, M\BPBI A.%
, {Weiss}, A\BPBI J.%
, {Amerstorfer}, U\BPBI V.%
, {Amerstorfer}, T.%
\BDBL {}{Leonhardt}, R.%
\end{APACrefauthors}%
\unskip\
\newblock
\APACrefYearMonthDay{2020}{{\APACmonth{05}}}{}.
\newblock
{\BBOQ}\APACrefatitle {{Prediction of Dst During Solar Minimum Using In Situ Measurements at L5}} {{Prediction of Dst During Solar Minimum Using In Situ Measurements at L5}}.{\BBCQ}
\newblock
\APACjournalVolNumPages{Space Weather}{18}{5}{e02424}.
\newblock
\begin{APACrefDOI} \doi{10.1029/2019SW002424} \end{APACrefDOI}
\PrintBackRefs{\CurrentBib}

\bibitem [\protect \citeauthoryear {%
{Banu}%
\ \protect \BOthers {.}}{%
{Banu}%
\ \protect \BOthers {.}}{%
{\protect \APACyear {2025}}%
}]{%
banu2025stereoa}
\APACinsertmetastar {%
banu2025stereoa}%
\begin{APACrefauthors}%
{Banu}, S\BPBI A.%
, {Lugaz}, N.%
, {Zhuang}, B.%
, {Al-Haddad}, N.%
, {Farrugia}, C\BPBI J.%
\BCBL {}\ \BBA {} {Galvin}, A\BPBI B.%
\end{APACrefauthors}%
\unskip\
\newblock
\APACrefYearMonthDay{2025}{{\APACmonth{03}}}{}.
\newblock
{\BBOQ}\APACrefatitle {{Investigating Coronal Mass Ejections through Multispacecraft Measurements: STEREO-A and L1 in 2022{\textendash}2023}} {{Investigating Coronal Mass Ejections through Multispacecraft Measurements: STEREO-A and L1 in 2022{\textendash}2023}}.{\BBCQ}
\newblock
\APACjournalVolNumPages{\apj}{982}{1}{47}.
\newblock
\begin{APACrefDOI} \doi{10.3847/1538-4357/adb60c} \end{APACrefDOI}
\PrintBackRefs{\CurrentBib}

\bibitem [\protect \citeauthoryear {%
Bartels%
}{%
Bartels%
}{%
{\protect \APACyear {1957}}%
}]{%
bartels1957geomagnetic}
\APACinsertmetastar {%
bartels1957geomagnetic}%
\begin{APACrefauthors}%
Bartels, J.%
\end{APACrefauthors}%
\unskip\
\newblock
\APACrefYearMonthDay{1957}{}{}.
\newblock
{\BBOQ}\APACrefatitle {The geomagnetic measures for the time-variations of solar corpuscular radiation, described for use in correlation studies in other geophysical fields} {The geomagnetic measures for the time-variations of solar corpuscular radiation, described for use in correlation studies in other geophysical fields}.{\BBCQ}
\newblock
\APACjournalVolNumPages{Ann. Intern. Geophys.}{4}{}{227--236}.
\PrintBackRefs{\CurrentBib}

\bibitem [\protect \citeauthoryear {%
{Benkhoff}%
\ \protect \BOthers {.}}{%
{Benkhoff}%
\ \protect \BOthers {.}}{%
{\protect \APACyear {2010}}%
}]{%
benkhoff2010bepicolombo}
\APACinsertmetastar {%
benkhoff2010bepicolombo}%
\begin{APACrefauthors}%
{Benkhoff}, J.%
, {van Casteren}, J.%
, {Hayakawa}, H.%
, {Fujimoto}, M.%
, {Laakso}, H.%
, {Novara}, M.%
\BDBL {}{Ziethe}, R.%
\end{APACrefauthors}%
\unskip\
\newblock
\APACrefYearMonthDay{2010}{{\APACmonth{01}}}{}.
\newblock
{\BBOQ}\APACrefatitle {{BepiColombo{\textemdash}Comprehensive exploration of Mercury: Mission overview and science goals}} {{BepiColombo{\textemdash}Comprehensive exploration of Mercury: Mission overview and science goals}}.{\BBCQ}
\newblock
\APACjournalVolNumPages{\planss}{58}{1-2}{2-20}.
\newblock
\begin{APACrefDOI} \doi{10.1016/j.pss.2009.09.020} \end{APACrefDOI}
\PrintBackRefs{\CurrentBib}

\bibitem [\protect \citeauthoryear {%
{Bolton}%
\ \protect \BOthers {.}}{%
{Bolton}%
\ \protect \BOthers {.}}{%
{\protect \APACyear {2017}}%
}]{%
bolton2017juno}
\APACinsertmetastar {%
bolton2017juno}%
\begin{APACrefauthors}%
{Bolton}, S\BPBI J.%
, {Lunine}, J.%
, {Stevenson}, D.%
, {Connerney}, J\BPBI E\BPBI P.%
, {Levin}, S.%
, {Owen}, T\BPBI C.%
\BDBL {}others%
\end{APACrefauthors}%
\unskip\
\newblock
\APACrefYearMonthDay{2017}{{\APACmonth{11}}}{}.
\newblock
{\BBOQ}\APACrefatitle {{The Juno Mission}} {{The Juno Mission}}.{\BBCQ}
\newblock
\APACjournalVolNumPages{\ssr}{213}{1-4}{5-37}.
\newblock
\begin{APACrefDOI} \doi{10.1007/s11214-017-0429-6} \end{APACrefDOI}
\PrintBackRefs{\CurrentBib}

\bibitem [\protect \citeauthoryear {%
Bolton%
\ \protect \BOthers {.}}{%
Bolton%
\ \protect \BOthers {.}}{%
{\protect \APACyear {2010}}%
}]{%
bolton2010juno}
\APACinsertmetastar {%
bolton2010juno}%
\begin{APACrefauthors}%
Bolton, S\BPBI J.%
\BCBT {}\ \BOthersPeriod {.}
\end{APACrefauthors}%
\unskip\
\newblock
\APACrefYearMonthDay{2010}{}{}.
\newblock
{\BBOQ}\APACrefatitle {{The Juno Mission}} {{The Juno Mission}}.{\BBCQ}
\newblock
\APACjournalVolNumPages{Proceedings of the International Astronomical Union}{6}{S269}{92--100}.
\newblock
\begin{APACrefDOI} \doi{10.1017/S1743921310007313} \end{APACrefDOI}
\PrintBackRefs{\CurrentBib}

\bibitem [\protect \citeauthoryear {%
{Bothmer}%
\ \BBA {} {Schwenn}%
}{%
{Bothmer}%
\ \BBA {} {Schwenn}%
}{%
{\protect \APACyear {1998}}%
}]{%
bothmer1998structure}
\APACinsertmetastar {%
bothmer1998structure}%
\begin{APACrefauthors}%
{Bothmer}, V.%
\BCBT {}\ \BBA {} {Schwenn}, R.%
\end{APACrefauthors}%
\unskip\
\newblock
\APACrefYearMonthDay{1998}{{\APACmonth{01}}}{}.
\newblock
{\BBOQ}\APACrefatitle {{The structure and origin of magnetic clouds in the solar wind}} {{The structure and origin of magnetic clouds in the solar wind}}.{\BBCQ}
\newblock
\APACjournalVolNumPages{\angeo}{16}{}{1-24}.
\newblock
\begin{APACrefDOI} \doi{10.1007/s00585-997-0001-x} \end{APACrefDOI}
\PrintBackRefs{\CurrentBib}

\bibitem [\protect \citeauthoryear {%
{Brueckner}%
\ \protect \BOthers {.}}{%
{Brueckner}%
\ \protect \BOthers {.}}{%
{\protect \APACyear {1995}}%
}]{%
brueckner1995lasco}
\APACinsertmetastar {%
brueckner1995lasco}%
\begin{APACrefauthors}%
{Brueckner}, G\BPBI E.%
, {Howard}, R\BPBI A.%
, {Koomen}, M\BPBI J.%
, {Korendyke}, C\BPBI M.%
, {Michels}, D\BPBI J.%
, {Moses}, J\BPBI D.%
\BDBL {}{Eyles}, C\BPBI J.%
\end{APACrefauthors}%
\unskip\
\newblock
\APACrefYearMonthDay{1995}{{\APACmonth{12}}}{}.
\newblock
{\BBOQ}\APACrefatitle {{The Large Angle Spectroscopic Coronagraph (LASCO)}} {{The Large Angle Spectroscopic Coronagraph (LASCO)}}.{\BBCQ}
\newblock
\APACjournalVolNumPages{\solphys}{162}{}{357-402}.
\newblock
\begin{APACrefDOI} \doi{10.1007/BF00733434} \end{APACrefDOI}
\PrintBackRefs{\CurrentBib}

\bibitem [\protect \citeauthoryear {%
{Burt}%
\ \BBA {} {Smith}%
}{%
{Burt}%
\ \BBA {} {Smith}%
}{%
{\protect \APACyear {2012}}%
}]{%
burt2012dscovr}
\APACinsertmetastar {%
burt2012dscovr}%
\begin{APACrefauthors}%
{Burt}, J.%
\BCBT {}\ \BBA {} {Smith}, B.%
\end{APACrefauthors}%
\unskip\
\newblock
\APACrefYear{2012}.
\newblock
\APACrefbtitle {{Deep Space Climate Observatory: The DSCOVR mission}} {{Deep Space Climate Observatory: The DSCOVR mission}}.
\newblock
\begin{APACrefDOI} \doi{10.1109/AERO.2012.6187025} \end{APACrefDOI}
\PrintBackRefs{\CurrentBib}

\bibitem [\protect \citeauthoryear {%
{Burton}%
, {McPherron}%
\BCBL {}\ \BBA {} {Russell}%
}{%
{Burton}%
\ \protect \BOthers {.}}{%
{\protect \APACyear {1975}}%
}]{%
burton1975empirical}
\APACinsertmetastar {%
burton1975empirical}%
\begin{APACrefauthors}%
{Burton}, R\BPBI K.%
, {McPherron}, R\BPBI L.%
\BCBL {}\ \BBA {} {Russell}, C\BPBI T.%
\end{APACrefauthors}%
\unskip\
\newblock
\APACrefYearMonthDay{1975}{{\APACmonth{11}}}{}.
\newblock
{\BBOQ}\APACrefatitle {{An empirical relationship between interplanetary conditions and Dst}} {{An empirical relationship between interplanetary conditions and Dst}}.{\BBCQ}
\newblock
\APACjournalVolNumPages{\jgr}{80}{31}{4204}.
\newblock
\begin{APACrefDOI} \doi{10.1029/JA080i031p04204} \end{APACrefDOI}
\PrintBackRefs{\CurrentBib}

\bibitem [\protect \citeauthoryear {%
Cicalò%
\ \protect \BOthers {.}}{%
Cicalò%
\ \protect \BOthers {.}}{%
{\protect \APACyear {2025}}%
}]{%
cicalò2025henon}
\APACinsertmetastar {%
cicalò2025henon}%
\begin{APACrefauthors}%
Cicalò, S.%
, Alessi, E\BPBI M.%
, Provinciali, L.%
, Amabili, P.%
, Saita, G.%
, Calcagno, D.%
\BDBL {}Khan, M.%
\end{APACrefauthors}%
\unskip\
\newblock
\APACrefYearMonthDay{2025}{}{}.
\newblock
\APACrefbtitle {Mission Analysis for the HENON CubeSat Mission to a Large Sun-Earth Distant Retrograde Orbit.} {Mission analysis for the henon cubesat mission to a large sun-earth distant retrograde orbit.}
\newblock
\begin{APACrefURL} \url{https://arxiv.org/abs/2508.02138} \end{APACrefURL}
\PrintBackRefs{\CurrentBib}

\bibitem [\protect \citeauthoryear {%
Davies%
}{%
Davies%
}{%
{\protect \APACyear {2025}}%
{\protect \APACexlab {{\protect \BCnt {1}}}}}]{%
davies_2025_17march_posts}
\APACinsertmetastar {%
davies_2025_17march_posts}%
\begin{APACrefauthors}%
Davies, E.%
\end{APACrefauthors}%
\unskip\
\newblock
\APACrefYearMonthDay{2025{\protect \BCnt {1}}}{{\APACmonth{08}}}{}.
\newblock
\APACrefbtitle {{17th March CME: Real-Time Posts}.} {{17th March CME: Real-Time Posts}.}
\newblock
\APACaddressPublisher{}{figshare}.
\newblock
\begin{APACrefURL} \url{https://figshare.com/articles/media/17th_March_CME_Real-Time_Posts/29155202/1} \end{APACrefURL}
\newblock
\begin{APACrefDOI} \doi{10.6084/m9.figshare.29155202.v1} \end{APACrefDOI}
\PrintBackRefs{\CurrentBib}

\bibitem [\protect \citeauthoryear {%
Davies%
}{%
Davies%
}{%
{\protect \APACyear {2025}}%
{\protect \APACexlab {{\protect \BCnt {2}}}}}]{%
davies_2025_23march_posts}
\APACinsertmetastar {%
davies_2025_23march_posts}%
\begin{APACrefauthors}%
Davies, E.%
\end{APACrefauthors}%
\unskip\
\newblock
\APACrefYearMonthDay{2025{\protect \BCnt {2}}}{{\APACmonth{08}}}{}.
\newblock
\APACrefbtitle {{23rd March CME: Real-Time Posts}.} {{23rd March CME: Real-Time Posts}.}
\newblock
\APACaddressPublisher{}{figshare}.
\newblock
\begin{APACrefURL} \url{https://figshare.com/articles/media/23rd_March_CME_Real-Time_Posts/29155256/1} \end{APACrefURL}
\newblock
\begin{APACrefDOI} \doi{10.6084/m9.figshare.29155256.v1} \end{APACrefDOI}
\PrintBackRefs{\CurrentBib}

\bibitem [\protect \citeauthoryear {%
Davies%
}{%
Davies%
}{%
{\protect \APACyear {2025}}%
{\protect \APACexlab {{\protect \BCnt {3}}}}}]{%
davies_2025_ccmc}
\APACinsertmetastar {%
davies_2025_ccmc}%
\begin{APACrefauthors}%
Davies, E.%
\end{APACrefauthors}%
\unskip\
\newblock
\APACrefYearMonthDay{2025{\protect \BCnt {3}}}{{\APACmonth{08}}}{}.
\newblock
\APACrefbtitle {{CCMC Scoreboard: Real-Time Posts}.} {{CCMC Scoreboard: Real-Time Posts}.}
\newblock
\APACaddressPublisher{}{figshare}.
\newblock
\begin{APACrefURL} \url{https://figshare.com/articles/media/CCMC_Scoreboard_Real-Time_Posts/29198441/1} \end{APACrefURL}
\newblock
\begin{APACrefDOI} \doi{10.6084/m9.figshare.29198441.v1} \end{APACrefDOI}
\PrintBackRefs{\CurrentBib}

\bibitem [\protect \citeauthoryear {%
Davies%
}{%
Davies%
}{%
{\protect \APACyear {2025}}%
{\protect \APACexlab {{\protect \BCnt {4}}}}}]{%
davies_2025_data}
\APACinsertmetastar {%
davies_2025_data}%
\begin{APACrefauthors}%
Davies, E.%
\end{APACrefauthors}%
\unskip\
\newblock
\APACrefYearMonthDay{2025{\protect \BCnt {4}}}{{\APACmonth{08}}}{}.
\newblock
\APACrefbtitle {{D}ata files used in real-time {S}olar {O}rbiter predictions.} {{D}ata files used in real-time {S}olar {O}rbiter predictions.}
\newblock
\APACaddressPublisher{}{figshare}.
\newblock
\begin{APACrefURL} \url{https://figshare.com/articles/dataset/Data_files_used_in_real-time_Solar_Orbiter_predictions/29930555/1} \end{APACrefURL}
\newblock
\begin{APACrefDOI} \doi{10.6084/m9.figshare.29930555.v1} \end{APACrefDOI}
\PrintBackRefs{\CurrentBib}

\bibitem [\protect \citeauthoryear {%
Davies%
}{%
Davies%
}{%
{\protect \APACyear {2026}}%
}]{%
davies_2026_github}
\APACinsertmetastar {%
davies_2026_github}%
\begin{APACrefauthors}%
Davies, E.%
\end{APACrefauthors}%
\unskip\
\newblock
\APACrefYearMonthDay{2026}{{\APACmonth{08}}}{}.
\newblock
\APACrefbtitle {Solar Orbiter Real-Time Predictions (Version v1.1) [Computer software].} {Solar orbiter real-time predictions (version v1.1) [computer software].}
\newblock
\APACaddressPublisher{}{Zenodo}.
\newblock
\begin{APACrefURL} \url{ee-davies/solar_orbiter_real_time_predictions} \end{APACrefURL}
\newblock
\begin{APACrefDOI} \doi{10.5281/zenodo.21874742} \end{APACrefDOI}
\PrintBackRefs{\CurrentBib}

\bibitem [\protect \citeauthoryear {%
{Davies}%
, {Forsyth}%
, {Good}%
\BCBL {}\ \BBA {} {Kilpua}%
}{%
{Davies}%
\ \protect \BOthers {.}}{%
{\protect \APACyear {2020}}%
}]{%
davies2020radial}
\APACinsertmetastar {%
davies2020radial}%
\begin{APACrefauthors}%
{Davies}, E\BPBI E.%
, {Forsyth}, R\BPBI J.%
, {Good}, S\BPBI W.%
\BCBL {}\ \BBA {} {Kilpua}, E\BPBI K\BPBI J.%
\end{APACrefauthors}%
\unskip\
\newblock
\APACrefYearMonthDay{2020}{{\APACmonth{11}}}{}.
\newblock
{\BBOQ}\APACrefatitle {{On the Radial and Longitudinal Variation of a Magnetic Cloud: ACE, Wind, ARTEMIS and Juno Observations}} {{On the Radial and Longitudinal Variation of a Magnetic Cloud: ACE, Wind, ARTEMIS and Juno Observations}}.{\BBCQ}
\newblock
\APACjournalVolNumPages{\solphys}{295}{11}{157}.
\newblock
\begin{APACrefDOI} \doi{10.1007/s11207-020-01714-z} \end{APACrefDOI}
\PrintBackRefs{\CurrentBib}

\bibitem [\protect \citeauthoryear {%
{Davies}%
, {Forsyth}%
, {Winslow}%
, {M{\"o}stl}%
\BCBL {}\ \BBA {} {Lugaz}%
}{%
{Davies}%
, {Forsyth}%
\BCBL {}\ \protect \BOthers {.}}{%
{\protect \APACyear {2021}}%
}]{%
davies2021catalogue}
\APACinsertmetastar {%
davies2021catalogue}%
\begin{APACrefauthors}%
{Davies}, E\BPBI E.%
, {Forsyth}, R\BPBI J.%
, {Winslow}, R\BPBI M.%
, {M{\"o}stl}, C.%
\BCBL {}\ \BBA {} {Lugaz}, N.%
\end{APACrefauthors}%
\unskip\
\newblock
\APACrefYearMonthDay{2021}{{\APACmonth{12}}}{}.
\newblock
{\BBOQ}\APACrefatitle {{A Catalog of Interplanetary Coronal Mass Ejections Observed by Juno between 1 and 5.4 au}} {{A Catalog of Interplanetary Coronal Mass Ejections Observed by Juno between 1 and 5.4 au}}.{\BBCQ}
\newblock
\APACjournalVolNumPages{\apj}{923}{2}{136}.
\newblock
\begin{APACrefDOI} \doi{10.3847/1538-4357/ac2ccb} \end{APACrefDOI}
\PrintBackRefs{\CurrentBib}

\bibitem [\protect \citeauthoryear {%
{Davies}%
, {M{\"o}stl}%
\BCBL {}\ \protect \BOthers {.}}{%
{Davies}%
, {M{\"o}stl}%
\BCBL {}\ \protect \BOthers {.}}{%
{\protect \APACyear {2021}}%
}]{%
davies2021solo}
\APACinsertmetastar {%
davies2021solo}%
\begin{APACrefauthors}%
{Davies}, E\BPBI E.%
, {M{\"o}stl}, C.%
, {Owens}, M\BPBI J.%
, {Weiss}, A\BPBI J.%
, {Amerstorfer}, T.%
, {Hinterreiter}, J.%
\BDBL {}{Harrison}, R\BPBI A.%
\end{APACrefauthors}%
\unskip\
\newblock
\APACrefYearMonthDay{2021}{{\APACmonth{12}}}{}.
\newblock
{\BBOQ}\APACrefatitle {{In situ multi-spacecraft and remote imaging observations of the first CME detected by Solar Orbiter and BepiColombo}} {{In situ multi-spacecraft and remote imaging observations of the first CME detected by Solar Orbiter and BepiColombo}}.{\BBCQ}
\newblock
\APACjournalVolNumPages{\aap}{656}{}{A2}.
\newblock
\begin{APACrefDOI} \doi{10.1051/0004-6361/202040113} \end{APACrefDOI}
\PrintBackRefs{\CurrentBib}

\bibitem [\protect \citeauthoryear {%
{Davies}%
\ \protect \BOthers {.}}{%
{Davies}%
\ \protect \BOthers {.}}{%
{\protect \APACyear {2024}}%
}]{%
davies2024fluxrope}
\APACinsertmetastar {%
davies2024fluxrope}%
\begin{APACrefauthors}%
{Davies}, E\BPBI E.%
, {R{\"u}disser}, H\BPBI T.%
, {Amerstorfer}, U\BPBI V.%
, {M{\"o}stl}, C.%
, {Bauer}, M.%
, {Weiler}, E.%
\BDBL {}{Kasper}, J\BPBI C.%
\end{APACrefauthors}%
\unskip\
\newblock
\APACrefYearMonthDay{2024}{{\APACmonth{09}}}{}.
\newblock
{\BBOQ}\APACrefatitle {{Flux Rope Modeling of the 2022 September 5 Coronal Mass Ejection Observed by Parker Solar Probe and Solar Orbiter from 0.07 to 0.69 au}} {{Flux Rope Modeling of the 2022 September 5 Coronal Mass Ejection Observed by Parker Solar Probe and Solar Orbiter from 0.07 to 0.69 au}}.{\BBCQ}
\newblock
\APACjournalVolNumPages{\apj}{973}{1}{51}.
\newblock
\begin{APACrefDOI} \doi{10.3847/1538-4357/ad64cb} \end{APACrefDOI}
\PrintBackRefs{\CurrentBib}

\bibitem [\protect \citeauthoryear {%
{Davies}%
\ \protect \BOthers {.}}{%
{Davies}%
\ \protect \BOthers {.}}{%
{\protect \APACyear {2022}}%
}]{%
davies2022multi}
\APACinsertmetastar {%
davies2022multi}%
\begin{APACrefauthors}%
{Davies}, E\BPBI E.%
, {Winslow}, R\BPBI M.%
, {Scolini}, C.%
, {Forsyth}, R\BPBI J.%
, {M{\"o}stl}, C.%
, {Lugaz}, N.%
\BCBL {}\ \BBA {} {Galvin}, A\BPBI B.%
\end{APACrefauthors}%
\unskip\
\newblock
\APACrefYearMonthDay{2022}{{\APACmonth{07}}}{}.
\newblock
{\BBOQ}\APACrefatitle {{Multi-spacecraft Observations of the Evolution of Interplanetary Coronal Mass Ejections between 0.3 and 2.2 au: Conjunctions with the Juno Spacecraft}} {{Multi-spacecraft Observations of the Evolution of Interplanetary Coronal Mass Ejections between 0.3 and 2.2 au: Conjunctions with the Juno Spacecraft}}.{\BBCQ}
\newblock
\APACjournalVolNumPages{\apj}{933}{2}{127}.
\newblock
\begin{APACrefDOI} \doi{10.3847/1538-4357/ac731a} \end{APACrefDOI}
\PrintBackRefs{\CurrentBib}

\bibitem [\protect \citeauthoryear {%
{D{\'e}moulin}%
\ \BBA {} {Dasso}%
}{%
{D{\'e}moulin}%
\ \BBA {} {Dasso}%
}{%
{\protect \APACyear {2009}}%
}]{%
demoulin2009causes}
\APACinsertmetastar {%
demoulin2009causes}%
\begin{APACrefauthors}%
{D{\'e}moulin}, P.%
\BCBT {}\ \BBA {} {Dasso}, S.%
\end{APACrefauthors}%
\unskip\
\newblock
\APACrefYearMonthDay{2009}{{\APACmonth{05}}}{}.
\newblock
{\BBOQ}\APACrefatitle {{Causes and consequences of magnetic cloud expansion}} {{Causes and consequences of magnetic cloud expansion}}.{\BBCQ}
\newblock
\APACjournalVolNumPages{\aap}{498}{2}{551-566}.
\newblock
\begin{APACrefDOI} \doi{10.1051/0004-6361/200810971} \end{APACrefDOI}
\PrintBackRefs{\CurrentBib}

\bibitem [\protect \citeauthoryear {%
{Domingo}%
, {Fleck}%
\BCBL {}\ \BBA {} {Poland}%
}{%
{Domingo}%
\ \protect \BOthers {.}}{%
{\protect \APACyear {1995}}%
}]{%
domingo1995soho}
\APACinsertmetastar {%
domingo1995soho}%
\begin{APACrefauthors}%
{Domingo}, V.%
, {Fleck}, B.%
\BCBL {}\ \BBA {} {Poland}, A\BPBI I.%
\end{APACrefauthors}%
\unskip\
\newblock
\APACrefYearMonthDay{1995}{{\APACmonth{04}}}{}.
\newblock
{\BBOQ}\APACrefatitle {{SOHO: The Solar and Heliospheric Observatory}} {{SOHO: The Solar and Heliospheric Observatory}}.{\BBCQ}
\newblock
\APACjournalVolNumPages{\ssr}{72}{1-2}{81-84}.
\newblock
\begin{APACrefDOI} \doi{10.1007/BF00768758} \end{APACrefDOI}
\PrintBackRefs{\CurrentBib}

\bibitem [\protect \citeauthoryear {%
{Eastwood}%
\ \protect \BOthers {.}}{%
{Eastwood}%
\ \protect \BOthers {.}}{%
{\protect \APACyear {2017}}%
}]{%
eastwood2017risk}
\APACinsertmetastar {%
eastwood2017risk}%
\begin{APACrefauthors}%
{Eastwood}, J\BPBI P.%
, {Biffis}, E.%
, {Hapgood}, M\BPBI A.%
, {Green}, L.%
, {Bisi}, M\BPBI M.%
, {Bentley}, R\BPBI D.%
\BDBL {}{Burnett}, C.%
\end{APACrefauthors}%
\unskip\
\newblock
\APACrefYearMonthDay{2017}{{\APACmonth{02}}}{}.
\newblock
{\BBOQ}\APACrefatitle {{The Economic Impact of Space Weather: Where Do We Stand?}} {{The Economic Impact of Space Weather: Where Do We Stand?}}{\BBCQ}
\newblock
\APACjournalVolNumPages{Risk Analysis}{37}{2}{206-218}.
\newblock
\begin{APACrefDOI} \doi{10.1111/risa.12765} \end{APACrefDOI}
\PrintBackRefs{\CurrentBib}

\bibitem [\protect \citeauthoryear {%
{Eastwood}%
, {Kataria}%
, {McInnes}%
, {Barnes}%
\BCBL {}\ \BBA {} {Mulligan}%
}{%
{Eastwood}%
\ \protect \BOthers {.}}{%
{\protect \APACyear {2015}}%
}]{%
eastwood2015sunjammer}
\APACinsertmetastar {%
eastwood2015sunjammer}%
\begin{APACrefauthors}%
{Eastwood}, J\BPBI P.%
, {Kataria}, D\BPBI O.%
, {McInnes}, C\BPBI R.%
, {Barnes}, N\BPBI C.%
\BCBL {}\ \BBA {} {Mulligan}, P.%
\end{APACrefauthors}%
\unskip\
\newblock
\APACrefYearMonthDay{2015}{{\APACmonth{01}}}{}.
\newblock
{\BBOQ}\APACrefatitle {{Sunjammer}} {{Sunjammer}}.{\BBCQ}
\newblock
\APACjournalVolNumPages{Weather}{70}{1}{27-30}.
\newblock
\begin{APACrefDOI} \doi{10.1002/wea.2438} \end{APACrefDOI}
\PrintBackRefs{\CurrentBib}

\bibitem [\protect \citeauthoryear {%
{Farrugia}%
\ \protect \BOthers {.}}{%
{Farrugia}%
\ \protect \BOthers {.}}{%
{\protect \APACyear {1993}}%
}]{%
farrugia1993study}
\APACinsertmetastar {%
farrugia1993study}%
\begin{APACrefauthors}%
{Farrugia}, C\BPBI J.%
, {Burlaga}, L\BPBI F.%
, {Osherovich}, V\BPBI A.%
, {Richardson}, I\BPBI G.%
, {Freeman}, M\BPBI P.%
, {Lepping}, R\BPBI P.%
\BCBL {}\ \BBA {} {Lazarus}, A\BPBI J.%
\end{APACrefauthors}%
\unskip\
\newblock
\APACrefYearMonthDay{1993}{{\APACmonth{05}}}{}.
\newblock
{\BBOQ}\APACrefatitle {{A study of an expanding interplanetary magnetic cloud and its interaction with the Earth's magnetosphere: The interplanetary aspect}} {{A study of an expanding interplanetary magnetic cloud and its interaction with the Earth's magnetosphere: The interplanetary aspect}}.{\BBCQ}
\newblock
\APACjournalVolNumPages{\jgr}{98}{A5}{7621-7632}.
\newblock
\begin{APACrefDOI} \doi{10.1029/92JA02349} \end{APACrefDOI}
\PrintBackRefs{\CurrentBib}

\bibitem [\protect \citeauthoryear {%
{Fox}%
\ \protect \BOthers {.}}{%
{Fox}%
\ \protect \BOthers {.}}{%
{\protect \APACyear {2016}}%
}]{%
fox2016psp}
\APACinsertmetastar {%
fox2016psp}%
\begin{APACrefauthors}%
{Fox}, N\BPBI J.%
, {Velli}, M\BPBI C.%
, {Bale}, S\BPBI D.%
, {Decker}, R.%
, {Driesman}, A.%
, {Howard}, R\BPBI A.%
\BDBL {}{Szabo}, A.%
\end{APACrefauthors}%
\unskip\
\newblock
\APACrefYearMonthDay{2016}{{\APACmonth{12}}}{}.
\newblock
{\BBOQ}\APACrefatitle {{The Solar Probe Plus Mission: Humanity's First Visit to Our Star}} {{The Solar Probe Plus Mission: Humanity's First Visit to Our Star}}.{\BBCQ}
\newblock
\APACjournalVolNumPages{\ssr}{204}{1-4}{7-48}.
\newblock
\begin{APACrefDOI} \doi{10.1007/s11214-015-0211-6} \end{APACrefDOI}
\PrintBackRefs{\CurrentBib}

\bibitem [\protect \citeauthoryear {%
{Good}%
\ \BBA {} {Forsyth}%
}{%
{Good}%
\ \BBA {} {Forsyth}%
}{%
{\protect \APACyear {2016}}%
}]{%
good2016interplanetary}
\APACinsertmetastar {%
good2016interplanetary}%
\begin{APACrefauthors}%
{Good}, S\BPBI W.%
\BCBT {}\ \BBA {} {Forsyth}, R\BPBI J.%
\end{APACrefauthors}%
\unskip\
\newblock
\APACrefYearMonthDay{2016}{{\APACmonth{01}}}{}.
\newblock
{\BBOQ}\APACrefatitle {{Interplanetary Coronal Mass Ejections Observed by MESSENGER and Venus Express}} {{Interplanetary Coronal Mass Ejections Observed by MESSENGER and Venus Express}}.{\BBCQ}
\newblock
\APACjournalVolNumPages{\solphys}{291}{1}{239-263}.
\newblock
\begin{APACrefDOI} \doi{10.1007/s11207-015-0828-3} \end{APACrefDOI}
\PrintBackRefs{\CurrentBib}

\bibitem [\protect \citeauthoryear {%
{Good}%
, {Forsyth}%
, {Eastwood}%
\BCBL {}\ \BBA {} {M{\"o}stl}%
}{%
{Good}%
\ \protect \BOthers {.}}{%
{\protect \APACyear {2018}}%
}]{%
good2018correlation}
\APACinsertmetastar {%
good2018correlation}%
\begin{APACrefauthors}%
{Good}, S\BPBI W.%
, {Forsyth}, R\BPBI J.%
, {Eastwood}, J\BPBI P.%
\BCBL {}\ \BBA {} {M{\"o}stl}, C.%
\end{APACrefauthors}%
\unskip\
\newblock
\APACrefYearMonthDay{2018}{{\APACmonth{03}}}{}.
\newblock
{\BBOQ}\APACrefatitle {{Correlation of ICME Magnetic Fields at Radially Aligned Spacecraft}} {{Correlation of ICME Magnetic Fields at Radially Aligned Spacecraft}}.{\BBCQ}
\newblock
\APACjournalVolNumPages{\solphys}{293}{3}{52}.
\newblock
\begin{APACrefDOI} \doi{10.1007/s11207-018-1264-y} \end{APACrefDOI}
\PrintBackRefs{\CurrentBib}

\bibitem [\protect \citeauthoryear {%
{Good}%
\ \protect \BOthers {.}}{%
{Good}%
\ \protect \BOthers {.}}{%
{\protect \APACyear {2015}}%
}]{%
good2015radial}
\APACinsertmetastar {%
good2015radial}%
\begin{APACrefauthors}%
{Good}, S\BPBI W.%
, {Forsyth}, R\BPBI J.%
, {Raines}, J\BPBI M.%
, {Gershman}, D\BPBI J.%
, {Slavin}, J\BPBI A.%
\BCBL {}\ \BBA {} {Zurbuchen}, T\BPBI H.%
\end{APACrefauthors}%
\unskip\
\newblock
\APACrefYearMonthDay{2015}{{\APACmonth{07}}}{}.
\newblock
{\BBOQ}\APACrefatitle {{Radial Evolution of a Magnetic Cloud: MESSENGER, STEREO, and Venus Express Observations}} {{Radial Evolution of a Magnetic Cloud: MESSENGER, STEREO, and Venus Express Observations}}.{\BBCQ}
\newblock
\APACjournalVolNumPages{\apj}{807}{2}{177}.
\newblock
\begin{APACrefDOI} \doi{10.1088/0004-637X/807/2/177} \end{APACrefDOI}
\PrintBackRefs{\CurrentBib}

\bibitem [\protect \citeauthoryear {%
{Good}%
\ \protect \BOthers {.}}{%
{Good}%
\ \protect \BOthers {.}}{%
{\protect \APACyear {2019}}%
}]{%
good2019self}
\APACinsertmetastar {%
good2019self}%
\begin{APACrefauthors}%
{Good}, S\BPBI W.%
, {Kilpua}, E\BPBI K\BPBI J.%
, {LaMoury}, A\BPBI T.%
, {Forsyth}, R\BPBI J.%
, {Eastwood}, J\BPBI P.%
\BCBL {}\ \BBA {} {M{\"o}stl}, C.%
\end{APACrefauthors}%
\unskip\
\newblock
\APACrefYearMonthDay{2019}{{\APACmonth{07}}}{}.
\newblock
{\BBOQ}\APACrefatitle {{Self-Similarity of ICME Flux Ropes: Observations by Radially Aligned Spacecraft in the Inner Heliosphere}} {{Self-Similarity of ICME Flux Ropes: Observations by Radially Aligned Spacecraft in the Inner Heliosphere}}.{\BBCQ}
\newblock
\APACjournalVolNumPages{\jgr (Space Physics)}{124}{7}{4960-4982}.
\newblock
\begin{APACrefDOI} \doi{10.1029/2019JA026475} \end{APACrefDOI}
\PrintBackRefs{\CurrentBib}

\bibitem [\protect \citeauthoryear {%
{Gopalswamy}%
\ \protect \BOthers {.}}{%
{Gopalswamy}%
\ \protect \BOthers {.}}{%
{\protect \APACyear {2000}}%
}]{%
gopalswamy2000interplanetary}
\APACinsertmetastar {%
gopalswamy2000interplanetary}%
\begin{APACrefauthors}%
{Gopalswamy}, N.%
, {Lara}, A.%
, {Lepping}, R\BPBI P.%
, {Kaiser}, M\BPBI L.%
, {Berdichevsky}, D.%
\BCBL {}\ \BBA {} {St. Cyr}, O\BPBI C.%
\end{APACrefauthors}%
\unskip\
\newblock
\APACrefYearMonthDay{2000}{{\APACmonth{01}}}{}.
\newblock
{\BBOQ}\APACrefatitle {{Interplanetary acceleration of coronal mass ejections}} {{Interplanetary acceleration of coronal mass ejections}}.{\BBCQ}
\newblock
\APACjournalVolNumPages{\grl}{27}{2}{145-148}.
\newblock
\begin{APACrefDOI} \doi{10.1029/1999GL003639} \end{APACrefDOI}
\PrintBackRefs{\CurrentBib}

\bibitem [\protect \citeauthoryear {%
{Gosling}%
}{%
{Gosling}%
}{%
{\protect \APACyear {1990}}%
}]{%
gosling1990coronal}
\APACinsertmetastar {%
gosling1990coronal}%
\begin{APACrefauthors}%
{Gosling}, J\BPBI T.%
\end{APACrefauthors}%
\unskip\
\newblock
\APACrefYearMonthDay{1990}{{\APACmonth{01}}}{}.
\newblock
{\BBOQ}\APACrefatitle {{Coronal mass ejections and magnetic flux ropes in interplanetary space}} {{Coronal mass ejections and magnetic flux ropes in interplanetary space}}.{\BBCQ}
\newblock
\APACjournalVolNumPages{Washington DC American Geophysical Union Geophysical Monograph Series}{58}{}{343-364}.
\newblock
\begin{APACrefDOI} \doi{10.1029/GM058p0343} \end{APACrefDOI}
\PrintBackRefs{\CurrentBib}

\bibitem [\protect \citeauthoryear {%
{Gulisano}%
, {D{\'e}moulin}%
, {Dasso}%
, {Ruiz}%
\BCBL {}\ \BBA {} {Marsch}%
}{%
{Gulisano}%
\ \protect \BOthers {.}}{%
{\protect \APACyear {2010}}%
}]{%
gulisano2010global}
\APACinsertmetastar {%
gulisano2010global}%
\begin{APACrefauthors}%
{Gulisano}, A\BPBI M.%
, {D{\'e}moulin}, P.%
, {Dasso}, S.%
, {Ruiz}, M\BPBI E.%
\BCBL {}\ \BBA {} {Marsch}, E.%
\end{APACrefauthors}%
\unskip\
\newblock
\APACrefYearMonthDay{2010}{{\APACmonth{01}}}{}.
\newblock
{\BBOQ}\APACrefatitle {{Global and local expansion of magnetic clouds in the inner heliosphere}} {{Global and local expansion of magnetic clouds in the inner heliosphere}}.{\BBCQ}
\newblock
\APACjournalVolNumPages{\aap}{509}{}{A39}.
\newblock
\begin{APACrefDOI} \doi{10.1051/0004-6361/200912375} \end{APACrefDOI}
\PrintBackRefs{\CurrentBib}

\bibitem [\protect \citeauthoryear {%
{Hapgood}%
\ \protect \BOthers {.}}{%
{Hapgood}%
\ \protect \BOthers {.}}{%
{\protect \APACyear {2021}}%
}]{%
hapgood2021development}
\APACinsertmetastar {%
hapgood2021development}%
\begin{APACrefauthors}%
{Hapgood}, M.%
, {Angling}, M\BPBI J.%
, {Attrill}, G.%
, {Bisi}, M.%
, {Cannon}, P\BPBI S.%
, {Dyer}, C.%
\BDBL {}others%
\end{APACrefauthors}%
\unskip\
\newblock
\APACrefYearMonthDay{2021}{}{}.
\newblock
{\BBOQ}\APACrefatitle {{Development of Space Weather Reasonable Worst Case Scenarios for the UK National Risk Assessment}} {{Development of Space Weather Reasonable Worst Case Scenarios for the UK National Risk Assessment}}.{\BBCQ}
\newblock
\APACjournalVolNumPages{Space Weather}{}{}{e2020SW002593}.
\newblock
\begin{APACrefDOI} \doi{10.1029/2020SW002593} \end{APACrefDOI}
\PrintBackRefs{\CurrentBib}

\bibitem [\protect \citeauthoryear {%
Harris%
\ \protect \BOthers {.}}{%
Harris%
\ \protect \BOthers {.}}{%
{\protect \APACyear {2020}}%
}]{%
harris2020numpy}
\APACinsertmetastar {%
harris2020numpy}%
\begin{APACrefauthors}%
Harris, C\BPBI R.%
, Millman, K\BPBI J.%
, van~der Walt, S\BPBI J.%
, Gommers, R.%
, Virtanen, P.%
, Cournapeau, D.%
\BDBL {}Oliphant, T\BPBI E.%
\end{APACrefauthors}%
\unskip\
\newblock
\APACrefYearMonthDay{2020}{{\APACmonth{09}}}{}.
\newblock
{\BBOQ}\APACrefatitle {Array programming with {NumPy}} {Array programming with {NumPy}}.{\BBCQ}
\newblock
\APACjournalVolNumPages{Nature}{585}{7825}{357--362}.
\newblock
\begin{APACrefURL} \url{https://doi.org/10.1038/s41586-020-2649-2} \end{APACrefURL}
\newblock
\begin{APACrefDOI} \doi{10.1038/s41586-020-2649-2} \end{APACrefDOI}
\PrintBackRefs{\CurrentBib}

\bibitem [\protect \citeauthoryear {%
{Henon}%
}{%
{Henon}%
}{%
{\protect \APACyear {1969}}%
}]{%
henon1969}
\APACinsertmetastar {%
henon1969}%
\begin{APACrefauthors}%
{Henon}, M.%
\end{APACrefauthors}%
\unskip\
\newblock
\APACrefYearMonthDay{1969}{{\APACmonth{02}}}{}.
\newblock
{\BBOQ}\APACrefatitle {{Numerical exploration of the restricted problem, V}} {{Numerical exploration of the restricted problem, V}}.{\BBCQ}
\newblock
\APACjournalVolNumPages{\aap}{1}{}{223-238}.
\PrintBackRefs{\CurrentBib}

\bibitem [\protect \citeauthoryear {%
{Horbury}%
\ \protect \BOthers {.}}{%
{Horbury}%
\ \protect \BOthers {.}}{%
{\protect \APACyear {2020}}%
}]{%
horbury2020mag}
\APACinsertmetastar {%
horbury2020mag}%
\begin{APACrefauthors}%
{Horbury}, T\BPBI S.%
, {O'Brien}, H.%
, {Carrasco Blazquez}, I.%
, {Bendyk}, M.%
, {Brown}, P.%
, {Hudson}, R.%
\BDBL {}others%
\end{APACrefauthors}%
\unskip\
\newblock
\APACrefYearMonthDay{2020}{{\APACmonth{10}}}{}.
\newblock
{\BBOQ}\APACrefatitle {{The Solar Orbiter magnetometer}} {{The Solar Orbiter magnetometer}}.{\BBCQ}
\newblock
\APACjournalVolNumPages{\aap}{642}{}{A9}.
\newblock
\begin{APACrefDOI} \doi{10.1051/0004-6361/201937257} \end{APACrefDOI}
\PrintBackRefs{\CurrentBib}

\bibitem [\protect \citeauthoryear {%
Hunter%
}{%
Hunter%
}{%
{\protect \APACyear {2007}}%
}]{%
hunter2007matplotlib}
\APACinsertmetastar {%
hunter2007matplotlib}%
\begin{APACrefauthors}%
Hunter, J\BPBI D.%
\end{APACrefauthors}%
\unskip\
\newblock
\APACrefYearMonthDay{2007}{}{}.
\newblock
{\BBOQ}\APACrefatitle {Matplotlib: A 2D graphics environment} {Matplotlib: A 2d graphics environment}.{\BBCQ}
\newblock
\APACjournalVolNumPages{Computing in Science \& Engineering}{9}{3}{90--95}.
\newblock
\begin{APACrefDOI} \doi{10.1109/MCSE.2007.55} \end{APACrefDOI}
\PrintBackRefs{\CurrentBib}

\bibitem [\protect \citeauthoryear {%
{Kaiser}%
\ \protect \BOthers {.}}{%
{Kaiser}%
\ \protect \BOthers {.}}{%
{\protect \APACyear {2008}}%
}]{%
kaiser2008stereo}
\APACinsertmetastar {%
kaiser2008stereo}%
\begin{APACrefauthors}%
{Kaiser}, M\BPBI L.%
, {Kucera}, T\BPBI A.%
, {Davila}, J\BPBI M.%
, {St. Cyr}, O\BPBI C.%
, {Guhathakurta}, M.%
\BCBL {}\ \BBA {} {Christian}, E.%
\end{APACrefauthors}%
\unskip\
\newblock
\APACrefYearMonthDay{2008}{{\APACmonth{04}}}{}.
\newblock
{\BBOQ}\APACrefatitle {{The STEREO Mission: An Introduction}} {{The STEREO Mission: An Introduction}}.{\BBCQ}
\newblock
\APACjournalVolNumPages{\ssr}{136}{1-4}{5-16}.
\newblock
\begin{APACrefDOI} \doi{10.1007/s11214-007-9277-0} \end{APACrefDOI}
\PrintBackRefs{\CurrentBib}

\bibitem [\protect \citeauthoryear {%
{Kay}%
, {Opher}%
\BCBL {}\ \BBA {} {Evans}%
}{%
{Kay}%
\ \protect \BOthers {.}}{%
{\protect \APACyear {2015}}%
}]{%
kay2015deflect}
\APACinsertmetastar {%
kay2015deflect}%
\begin{APACrefauthors}%
{Kay}, C.%
, {Opher}, M.%
\BCBL {}\ \BBA {} {Evans}, R\BPBI M.%
\end{APACrefauthors}%
\unskip\
\newblock
\APACrefYearMonthDay{2015}{{\APACmonth{06}}}{}.
\newblock
{\BBOQ}\APACrefatitle {{Global Trends of CME Deflections Based on CME and Solar Parameters}} {{Global Trends of CME Deflections Based on CME and Solar Parameters}}.{\BBCQ}
\newblock
\APACjournalVolNumPages{\apj}{805}{2}{168}.
\newblock
\begin{APACrefDOI} \doi{10.1088/0004-637X/805/2/168} \end{APACrefDOI}
\PrintBackRefs{\CurrentBib}

\bibitem [\protect \citeauthoryear {%
{Kay}%
\ \BBA {} {Palmerio}%
}{%
{Kay}%
\ \BBA {} {Palmerio}%
}{%
{\protect \APACyear {2024}}%
}]{%
kay2024llamacore}
\APACinsertmetastar {%
kay2024llamacore}%
\begin{APACrefauthors}%
{Kay}, C.%
\BCBT {}\ \BBA {} {Palmerio}, E.%
\end{APACrefauthors}%
\unskip\
\newblock
\APACrefYearMonthDay{2024}{{\APACmonth{01}}}{}.
\newblock
{\BBOQ}\APACrefatitle {{Collection, Collation, and Comparison of 3D Coronal CME Reconstructions}} {{Collection, Collation, and Comparison of 3D Coronal CME Reconstructions}}.{\BBCQ}
\newblock
\APACjournalVolNumPages{Space Weather}{22}{1}{e2023SW003796}.
\newblock
\begin{APACrefDOI} \doi{10.1029/2023SW003796} \end{APACrefDOI}
\PrintBackRefs{\CurrentBib}

\bibitem [\protect \citeauthoryear {%
{Kay}%
\ \protect \BOthers {.}}{%
{Kay}%
\ \protect \BOthers {.}}{%
{\protect \APACyear {2024}}%
}]{%
kay2024updating}
\APACinsertmetastar {%
kay2024updating}%
\begin{APACrefauthors}%
{Kay}, C.%
, {Palmerio}, E.%
, {Riley}, P.%
, {Mays}, M\BPBI L.%
, {Nieves-Chinchilla}, T.%
, {Romano}, M.%
\BDBL {}{Chulaki}, A.%
\end{APACrefauthors}%
\unskip\
\newblock
\APACrefYearMonthDay{2024}{{\APACmonth{07}}}{}.
\newblock
{\BBOQ}\APACrefatitle {{Updating Measures of CME Arrival Time Errors}} {{Updating Measures of CME Arrival Time Errors}}.{\BBCQ}
\newblock
\APACjournalVolNumPages{Space Weather}{22}{7}{e2024SW003951}.
\newblock
\begin{APACrefDOI} \doi{10.1029/2024SW003951} \end{APACrefDOI}
\PrintBackRefs{\CurrentBib}

\bibitem [\protect \citeauthoryear {%
{Kilpua}%
, {Balogh}%
, {von Steiger}%
\BCBL {}\ \BBA {} {Liu}%
}{%
{Kilpua}%
\ \protect \BOthers {.}}{%
{\protect \APACyear {2017}}%
}]{%
kilpua2017geoeffective}
\APACinsertmetastar {%
kilpua2017geoeffective}%
\begin{APACrefauthors}%
{Kilpua}, E\BPBI K\BPBI J.%
, {Balogh}, A.%
, {von Steiger}, R.%
\BCBL {}\ \BBA {} {Liu}, Y\BPBI D.%
\end{APACrefauthors}%
\unskip\
\newblock
\APACrefYearMonthDay{2017}{{\APACmonth{11}}}{}.
\newblock
{\BBOQ}\APACrefatitle {{Geoeffective Properties of Solar Transients and Stream Interaction Regions}} {{Geoeffective Properties of Solar Transients and Stream Interaction Regions}}.{\BBCQ}
\newblock
\APACjournalVolNumPages{\ssr}{212}{3-4}{1271-1314}.
\newblock
\begin{APACrefDOI} \doi{10.1007/s11214-017-0411-3} \end{APACrefDOI}
\PrintBackRefs{\CurrentBib}

\bibitem [\protect \citeauthoryear {%
{Kilpua}%
, {Lugaz}%
, {Mays}%
\BCBL {}\ \BBA {} {Temmer}%
}{%
{Kilpua}%
\ \protect \BOthers {.}}{%
{\protect \APACyear {2019}}%
}]{%
kilpua2019forecasting}
\APACinsertmetastar {%
kilpua2019forecasting}%
\begin{APACrefauthors}%
{Kilpua}, E\BPBI K\BPBI J.%
, {Lugaz}, N.%
, {Mays}, M\BPBI L.%
\BCBL {}\ \BBA {} {Temmer}, M.%
\end{APACrefauthors}%
\unskip\
\newblock
\APACrefYearMonthDay{2019}{{\APACmonth{04}}}{}.
\newblock
{\BBOQ}\APACrefatitle {{Forecasting the Structure and Orientation of Earthbound Coronal Mass Ejections}} {{Forecasting the Structure and Orientation of Earthbound Coronal Mass Ejections}}.{\BBCQ}
\newblock
\APACjournalVolNumPages{Space Weather}{17}{4}{498-526}.
\newblock
\begin{APACrefDOI} \doi{10.1029/2018SW001944} \end{APACrefDOI}
\PrintBackRefs{\CurrentBib}

\bibitem [\protect \citeauthoryear {%
{Krieger}%
, {Timothy}%
\BCBL {}\ \BBA {} {Roelof}%
}{%
{Krieger}%
\ \protect \BOthers {.}}{%
{\protect \APACyear {1973}}%
}]{%
krieger1973coronal}
\APACinsertmetastar {%
krieger1973coronal}%
\begin{APACrefauthors}%
{Krieger}, A\BPBI S.%
, {Timothy}, A\BPBI F.%
\BCBL {}\ \BBA {} {Roelof}, E\BPBI C.%
\end{APACrefauthors}%
\unskip\
\newblock
\APACrefYearMonthDay{1973}{{\APACmonth{04}}}{}.
\newblock
{\BBOQ}\APACrefatitle {{A Coronal Hole and Its Identification as the Source of a High Velocity Solar Wind Stream}} {{A Coronal Hole and Its Identification as the Source of a High Velocity Solar Wind Stream}}.{\BBCQ}
\newblock
\APACjournalVolNumPages{\solphys}{29}{2}{505-525}.
\newblock
\begin{APACrefDOI} \doi{10.1007/BF00150828} \end{APACrefDOI}
\PrintBackRefs{\CurrentBib}

\bibitem [\protect \citeauthoryear {%
Kruchten%
, Seier%
\BCBL {}\ \BBA {} Parmer%
}{%
Kruchten%
\ \protect \BOthers {.}}{%
{\protect \APACyear {2026}}%
}]{%
plotly2026}
\APACinsertmetastar {%
plotly2026}%
\begin{APACrefauthors}%
Kruchten, N.%
, Seier, A.%
\BCBL {}\ \BBA {} Parmer, C.%
\end{APACrefauthors}%
\unskip\
\newblock
\APACrefYearMonthDay{2026}{}{}.
\newblock
\APACrefbtitle {An interactive, open-source, and browser-based graphing library for Python.} {An interactive, open-source, and browser-based graphing library for python.}
\newblock
\begin{APACrefURL} \url{"https://github.com/plotly/plotly.py"} \end{APACrefURL}
\newblock
\begin{APACrefDOI} \doi{10.5281/zenodo.14503524} \end{APACrefDOI}
\PrintBackRefs{\CurrentBib}

\bibitem [\protect \citeauthoryear {%
{Kubicka}%
\ \protect \BOthers {.}}{%
{Kubicka}%
\ \protect \BOthers {.}}{%
{\protect \APACyear {2016}}%
}]{%
kubicka2016dst}
\APACinsertmetastar {%
kubicka2016dst}%
\begin{APACrefauthors}%
{Kubicka}, M.%
, {M{\"o}stl}, C.%
, {Amerstorfer}, T.%
, {Boakes}, P\BPBI D.%
, {Feng}, L.%
, {Eastwood}, J\BPBI P.%
\BCBL {}\ \BBA {} {T{\"o}rm{\"a}nen}, O.%
\end{APACrefauthors}%
\unskip\
\newblock
\APACrefYearMonthDay{2016}{{\APACmonth{12}}}{}.
\newblock
{\BBOQ}\APACrefatitle {{Prediction of Geomagnetic Storm Strength from Inner Heliospheric In Situ Observations}} {{Prediction of Geomagnetic Storm Strength from Inner Heliospheric In Situ Observations}}.{\BBCQ}
\newblock
\APACjournalVolNumPages{\apj}{833}{2}{255}.
\newblock
\begin{APACrefDOI} \doi{10.3847/1538-4357/833/2/255} \end{APACrefDOI}
\PrintBackRefs{\CurrentBib}

\bibitem [\protect \citeauthoryear {%
{Kumar}%
\ \BBA {} {Rust}%
}{%
{Kumar}%
\ \BBA {} {Rust}%
}{%
{\protect \APACyear {1996}}%
}]{%
kumar1996magnetic}
\APACinsertmetastar {%
kumar1996magnetic}%
\begin{APACrefauthors}%
{Kumar}, A.%
\BCBT {}\ \BBA {} {Rust}, D\BPBI M.%
\end{APACrefauthors}%
\unskip\
\newblock
\APACrefYearMonthDay{1996}{{\APACmonth{07}}}{}.
\newblock
{\BBOQ}\APACrefatitle {{Interplanetary magnetic clouds, helicity conservation, and current-core flux-ropes}} {{Interplanetary magnetic clouds, helicity conservation, and current-core flux-ropes}}.{\BBCQ}
\newblock
\APACjournalVolNumPages{\jgr}{101}{A7}{15667-15684}.
\newblock
\begin{APACrefDOI} \doi{10.1029/96JA00544} \end{APACrefDOI}
\PrintBackRefs{\CurrentBib}

\bibitem [\protect \citeauthoryear {%
{Laker}%
\ \protect \BOthers {.}}{%
{Laker}%
\ \protect \BOthers {.}}{%
{\protect \APACyear {2024}}%
}]{%
laker2024upstream}
\APACinsertmetastar {%
laker2024upstream}%
\begin{APACrefauthors}%
{Laker}, R.%
, {Horbury}, T\BPBI S.%
, {O'Brien}, H.%
, {Fauchon-Jones}, E\BPBI J.%
, {Angelini}, V.%
, {Fargette}, N.%
\BDBL {}{Dumbovi{\'c}}, M.%
\end{APACrefauthors}%
\unskip\
\newblock
\APACrefYearMonthDay{2024}{{\APACmonth{02}}}{}.
\newblock
{\BBOQ}\APACrefatitle {{Using Solar Orbiter as an Upstream Solar Wind Monitor for Real Time Space Weather Predictions}} {{Using Solar Orbiter as an Upstream Solar Wind Monitor for Real Time Space Weather Predictions}}.{\BBCQ}
\newblock
\APACjournalVolNumPages{Space Weather}{22}{2}{e2023SW003628}.
\newblock
\begin{APACrefDOI} \doi{10.1029/2023SW003628} \end{APACrefDOI}
\PrintBackRefs{\CurrentBib}

\bibitem [\protect \citeauthoryear {%
{Larrodera}%
\ \BBA {} {Temmer}%
}{%
{Larrodera}%
\ \BBA {} {Temmer}%
}{%
{\protect \APACyear {2024}}%
}]{%
larrodera2024sheaths}
\APACinsertmetastar {%
larrodera2024sheaths}%
\begin{APACrefauthors}%
{Larrodera}, C.%
\BCBT {}\ \BBA {} {Temmer}, M.%
\end{APACrefauthors}%
\unskip\
\newblock
\APACrefYearMonthDay{2024}{{\APACmonth{05}}}{}.
\newblock
{\BBOQ}\APACrefatitle {{Evolution of coronal mass ejections with and without sheaths from the inner to the outer heliosphere: Statistical investigation for 1975 to 2022}} {{Evolution of coronal mass ejections with and without sheaths from the inner to the outer heliosphere: Statistical investigation for 1975 to 2022}}.{\BBCQ}
\newblock
\APACjournalVolNumPages{\aap}{685}{}{A89}.
\newblock
\begin{APACrefDOI} \doi{10.1051/0004-6361/202348641} \end{APACrefDOI}
\PrintBackRefs{\CurrentBib}

\bibitem [\protect \citeauthoryear {%
{Leitner}%
\ \protect \BOthers {.}}{%
{Leitner}%
\ \protect \BOthers {.}}{%
{\protect \APACyear {2007}}%
}]{%
leitner2007consequences}
\APACinsertmetastar {%
leitner2007consequences}%
\begin{APACrefauthors}%
{Leitner}, M.%
, {Farrugia}, C\BPBI J.%
, {M{\"o}stl}, C.%
, {Ogilvie}, K\BPBI W.%
, {Galvin}, A\BPBI B.%
, {Schwenn}, R.%
\BCBL {}\ \BBA {} {Biernat}, H\BPBI K.%
\end{APACrefauthors}%
\unskip\
\newblock
\APACrefYearMonthDay{2007}{{\APACmonth{06}}}{}.
\newblock
{\BBOQ}\APACrefatitle {{Consequences of the force-free model of magnetic clouds for their heliospheric evolution}} {{Consequences of the force-free model of magnetic clouds for their heliospheric evolution}}.{\BBCQ}
\newblock
\APACjournalVolNumPages{\jgr (Space Physics)}{112}{A6}{A06113}.
\newblock
\begin{APACrefDOI} \doi{10.1029/2006JA011940} \end{APACrefDOI}
\PrintBackRefs{\CurrentBib}

\bibitem [\protect \citeauthoryear {%
{Lemen}%
\ \protect \BOthers {.}}{%
{Lemen}%
\ \protect \BOthers {.}}{%
{\protect \APACyear {2012}}%
}]{%
lemen2011atmospheric}
\APACinsertmetastar {%
lemen2011atmospheric}%
\begin{APACrefauthors}%
{Lemen}, J\BPBI R.%
, {Title}, A\BPBI M.%
, {Akin}, D\BPBI J.%
, {Boerner}, P\BPBI F.%
, {Chou}, C.%
, {Drake}, J\BPBI F.%
\BDBL {}others%
\end{APACrefauthors}%
\unskip\
\newblock
\APACrefYearMonthDay{2012}{{\APACmonth{01}}}{}.
\newblock
{\BBOQ}\APACrefatitle {{The Atmospheric Imaging Assembly (AIA) on the Solar Dynamics Observatory (SDO)}} {{The Atmospheric Imaging Assembly (AIA) on the Solar Dynamics Observatory (SDO)}}.{\BBCQ}
\newblock
\APACjournalVolNumPages{\solphys}{275}{1-2}{17-40}.
\newblock
\begin{APACrefDOI} \doi{10.1007/s11207-011-9776-8} \end{APACrefDOI}
\PrintBackRefs{\CurrentBib}

\bibitem [\protect \citeauthoryear {%
{Lindsay}%
, {Russell}%
\BCBL {}\ \BBA {} {Luhmann}%
}{%
{Lindsay}%
\ \protect \BOthers {.}}{%
{\protect \APACyear {1999}}%
}]{%
lindsay1999dst}
\APACinsertmetastar {%
lindsay1999dst}%
\begin{APACrefauthors}%
{Lindsay}, G\BPBI M.%
, {Russell}, C\BPBI T.%
\BCBL {}\ \BBA {} {Luhmann}, J\BPBI G.%
\end{APACrefauthors}%
\unskip\
\newblock
\APACrefYearMonthDay{1999}{{\APACmonth{05}}}{}.
\newblock
{\BBOQ}\APACrefatitle {{Predictability of Dst index based upon solar wind conditions monitored inside 1 AU}} {{Predictability of Dst index based upon solar wind conditions monitored inside 1 AU}}.{\BBCQ}
\newblock
\APACjournalVolNumPages{\jgr}{104}{A5}{10335-10344}.
\newblock
\begin{APACrefDOI} \doi{10.1029/1999JA900010} \end{APACrefDOI}
\PrintBackRefs{\CurrentBib}

\bibitem [\protect \citeauthoryear {%
Y.~{Liu}%
, {Richardson}%
\BCBL {}\ \BBA {} {Belcher}%
}{%
Y.~{Liu}%
\ \protect \BOthers {.}}{%
{\protect \APACyear {2005}}%
}]{%
liu2005statistical}
\APACinsertmetastar {%
liu2005statistical}%
\begin{APACrefauthors}%
{Liu}, Y.%
, {Richardson}, J\BPBI D.%
\BCBL {}\ \BBA {} {Belcher}, J\BPBI W.%
\end{APACrefauthors}%
\unskip\
\newblock
\APACrefYearMonthDay{2005}{{\APACmonth{01}}}{}.
\newblock
{\BBOQ}\APACrefatitle {{A statistical study of the properties of interplanetary coronal mass ejections from 0.3 to 5.4 AU}} {{A statistical study of the properties of interplanetary coronal mass ejections from 0.3 to 5.4 AU}}.{\BBCQ}
\newblock
\APACjournalVolNumPages{\planss}{53}{1-3}{3-17}.
\newblock
\begin{APACrefDOI} \doi{10.1016/j.pss.2004.09.023} \end{APACrefDOI}
\PrintBackRefs{\CurrentBib}

\bibitem [\protect \citeauthoryear {%
Y.~{Liu}%
\ \protect \BOthers {.}}{%
Y.~{Liu}%
\ \protect \BOthers {.}}{%
{\protect \APACyear {2006}}%
}]{%
liu2006curvature}
\APACinsertmetastar {%
liu2006curvature}%
\begin{APACrefauthors}%
{Liu}, Y.%
, {Richardson}, J\BPBI D.%
, {Belcher}, J\BPBI W.%
, {Wang}, C.%
, {Hu}, Q.%
\BCBL {}\ \BBA {} {Kasper}, J\BPBI C.%
\end{APACrefauthors}%
\unskip\
\newblock
\APACrefYearMonthDay{2006}{{\APACmonth{12}}}{}.
\newblock
{\BBOQ}\APACrefatitle {{Constraints on the global structure of magnetic clouds: Transverse size and curvature}} {{Constraints on the global structure of magnetic clouds: Transverse size and curvature}}.{\BBCQ}
\newblock
\APACjournalVolNumPages{Journal of Geophysical Research (Space Physics)}{111}{A12}{A12S03}.
\newblock
\begin{APACrefDOI} \doi{10.1029/2006JA011890} \end{APACrefDOI}
\PrintBackRefs{\CurrentBib}

\bibitem [\protect \citeauthoryear {%
Y\BPBI D.~{Liu}%
, {Hu}%
, {Zhao}%
, {Chen}%
\BCBL {}\ \BBA {} {Wang}%
}{%
Y\BPBI D.~{Liu}%
\ \protect \BOthers {.}}{%
{\protect \APACyear {2024}}%
}]{%
liu2024pileup}
\APACinsertmetastar {%
liu2024pileup}%
\begin{APACrefauthors}%
{Liu}, Y\BPBI D.%
, {Hu}, H.%
, {Zhao}, X.%
, {Chen}, C.%
\BCBL {}\ \BBA {} {Wang}, R.%
\end{APACrefauthors}%
\unskip\
\newblock
\APACrefYearMonthDay{2024}{{\APACmonth{10}}}{}.
\newblock
{\BBOQ}\APACrefatitle {{A Pileup of Coronal Mass Ejections Produced the Largest Geomagnetic Storm in Two Decades}} {{A Pileup of Coronal Mass Ejections Produced the Largest Geomagnetic Storm in Two Decades}}.{\BBCQ}
\newblock
\APACjournalVolNumPages{\apjl}{974}{1}{L8}.
\newblock
\begin{APACrefDOI} \doi{10.3847/2041-8213/ad7ba4} \end{APACrefDOI}
\PrintBackRefs{\CurrentBib}

\bibitem [\protect \citeauthoryear {%
{Lugaz}%
\ \protect \BOthers {.}}{%
{Lugaz}%
\ \protect \BOthers {.}}{%
{\protect \APACyear {2025}}%
}]{%
lugaz2025need}
\APACinsertmetastar {%
lugaz2025need}%
\begin{APACrefauthors}%
{Lugaz}, N.%
, {Al-Haddad}, N.%
, {Zhuang}, B.%
, {M{\"o}stl}, C.%
, {Davies}, E\BPBI E.%
, {Farrugia}, C\BPBI J.%
\BDBL {}{Galvin}, A\BPBI B.%
\end{APACrefauthors}%
\unskip\
\newblock
\APACrefYearMonthDay{2025}{{\APACmonth{02}}}{}.
\newblock
{\BBOQ}\APACrefatitle {{The Need for a Sub-L1 Space Weather Research Mission: Current Knowledge Gaps on Coronal Mass Ejections}} {{The Need for a Sub-L1 Space Weather Research Mission: Current Knowledge Gaps on Coronal Mass Ejections}}.{\BBCQ}
\newblock
\APACjournalVolNumPages{Space Weather}{23}{2}{2024SW004189}.
\newblock
\begin{APACrefDOI} \doi{10.1029/2024SW004189} \end{APACrefDOI}
\PrintBackRefs{\CurrentBib}

\bibitem [\protect \citeauthoryear {%
{Lugaz}%
, {Lee}%
\BCBL {}\ \protect \BOthers {.}}{%
{Lugaz}%
, {Lee}%
\BCBL {}\ \protect \BOthers {.}}{%
{\protect \APACyear {2024}}%
}]{%
lugaz2024miist}
\APACinsertmetastar {%
lugaz2024miist}%
\begin{APACrefauthors}%
{Lugaz}, N.%
, {Lee}, C\BPBI O.%
, {Al-Haddad}, N.%
, {Lillis}, R\BPBI J.%
, {Jian}, L\BPBI K.%
, {Curtis}, D\BPBI W.%
\BDBL {}{Nieves-Chinchilla}, T.%
\end{APACrefauthors}%
\unskip\
\newblock
\APACrefYearMonthDay{2024}{{\APACmonth{10}}}{}.
\newblock
{\BBOQ}\APACrefatitle {{The Need for Near-Earth Multi-Spacecraft Heliospheric Measurements and an Explorer Mission to Investigate Interplanetary Structures and Transients in the Near-Earth Heliosphere}} {{The Need for Near-Earth Multi-Spacecraft Heliospheric Measurements and an Explorer Mission to Investigate Interplanetary Structures and Transients in the Near-Earth Heliosphere}}.{\BBCQ}
\newblock
\APACjournalVolNumPages{\ssr}{220}{7}{73}.
\newblock
\begin{APACrefDOI} \doi{10.1007/s11214-024-01108-8} \end{APACrefDOI}
\PrintBackRefs{\CurrentBib}

\bibitem [\protect \citeauthoryear {%
{Lugaz}%
\ \protect \BOthers {.}}{%
{Lugaz}%
\ \protect \BOthers {.}}{%
{\protect \APACyear {2020}}%
}]{%
lugaz2020inconsistencies}
\APACinsertmetastar {%
lugaz2020inconsistencies}%
\begin{APACrefauthors}%
{Lugaz}, N.%
, {Salman}, T\BPBI M.%
, {Winslow}, R\BPBI M.%
, {Al-Haddad}, N.%
, {Farrugia}, C\BPBI J.%
, {Zhuang}, B.%
\BCBL {}\ \BBA {} {Galvin}, A\BPBI B.%
\end{APACrefauthors}%
\unskip\
\newblock
\APACrefYearMonthDay{2020}{{\APACmonth{08}}}{}.
\newblock
{\BBOQ}\APACrefatitle {{Inconsistencies Between Local and Global Measures of CME Radial Expansion as Revealed by Spacecraft Conjunctions}} {{Inconsistencies Between Local and Global Measures of CME Radial Expansion as Revealed by Spacecraft Conjunctions}}.{\BBCQ}
\newblock
\APACjournalVolNumPages{\apj}{899}{2}{119}.
\newblock
\begin{APACrefDOI} \doi{10.3847/1538-4357/aba26b} \end{APACrefDOI}
\PrintBackRefs{\CurrentBib}

\bibitem [\protect \citeauthoryear {%
{Lugaz}%
, {Temmer}%
, {Wang}%
\BCBL {}\ \BBA {} {Farrugia}%
}{%
{Lugaz}%
\ \protect \BOthers {.}}{%
{\protect \APACyear {2017}}%
}]{%
lugaz2017interaction}
\APACinsertmetastar {%
lugaz2017interaction}%
\begin{APACrefauthors}%
{Lugaz}, N.%
, {Temmer}, M.%
, {Wang}, Y.%
\BCBL {}\ \BBA {} {Farrugia}, C\BPBI J.%
\end{APACrefauthors}%
\unskip\
\newblock
\APACrefYearMonthDay{2017}{{\APACmonth{04}}}{}.
\newblock
{\BBOQ}\APACrefatitle {{The Interaction of Successive Coronal Mass Ejections: A Review}} {{The Interaction of Successive Coronal Mass Ejections: A Review}}.{\BBCQ}
\newblock
\APACjournalVolNumPages{\solphys}{292}{4}{64}.
\newblock
\begin{APACrefDOI} \doi{10.1007/s11207-017-1091-6} \end{APACrefDOI}
\PrintBackRefs{\CurrentBib}

\bibitem [\protect \citeauthoryear {%
{Lugaz}%
, {Zhuang}%
\BCBL {}\ \protect \BOthers {.}}{%
{Lugaz}%
, {Zhuang}%
\BCBL {}\ \protect \BOthers {.}}{%
{\protect \APACyear {2024}}%
}]{%
lugaz2024width}
\APACinsertmetastar {%
lugaz2024width}%
\begin{APACrefauthors}%
{Lugaz}, N.%
, {Zhuang}, B.%
, {Scolini}, C.%
, {Al-Haddad}, N.%
, {Farrugia}, C\BPBI J.%
, {Winslow}, R\BPBI M.%
\BDBL {}{Galvin}, A\BPBI B.%
\end{APACrefauthors}%
\unskip\
\newblock
\APACrefYearMonthDay{2024}{{\APACmonth{02}}}{}.
\newblock
{\BBOQ}\APACrefatitle {{The Width of Magnetic Ejecta Measured near 1 au: Lessons from STEREO-A Measurements in 2021{\textendash}2022}} {{The Width of Magnetic Ejecta Measured near 1 au: Lessons from STEREO-A Measurements in 2021{\textendash}2022}}.{\BBCQ}
\newblock
\APACjournalVolNumPages{\apj}{962}{2}{193}.
\newblock
\begin{APACrefDOI} \doi{10.3847/1538-4357/ad17b9} \end{APACrefDOI}
\PrintBackRefs{\CurrentBib}

\bibitem [\protect \citeauthoryear {%
{Majumdar}%
, {Pant}%
, {Patel}%
\BCBL {}\ \BBA {} {Banerjee}%
}{%
{Majumdar}%
\ \protect \BOthers {.}}{%
{\protect \APACyear {2020}}%
}]{%
majumdar2020kinem}
\APACinsertmetastar {%
majumdar2020kinem}%
\begin{APACrefauthors}%
{Majumdar}, S.%
, {Pant}, V.%
, {Patel}, R.%
\BCBL {}\ \BBA {} {Banerjee}, D.%
\end{APACrefauthors}%
\unskip\
\newblock
\APACrefYearMonthDay{2020}{{\APACmonth{08}}}{}.
\newblock
{\BBOQ}\APACrefatitle {{Connecting 3D Evolution of Coronal Mass Ejections to Their Source Regions}} {{Connecting 3D Evolution of Coronal Mass Ejections to Their Source Regions}}.{\BBCQ}
\newblock
\APACjournalVolNumPages{\apj}{899}{1}{6}.
\newblock
\begin{APACrefDOI} \doi{10.3847/1538-4357/aba1f2} \end{APACrefDOI}
\PrintBackRefs{\CurrentBib}

\bibitem [\protect \citeauthoryear {%
{Majumdar}%
\ \protect \BOthers {.}}{%
{Majumdar}%
\ \protect \BOthers {.}}{%
{\protect \APACyear {2023}}%
}]{%
majumdar2023source}
\APACinsertmetastar {%
majumdar2023source}%
\begin{APACrefauthors}%
{Majumdar}, S.%
, {Patel}, R.%
, {Pant}, V.%
, {Banerjee}, D.%
, {Rawat}, A.%
, {Pradhan}, A.%
\BCBL {}\ \BBA {} {Singh}, P.%
\end{APACrefauthors}%
\unskip\
\newblock
\APACrefYearMonthDay{2023}{{\APACmonth{09}}}{}.
\newblock
{\BBOQ}\APACrefatitle {{A Coronal Mass Ejection Source Region Catalog and Their Associated Properties}} {{A Coronal Mass Ejection Source Region Catalog and Their Associated Properties}}.{\BBCQ}
\newblock
\APACjournalVolNumPages{\apjs}{268}{1}{38}.
\newblock
\begin{APACrefDOI} \doi{10.3847/1538-4365/aceb62} \end{APACrefDOI}
\PrintBackRefs{\CurrentBib}

\bibitem [\protect \citeauthoryear {%
{Majumdar}%
, {Reiss}%
, {Muglach}%
\BCBL {}\ \BBA {} {Arge}%
}{%
{Majumdar}%
\ \protect \BOthers {.}}{%
{\protect \APACyear {2025}}%
}]{%
majumdar2025solwind}
\APACinsertmetastar {%
majumdar2025solwind}%
\begin{APACrefauthors}%
{Majumdar}, S.%
, {Reiss}, M\BPBI A.%
, {Muglach}, K.%
\BCBL {}\ \BBA {} {Arge}, C\BPBI N.%
\end{APACrefauthors}%
\unskip\
\newblock
\APACrefYearMonthDay{2025}{{\APACmonth{08}}}{}.
\newblock
{\BBOQ}\APACrefatitle {{What Causes Errors in Wang{\textendash}Sheeley{\textendash}Arge Solar Wind Modeling at L1?}} {{What Causes Errors in Wang{\textendash}Sheeley{\textendash}Arge Solar Wind Modeling at L1?}}{\BBCQ}
\newblock
\APACjournalVolNumPages{\apj}{988}{2}{239}.
\newblock
\begin{APACrefDOI} \doi{10.3847/1538-4357/ade3d5} \end{APACrefDOI}
\PrintBackRefs{\CurrentBib}

\bibitem [\protect \citeauthoryear {%
{Manchester}%
\ \protect \BOthers {.}}{%
{Manchester}%
\ \protect \BOthers {.}}{%
{\protect \APACyear {2017}}%
}]{%
manchester2017physical}
\APACinsertmetastar {%
manchester2017physical}%
\begin{APACrefauthors}%
{Manchester}, W.%
, {Kilpua}, E\BPBI K\BPBI J.%
, {Liu}, Y\BPBI D.%
, {Lugaz}, N.%
, {Riley}, P.%
, {T{\"o}r{\"o}k}, T.%
\BCBL {}\ \BBA {} {Vr{\v{s}}nak}, B.%
\end{APACrefauthors}%
\unskip\
\newblock
\APACrefYearMonthDay{2017}{{\APACmonth{11}}}{}.
\newblock
{\BBOQ}\APACrefatitle {{The Physical Processes of CME/ICME Evolution}} {{The Physical Processes of CME/ICME Evolution}}.{\BBCQ}
\newblock
\APACjournalVolNumPages{\ssr}{212}{3-4}{1159-1219}.
\newblock
\begin{APACrefDOI} \doi{10.1007/s11214-017-0394-0} \end{APACrefDOI}
\PrintBackRefs{\CurrentBib}

\bibitem [\protect \citeauthoryear {%
{Mayaud}%
}{%
{Mayaud}%
}{%
{\protect \APACyear {1980}}%
}]{%
mayaud1980geomagnetic}
\APACinsertmetastar {%
mayaud1980geomagnetic}%
\begin{APACrefauthors}%
{Mayaud}, P\BPBI N.%
\end{APACrefauthors}%
\unskip\
\newblock
\APACrefYearMonthDay{1980}{{\APACmonth{01}}}{}.
\newblock
{\BBOQ}\APACrefatitle {{Derivation, Meaning, and Use of Geomagnetic Indices}} {{Derivation, Meaning, and Use of Geomagnetic Indices}}.{\BBCQ}
\newblock
\APACjournalVolNumPages{Geophysical Monograph Series}{22}{}{607}.
\newblock
\begin{APACrefDOI} \doi{10.1029/GM022} \end{APACrefDOI}
\PrintBackRefs{\CurrentBib}

\bibitem [\protect \citeauthoryear {%
{Mays}%
\ \protect \BOthers {.}}{%
{Mays}%
\ \protect \BOthers {.}}{%
{\protect \APACyear {2015}}%
}]{%
mays2015ensemble}
\APACinsertmetastar {%
mays2015ensemble}%
\begin{APACrefauthors}%
{Mays}, M\BPBI L.%
, {Taktakishvili}, A.%
, {Pulkkinen}, A.%
, {MacNeice}, P\BPBI J.%
, {Rast{\"a}tter}, L.%
, {Odstrcil}, D.%
\BDBL {}{Kuznetsova}, M\BPBI M.%
\end{APACrefauthors}%
\unskip\
\newblock
\APACrefYearMonthDay{2015}{{\APACmonth{06}}}{}.
\newblock
{\BBOQ}\APACrefatitle {{Ensemble Modeling of CMEs Using the WSA-ENLIL+Cone Model}} {{Ensemble Modeling of CMEs Using the WSA-ENLIL+Cone Model}}.{\BBCQ}
\newblock
\APACjournalVolNumPages{\solphys}{290}{6}{1775-1814}.
\newblock
\begin{APACrefDOI} \doi{10.1007/s11207-015-0692-1} \end{APACrefDOI}
\PrintBackRefs{\CurrentBib}

\bibitem [\protect \citeauthoryear {%
{McComas}%
\ \protect \BOthers {.}}{%
{McComas}%
\ \protect \BOthers {.}}{%
{\protect \APACyear {1998}}%
}]{%
mccomas1998solar}
\APACinsertmetastar {%
mccomas1998solar}%
\begin{APACrefauthors}%
{McComas}, D\BPBI J.%
, {Bame}, S\BPBI J.%
, {Barker}, P.%
, {Feldman}, W\BPBI C.%
, {Phillips}, J\BPBI L.%
, {Riley}, P.%
\BCBL {}\ \BBA {} {Griffee}, J\BPBI W.%
\end{APACrefauthors}%
\unskip\
\newblock
\APACrefYearMonthDay{1998}{{\APACmonth{07}}}{}.
\newblock
{\BBOQ}\APACrefatitle {{Solar Wind Electron Proton Alpha Monitor (SWEPAM) for the Advanced Composition Explorer}} {{Solar Wind Electron Proton Alpha Monitor (SWEPAM) for the Advanced Composition Explorer}}.{\BBCQ}
\newblock
\APACjournalVolNumPages{\ssr}{86}{}{563-612}.
\newblock
\begin{APACrefDOI} \doi{10.1023/A:1005040232597} \end{APACrefDOI}
\PrintBackRefs{\CurrentBib}

\bibitem [\protect \citeauthoryear {%
{M{\"o}stl}%
\ \protect \BOthers {.}}{%
{M{\"o}stl}%
\ \protect \BOthers {.}}{%
{\protect \APACyear {2015}}%
}]{%
moestl2015elevo}
\APACinsertmetastar {%
moestl2015elevo}%
\begin{APACrefauthors}%
{M{\"o}stl}, C.%
, {Rollett}, T.%
, {Frahm}, R\BPBI A.%
, {Liu}, Y\BPBI D.%
, {Long}, D\BPBI M.%
, {Colaninno}, R\BPBI C.%
\BDBL {}{Vr{\v{s}}nak}, B.%
\end{APACrefauthors}%
\unskip\
\newblock
\APACrefYearMonthDay{2015}{{\APACmonth{05}}}{}.
\newblock
{\BBOQ}\APACrefatitle {{Strong coronal channelling and interplanetary evolution of a solar storm up to Earth and Mars}} {{Strong coronal channelling and interplanetary evolution of a solar storm up to Earth and Mars}}.{\BBCQ}
\newblock
\APACjournalVolNumPages{Nature Communications}{6}{}{7135}.
\newblock
\begin{APACrefDOI} \doi{10.1038/ncomms8135} \end{APACrefDOI}
\PrintBackRefs{\CurrentBib}

\bibitem [\protect \citeauthoryear {%
{M{\"o}stl}%
\ \protect \BOthers {.}}{%
{M{\"o}stl}%
\ \protect \BOthers {.}}{%
{\protect \APACyear {2022}}%
}]{%
moestl2022multi}
\APACinsertmetastar {%
moestl2022multi}%
\begin{APACrefauthors}%
{M{\"o}stl}, C.%
, {Weiss}, A\BPBI J.%
, {Reiss}, M\BPBI A.%
, {Amerstorfer}, T.%
, {Bailey}, R\BPBI L.%
, {Hinterreiter}, J.%
\BDBL {}{Bale}, S\BPBI D.%
\end{APACrefauthors}%
\unskip\
\newblock
\APACrefYearMonthDay{2022}{{\APACmonth{01}}}{}.
\newblock
{\BBOQ}\APACrefatitle {{Multipoint Interplanetary Coronal Mass Ejections Observed with Solar Orbiter, BepiColombo, Parker Solar Probe, Wind, and STEREO-A}} {{Multipoint Interplanetary Coronal Mass Ejections Observed with Solar Orbiter, BepiColombo, Parker Solar Probe, Wind, and STEREO-A}}.{\BBCQ}
\newblock
\APACjournalVolNumPages{\apjl}{924}{1}{L6}.
\newblock
\begin{APACrefDOI} \doi{10.3847/2041-8213/ac42d0} \end{APACrefDOI}
\PrintBackRefs{\CurrentBib}

\bibitem [\protect \citeauthoryear {%
{M{\"u}ller}%
, {Marsden}%
, {St. Cyr}%
\BCBL {}\ \BBA {} {Gilbert}%
}{%
{M{\"u}ller}%
\ \protect \BOthers {.}}{%
{\protect \APACyear {2013}}%
}]{%
mueller2013solar}
\APACinsertmetastar {%
mueller2013solar}%
\begin{APACrefauthors}%
{M{\"u}ller}, D.%
, {Marsden}, R\BPBI G.%
, {St. Cyr}, O\BPBI C.%
\BCBL {}\ \BBA {} {Gilbert}, H\BPBI R.%
\end{APACrefauthors}%
\unskip\
\newblock
\APACrefYearMonthDay{2013}{{\APACmonth{07}}}{}.
\newblock
{\BBOQ}\APACrefatitle {{Solar Orbiter . Exploring the Sun-Heliosphere Connection}} {{Solar Orbiter . Exploring the Sun-Heliosphere Connection}}.{\BBCQ}
\newblock
\APACjournalVolNumPages{\solphys}{285}{1-2}{25-70}.
\newblock
\begin{APACrefDOI} \doi{10.1007/s11207-012-0085-7} \end{APACrefDOI}
\PrintBackRefs{\CurrentBib}

\bibitem [\protect \citeauthoryear {%
{M{\"u}ller}%
\ \protect \BOthers {.}}{%
{M{\"u}ller}%
\ \protect \BOthers {.}}{%
{\protect \APACyear {2017}}%
}]{%
mueller2017jhelioviewer}
\APACinsertmetastar {%
mueller2017jhelioviewer}%
\begin{APACrefauthors}%
{M{\"u}ller}, D.%
, {Nicula}, B.%
, {Felix}, S.%
, {Verstringe}, F.%
, {Bourgoignie}, B.%
, {Csillaghy}, A.%
\BDBL {}{Fleck}, B.%
\end{APACrefauthors}%
\unskip\
\newblock
\APACrefYearMonthDay{2017}{{\APACmonth{09}}}{}.
\newblock
{\BBOQ}\APACrefatitle {{JHelioviewer. Time-dependent 3D visualisation of solar and heliospheric data}} {{JHelioviewer. Time-dependent 3D visualisation of solar and heliospheric data}}.{\BBCQ}
\newblock
\APACjournalVolNumPages{\aap}{606}{}{A10}.
\newblock
\begin{APACrefDOI} \doi{10.1051/0004-6361/201730893} \end{APACrefDOI}
\PrintBackRefs{\CurrentBib}

\bibitem [\protect \citeauthoryear {%
{M{\"u}ller}%
\ \protect \BOthers {.}}{%
{M{\"u}ller}%
\ \protect \BOthers {.}}{%
{\protect \APACyear {2020}}%
}]{%
mueller2020solar}
\APACinsertmetastar {%
mueller2020solar}%
\begin{APACrefauthors}%
{M{\"u}ller}, D.%
, {St. Cyr}, O\BPBI C.%
, {Zouganelis}, I.%
, {Gilbert}, H\BPBI R.%
, {Marsden}, R.%
, {Nieves-Chinchilla}, T.%
\BDBL {}{Williams}, D.%
\end{APACrefauthors}%
\unskip\
\newblock
\APACrefYearMonthDay{2020}{{\APACmonth{10}}}{}.
\newblock
{\BBOQ}\APACrefatitle {{The Solar Orbiter mission. Science overview}} {{The Solar Orbiter mission. Science overview}}.{\BBCQ}
\newblock
\APACjournalVolNumPages{\aap}{642}{}{A1}.
\newblock
\begin{APACrefDOI} \doi{10.1051/0004-6361/202038467} \end{APACrefDOI}
\PrintBackRefs{\CurrentBib}

\bibitem [\protect \citeauthoryear {%
Niehof%
, Morley%
, Welling%
\BCBL {}\ \BBA {} Larsen%
}{%
Niehof%
\ \protect \BOthers {.}}{%
{\protect \APACyear {2022}}%
}]{%
niehof2022spacepy}
\APACinsertmetastar {%
niehof2022spacepy}%
\begin{APACrefauthors}%
Niehof, J\BPBI T.%
, Morley, S\BPBI K.%
, Welling, D\BPBI T.%
\BCBL {}\ \BBA {} Larsen, B\BPBI A.%
\end{APACrefauthors}%
\unskip\
\newblock
\APACrefYearMonthDay{2022}{}{}.
\newblock
{\BBOQ}\APACrefatitle {The SpacePy space science package at 12 years} {The spacepy space science package at 12 years}.{\BBCQ}
\newblock
\APACjournalVolNumPages{Frontiers in Astronomy and Space Sciences}{9}{}{}.
\newblock
\begin{APACrefDOI} \doi{10.3389/fspas.2022.1023612} \end{APACrefDOI}
\PrintBackRefs{\CurrentBib}

\bibitem [\protect \citeauthoryear {%
{Nieves-Chinchilla}%
\ \protect \BOthers {.}}{%
{Nieves-Chinchilla}%
\ \protect \BOthers {.}}{%
{\protect \APACyear {2013}}%
}]{%
nieves2013inner}
\APACinsertmetastar {%
nieves2013inner}%
\begin{APACrefauthors}%
{Nieves-Chinchilla}, T.%
, {Vourlidas}, A.%
, {Stenborg}, G.%
, {Savani}, N\BPBI P.%
, {Koval}, A.%
, {Szabo}, A.%
\BCBL {}\ \BBA {} {Jian}, L\BPBI K.%
\end{APACrefauthors}%
\unskip\
\newblock
\APACrefYearMonthDay{2013}{{\APACmonth{12}}}{}.
\newblock
{\BBOQ}\APACrefatitle {{Inner Heliospheric Evolution of a ``Stealth'' CME Derived from Multi-view Imaging and Multipoint in Situ observations. I. Propagation to 1 AU}} {{Inner Heliospheric Evolution of a ``Stealth'' CME Derived from Multi-view Imaging and Multipoint in Situ observations. I. Propagation to 1 AU}}.{\BBCQ}
\newblock
\APACjournalVolNumPages{\apj}{779}{1}{55}.
\newblock
\begin{APACrefDOI} \doi{10.1088/0004-637X/779/1/55} \end{APACrefDOI}
\PrintBackRefs{\CurrentBib}

\bibitem [\protect \citeauthoryear {%
{Nose}%
, {Iyemori}%
, {Sugiura}%
\BCBL {}\ \BBA {} {Kamei}%
}{%
{Nose}%
\ \protect \BOthers {.}}{%
{\protect \APACyear {2015}}%
}]{%
kyoto_dst}
\APACinsertmetastar {%
kyoto_dst}%
\begin{APACrefauthors}%
{Nose}, M.%
, {Iyemori}, T.%
, {Sugiura}, M.%
\BCBL {}\ \BBA {} {Kamei}, T.%
\end{APACrefauthors}%
\unskip\
\newblock
\APACrefYearMonthDay{2015}{}{}.
\newblock
\APACrefbtitle {Geomagnetic Dst index.} {Geomagnetic dst index.}
\newblock
\APACaddressPublisher{}{World Data Center for Geomagnetism, Kyoto}.
\newblock
\begin{APACrefURL} \url{https://isds-datadoi.nict.go.jp/wds/10.17593__14515-74000.html} \end{APACrefURL}
\newblock
\begin{APACrefDOI} \doi{10.17593/14515-74000} \end{APACrefDOI}
\PrintBackRefs{\CurrentBib}

\bibitem [\protect \citeauthoryear {%
{O'Brien}%
\ \BBA {} {McPherron}%
}{%
{O'Brien}%
\ \BBA {} {McPherron}%
}{%
{\protect \APACyear {2000}}%
}]{%
obrien2000empirical}
\APACinsertmetastar {%
obrien2000empirical}%
\begin{APACrefauthors}%
{O'Brien}, T\BPBI P.%
\BCBT {}\ \BBA {} {McPherron}, R\BPBI L.%
\end{APACrefauthors}%
\unskip\
\newblock
\APACrefYearMonthDay{2000}{{\APACmonth{04}}}{}.
\newblock
{\BBOQ}\APACrefatitle {{An empirical phase space analysis of ring current dynamics: Solar wind control of injection and decay}} {{An empirical phase space analysis of ring current dynamics: Solar wind control of injection and decay}}.{\BBCQ}
\newblock
\APACjournalVolNumPages{\jgr}{105}{A4}{7707-7720}.
\newblock
\begin{APACrefDOI} \doi{10.1029/1998JA000437} \end{APACrefDOI}
\PrintBackRefs{\CurrentBib}

\bibitem [\protect \citeauthoryear {%
{Odstrcil}%
}{%
{Odstrcil}%
}{%
{\protect \APACyear {2003}}%
}]{%
odstrcil2003enlil}
\APACinsertmetastar {%
odstrcil2003enlil}%
\begin{APACrefauthors}%
{Odstrcil}, D.%
\end{APACrefauthors}%
\unskip\
\newblock
\APACrefYearMonthDay{2003}{{\APACmonth{08}}}{}.
\newblock
{\BBOQ}\APACrefatitle {{Modeling 3-D solar wind structure}} {{Modeling 3-D solar wind structure}}.{\BBCQ}
\newblock
\APACjournalVolNumPages{Advances in Space Research}{32}{4}{497-506}.
\newblock
\begin{APACrefDOI} \doi{10.1016/S0273-1177(03)00332-6} \end{APACrefDOI}
\PrintBackRefs{\CurrentBib}

\bibitem [\protect \citeauthoryear {%
{Ogilvie}%
\ \BBA {} {Desch}%
}{%
{Ogilvie}%
\ \BBA {} {Desch}%
}{%
{\protect \APACyear {1997}}%
}]{%
ogilvie1997wind}
\APACinsertmetastar {%
ogilvie1997wind}%
\begin{APACrefauthors}%
{Ogilvie}, K\BPBI W.%
\BCBT {}\ \BBA {} {Desch}, M\BPBI D.%
\end{APACrefauthors}%
\unskip\
\newblock
\APACrefYearMonthDay{1997}{{\APACmonth{01}}}{}.
\newblock
{\BBOQ}\APACrefatitle {{The wind spacecraft and its early scientific results}} {{The wind spacecraft and its early scientific results}}.{\BBCQ}
\newblock
\APACjournalVolNumPages{Advances in Space Research}{20}{4-5}{559-568}.
\newblock
\begin{APACrefDOI} \doi{10.1016/S0273-1177(97)00439-0} \end{APACrefDOI}
\PrintBackRefs{\CurrentBib}

\bibitem [\protect \citeauthoryear {%
{Owen}%
\ \protect \BOthers {.}}{%
{Owen}%
\ \protect \BOthers {.}}{%
{\protect \APACyear {2020}}%
}]{%
owen2020swa}
\APACinsertmetastar {%
owen2020swa}%
\begin{APACrefauthors}%
{Owen}, C\BPBI J.%
, {Bruno}, R.%
, {Livi}, S.%
, {Louarn}, P.%
, {Al Janabi}, K.%
, {Allegrini}, F.%
\BDBL {}others%
\end{APACrefauthors}%
\unskip\
\newblock
\APACrefYearMonthDay{2020}{{\APACmonth{10}}}{}.
\newblock
{\BBOQ}\APACrefatitle {{The Solar Orbiter Solar Wind Analyser (SWA) suite}} {{The Solar Orbiter Solar Wind Analyser (SWA) suite}}.{\BBCQ}
\newblock
\APACjournalVolNumPages{\aap}{642}{}{A16}.
\newblock
\begin{APACrefDOI} \doi{10.1051/0004-6361/201937259} \end{APACrefDOI}
\PrintBackRefs{\CurrentBib}

\bibitem [\protect \citeauthoryear {%
{Owens}%
}{%
{Owens}%
}{%
{\protect \APACyear {2006}}%
}]{%
owens2006magnetic}
\APACinsertmetastar {%
owens2006magnetic}%
\begin{APACrefauthors}%
{Owens}, M\BPBI J.%
\end{APACrefauthors}%
\unskip\
\newblock
\APACrefYearMonthDay{2006}{{\APACmonth{12}}}{}.
\newblock
{\BBOQ}\APACrefatitle {{Magnetic cloud distortion resulting from propagation through a structured solar wind: Models and observations}} {{Magnetic cloud distortion resulting from propagation through a structured solar wind: Models and observations}}.{\BBCQ}
\newblock
\APACjournalVolNumPages{\jgr (Space Physics)}{111}{A12}{A12109}.
\newblock
\begin{APACrefDOI} \doi{10.1029/2006JA011903} \end{APACrefDOI}
\PrintBackRefs{\CurrentBib}

\bibitem [\protect \citeauthoryear {%
{Owens}%
, {Lockwood}%
\BCBL {}\ \BBA {} {Barnard}%
}{%
{Owens}%
\ \protect \BOthers {.}}{%
{\protect \APACyear {2017}}%
}]{%
owens2017coronal}
\APACinsertmetastar {%
owens2017coronal}%
\begin{APACrefauthors}%
{Owens}, M\BPBI J.%
, {Lockwood}, M.%
\BCBL {}\ \BBA {} {Barnard}, L\BPBI A.%
\end{APACrefauthors}%
\unskip\
\newblock
\APACrefYearMonthDay{2017}{{\APACmonth{06}}}{}.
\newblock
{\BBOQ}\APACrefatitle {{Coronal mass ejections are not coherent magnetohydrodynamic structures}} {{Coronal mass ejections are not coherent magnetohydrodynamic structures}}.{\BBCQ}
\newblock
\APACjournalVolNumPages{Nature Scientific Reports}{7}{}{4152}.
\newblock
\begin{APACrefDOI} \doi{10.1038/s41598-017-04546-3} \end{APACrefDOI}
\PrintBackRefs{\CurrentBib}

\bibitem [\protect \citeauthoryear {%
{Owens}%
, {Lockwood}%
\BCBL {}\ \BBA {} {Barnard}%
}{%
{Owens}%
\ \protect \BOthers {.}}{%
{\protect \APACyear {2020}}%
}]{%
owens2020value}
\APACinsertmetastar {%
owens2020value}%
\begin{APACrefauthors}%
{Owens}, M\BPBI J.%
, {Lockwood}, M.%
\BCBL {}\ \BBA {} {Barnard}, L\BPBI A.%
\end{APACrefauthors}%
\unskip\
\newblock
\APACrefYearMonthDay{2020}{{\APACmonth{09}}}{}.
\newblock
{\BBOQ}\APACrefatitle {{The Value of CME Arrival Time Forecasts for Space Weather Mitigation}} {{The Value of CME Arrival Time Forecasts for Space Weather Mitigation}}.{\BBCQ}
\newblock
\APACjournalVolNumPages{Space Weather}{18}{9}{e02507}.
\newblock
\begin{APACrefDOI} \doi{10.1029/2020SW002507} \end{APACrefDOI}
\PrintBackRefs{\CurrentBib}

\bibitem [\protect \citeauthoryear {%
{Palmerio}%
\ \protect \BOthers {.}}{%
{Palmerio}%
\ \protect \BOthers {.}}{%
{\protect \APACyear {2017}}%
}]{%
palmerio2017determining}
\APACinsertmetastar {%
palmerio2017determining}%
\begin{APACrefauthors}%
{Palmerio}, E.%
, {Kilpua}, E\BPBI K\BPBI J.%
, {James}, A\BPBI W.%
, {Green}, L\BPBI M.%
, {Pomoell}, J.%
, {Isavnin}, A.%
\BCBL {}\ \BBA {} {Valori}, G.%
\end{APACrefauthors}%
\unskip\
\newblock
\APACrefYearMonthDay{2017}{{\APACmonth{02}}}{}.
\newblock
{\BBOQ}\APACrefatitle {{Determining the Intrinsic CME Flux Rope Type Using Remote-sensing Solar Disk Observations}} {{Determining the Intrinsic CME Flux Rope Type Using Remote-sensing Solar Disk Observations}}.{\BBCQ}
\newblock
\APACjournalVolNumPages{\solphys}{292}{2}{39}.
\newblock
\begin{APACrefDOI} \doi{10.1007/s11207-017-1063-x} \end{APACrefDOI}
\PrintBackRefs{\CurrentBib}

\bibitem [\protect \citeauthoryear {%
{Palmerio}%
\ \protect \BOthers {.}}{%
{Palmerio}%
\ \protect \BOthers {.}}{%
{\protect \APACyear {2018}}%
}]{%
palmerio2018coronal}
\APACinsertmetastar {%
palmerio2018coronal}%
\begin{APACrefauthors}%
{Palmerio}, E.%
, {Kilpua}, E\BPBI K\BPBI J.%
, {M{\"o}stl}, C.%
, {Bothmer}, V.%
, {James}, A\BPBI W.%
, {Green}, L\BPBI M.%
\BDBL {}{Harrison}, R\BPBI A.%
\end{APACrefauthors}%
\unskip\
\newblock
\APACrefYearMonthDay{2018}{{\APACmonth{05}}}{}.
\newblock
{\BBOQ}\APACrefatitle {{Coronal Magnetic Structure of Earthbound CMEs and In Situ Comparison}} {{Coronal Magnetic Structure of Earthbound CMEs and In Situ Comparison}}.{\BBCQ}
\newblock
\APACjournalVolNumPages{Space Weather}{16}{5}{442-460}.
\newblock
\begin{APACrefDOI} \doi{10.1002/2017SW001767} \end{APACrefDOI}
\PrintBackRefs{\CurrentBib}

\bibitem [\protect \citeauthoryear {%
{Parker}%
}{%
{Parker}%
}{%
{\protect \APACyear {1958}}%
}]{%
parker1958dynamics}
\APACinsertmetastar {%
parker1958dynamics}%
\begin{APACrefauthors}%
{Parker}, E\BPBI N.%
\end{APACrefauthors}%
\unskip\
\newblock
\APACrefYearMonthDay{1958}{{\APACmonth{11}}}{}.
\newblock
{\BBOQ}\APACrefatitle {{Dynamics of the Interplanetary Gas and Magnetic Fields.}} {{Dynamics of the Interplanetary Gas and Magnetic Fields.}}{\BBCQ}
\newblock
\APACjournalVolNumPages{\apj}{128}{}{664}.
\newblock
\begin{APACrefDOI} \doi{10.1086/146579} \end{APACrefDOI}
\PrintBackRefs{\CurrentBib}

\bibitem [\protect \citeauthoryear {%
{Pesnell}%
, {Thompson}%
\BCBL {}\ \BBA {} {Chamberlin}%
}{%
{Pesnell}%
\ \protect \BOthers {.}}{%
{\protect \APACyear {2012}}%
}]{%
pesnell2012sdo}
\APACinsertmetastar {%
pesnell2012sdo}%
\begin{APACrefauthors}%
{Pesnell}, W\BPBI D.%
, {Thompson}, B\BPBI J.%
\BCBL {}\ \BBA {} {Chamberlin}, P\BPBI C.%
\end{APACrefauthors}%
\unskip\
\newblock
\APACrefYearMonthDay{2012}{{\APACmonth{01}}}{}.
\newblock
{\BBOQ}\APACrefatitle {{The Solar Dynamics Observatory (SDO)}} {{The Solar Dynamics Observatory (SDO)}}.{\BBCQ}
\newblock
\APACjournalVolNumPages{\solphys}{275}{1-2}{3-15}.
\newblock
\begin{APACrefDOI} \doi{10.1007/s11207-011-9841-3} \end{APACrefDOI}
\PrintBackRefs{\CurrentBib}

\bibitem [\protect \citeauthoryear {%
{Pevtsov}%
\ \BBA {} {Balasubramaniam}%
}{%
{Pevtsov}%
\ \BBA {} {Balasubramaniam}%
}{%
{\protect \APACyear {2003}}%
}]{%
pevtsov2003helicity}
\APACinsertmetastar {%
pevtsov2003helicity}%
\begin{APACrefauthors}%
{Pevtsov}, A\BPBI A.%
\BCBT {}\ \BBA {} {Balasubramaniam}, K\BPBI S.%
\end{APACrefauthors}%
\unskip\
\newblock
\APACrefYearMonthDay{2003}{{\APACmonth{01}}}{}.
\newblock
{\BBOQ}\APACrefatitle {{Helicity patterns on the sun}} {{Helicity patterns on the sun}}.{\BBCQ}
\newblock
\APACjournalVolNumPages{Advances in Space Research}{32}{10}{1867-1874}.
\newblock
\begin{APACrefDOI} \doi{10.1016/S0273-1177(03)90620-X} \end{APACrefDOI}
\PrintBackRefs{\CurrentBib}

\bibitem [\protect \citeauthoryear {%
{Pizzo}%
\ \protect \BOthers {.}}{%
{Pizzo}%
\ \protect \BOthers {.}}{%
{\protect \APACyear {2011}}%
}]{%
pizzo2011wsaenlil}
\APACinsertmetastar {%
pizzo2011wsaenlil}%
\begin{APACrefauthors}%
{Pizzo}, V.%
, {Millward}, G.%
, {Parsons}, A.%
, {Biesecker}, D.%
, {Hill}, S.%
\BCBL {}\ \BBA {} {Odstrcil}, D.%
\end{APACrefauthors}%
\unskip\
\newblock
\APACrefYearMonthDay{2011}{{\APACmonth{03}}}{}.
\newblock
{\BBOQ}\APACrefatitle {{Wang-Sheeley-Arge-Enlil Cone Model Transitions to Operations}} {{Wang-Sheeley-Arge-Enlil Cone Model Transitions to Operations}}.{\BBCQ}
\newblock
\APACjournalVolNumPages{Space Weather}{9}{3}{03004}.
\newblock
\begin{APACrefDOI} \doi{10.1029/2011SW000663} \end{APACrefDOI}
\PrintBackRefs{\CurrentBib}

\bibitem [\protect \citeauthoryear {%
{Regnault}%
\ \protect \BOthers {.}}{%
{Regnault}%
\ \protect \BOthers {.}}{%
{\protect \APACyear {2024}}%
}]{%
regnault2024discrepancies}
\APACinsertmetastar {%
regnault2024discrepancies}%
\begin{APACrefauthors}%
{Regnault}, F.%
, {Al-Haddad}, N.%
, {Lugaz}, N.%
, {Farrugia}, C\BPBI J.%
, {Yu}, W.%
, {Zhuang}, B.%
\BCBL {}\ \BBA {} {Davies}, E\BPBI E.%
\end{APACrefauthors}%
\unskip\
\newblock
\APACrefYearMonthDay{2024}{{\APACmonth{02}}}{}.
\newblock
{\BBOQ}\APACrefatitle {{Discrepancies in the Properties of a Coronal Mass Ejection on Scales of 0.03 au as Revealed by Simultaneous Measurements at Solar Orbiter and Wind: The 2021 November 3{\textendash}5 Event}} {{Discrepancies in the Properties of a Coronal Mass Ejection on Scales of 0.03 au as Revealed by Simultaneous Measurements at Solar Orbiter and Wind: The 2021 November 3{\textendash}5 Event}}.{\BBCQ}
\newblock
\APACjournalVolNumPages{\apj}{962}{2}{190}.
\newblock
\begin{APACrefDOI} \doi{10.3847/1538-4357/ad1883} \end{APACrefDOI}
\PrintBackRefs{\CurrentBib}

\bibitem [\protect \citeauthoryear {%
{Reiss}%
\ \protect \BOthers {.}}{%
{Reiss}%
\ \protect \BOthers {.}}{%
{\protect \APACyear {2023}}%
}]{%
reiss2023validation}
\APACinsertmetastar {%
reiss2023validation}%
\begin{APACrefauthors}%
{Reiss}, M\BPBI A.%
, {Muglach}, K.%
, {Mullinix}, R.%
, {Kuznetsova}, M\BPBI M.%
, {Wiegand}, C.%
, {Temmer}, M.%
\BDBL {}{Samara}, E.%
\end{APACrefauthors}%
\unskip\
\newblock
\APACrefYearMonthDay{2023}{{\APACmonth{12}}}{}.
\newblock
{\BBOQ}\APACrefatitle {{Unifying the validation of ambient solar wind models}} {{Unifying the validation of ambient solar wind models}}.{\BBCQ}
\newblock
\APACjournalVolNumPages{Advances in Space Research}{72}{12}{5275-5286}.
\newblock
\begin{APACrefDOI} \doi{10.1016/j.asr.2022.05.026} \end{APACrefDOI}
\PrintBackRefs{\CurrentBib}

\bibitem [\protect \citeauthoryear {%
{Richardson}%
}{%
{Richardson}%
}{%
{\protect \APACyear {2014}}%
}]{%
richardson2014identification}
\APACinsertmetastar {%
richardson2014identification}%
\begin{APACrefauthors}%
{Richardson}, I\BPBI G.%
\end{APACrefauthors}%
\unskip\
\newblock
\APACrefYearMonthDay{2014}{{\APACmonth{10}}}{}.
\newblock
{\BBOQ}\APACrefatitle {{Identification of Interplanetary Coronal Mass Ejections at Ulysses Using Multiple Solar Wind Signatures}} {{Identification of Interplanetary Coronal Mass Ejections at Ulysses Using Multiple Solar Wind Signatures}}.{\BBCQ}
\newblock
\APACjournalVolNumPages{\solphys}{289}{10}{3843-3894}.
\newblock
\begin{APACrefDOI} \doi{10.1007/s11207-014-0540-8} \end{APACrefDOI}
\PrintBackRefs{\CurrentBib}

\bibitem [\protect \citeauthoryear {%
{Riley}%
\ \BBA {} {Crooker}%
}{%
{Riley}%
\ \BBA {} {Crooker}%
}{%
{\protect \APACyear {2004}}%
}]{%
riley2004kinematic}
\APACinsertmetastar {%
riley2004kinematic}%
\begin{APACrefauthors}%
{Riley}, P.%
\BCBT {}\ \BBA {} {Crooker}, N\BPBI U.%
\end{APACrefauthors}%
\unskip\
\newblock
\APACrefYearMonthDay{2004}{{\APACmonth{01}}}{}.
\newblock
{\BBOQ}\APACrefatitle {{Kinematic Treatment of Coronal Mass Ejection Evolution in the Solar Wind}} {{Kinematic Treatment of Coronal Mass Ejection Evolution in the Solar Wind}}.{\BBCQ}
\newblock
\APACjournalVolNumPages{\apj}{600}{2}{1035-1042}.
\newblock
\begin{APACrefDOI} \doi{10.1086/379974} \end{APACrefDOI}
\PrintBackRefs{\CurrentBib}

\bibitem [\protect \citeauthoryear {%
{Riley}%
\ \protect \BOthers {.}}{%
{Riley}%
\ \protect \BOthers {.}}{%
{\protect \APACyear {2018}}%
}]{%
riley2018forecasting}
\APACinsertmetastar {%
riley2018forecasting}%
\begin{APACrefauthors}%
{Riley}, P.%
, {Mays}, M\BPBI L.%
, {Andries}, J.%
, {Amerstorfer}, T.%
, {Biesecker}, D.%
, {Delouille}, V.%
\BDBL {}{Zhao}, X.%
\end{APACrefauthors}%
\unskip\
\newblock
\APACrefYearMonthDay{2018}{{\APACmonth{09}}}{}.
\newblock
{\BBOQ}\APACrefatitle {{Forecasting the Arrival Time of Coronal Mass Ejections: Analysis of the CCMC CME Scoreboard}} {{Forecasting the Arrival Time of Coronal Mass Ejections: Analysis of the CCMC CME Scoreboard}}.{\BBCQ}
\newblock
\APACjournalVolNumPages{Space Weather}{16}{9}{1245-1260}.
\newblock
\begin{APACrefDOI} \doi{10.1029/2018SW001962} \end{APACrefDOI}
\PrintBackRefs{\CurrentBib}

\bibitem [\protect \citeauthoryear {%
{Ritter}%
\ \protect \BOthers {.}}{%
{Ritter}%
\ \protect \BOthers {.}}{%
{\protect \APACyear {2015}}%
}]{%
ritter2015venus}
\APACinsertmetastar {%
ritter2015venus}%
\begin{APACrefauthors}%
{Ritter}, B.%
, {Meskers}, A\BPBI J\BPBI H.%
, {Miles}, O.%
, {Ru{\ss}wurm}, M.%
, {Scully}, S.%
, {Rold{\'a}n}, A.%
\BDBL {}{Ruffenach}, A.%
\end{APACrefauthors}%
\unskip\
\newblock
\APACrefYearMonthDay{2015}{{\APACmonth{02}}}{}.
\newblock
{\BBOQ}\APACrefatitle {{A Space weather information service based upon remote and in-situ measurements of coronal mass ejections heading for Earth. A concept mission consisting of six spacecraft in a heliocentric orbit at 0.72 AU}} {{A Space weather information service based upon remote and in-situ measurements of coronal mass ejections heading for Earth. A concept mission consisting of six spacecraft in a heliocentric orbit at 0.72 AU}}.{\BBCQ}
\newblock
\APACjournalVolNumPages{Journal of Space Weather and Space Climate}{5}{}{A3}.
\newblock
\begin{APACrefDOI} \doi{10.1051/swsc/2015006} \end{APACrefDOI}
\PrintBackRefs{\CurrentBib}

\bibitem [\protect \citeauthoryear {%
{R{\"u}disser}%
\ \protect \BOthers {.}}{%
{R{\"u}disser}%
\ \protect \BOthers {.}}{%
{\protect \APACyear {2024}}%
}]{%
ruedisser2024understanding}
\APACinsertmetastar {%
ruedisser2024understanding}%
\begin{APACrefauthors}%
{R{\"u}disser}, H\BPBI T.%
, {Weiss}, A\BPBI J.%
, {Le Lou{\"e}dec}, J.%
, {Amerstorfer}, U\BPBI V.%
, {M{\"o}stl}, C.%
, {Davies}, E\BPBI E.%
\BCBL {}\ \BBA {} {Lammer}, H.%
\end{APACrefauthors}%
\unskip\
\newblock
\APACrefYearMonthDay{2024}{{\APACmonth{10}}}{}.
\newblock
{\BBOQ}\APACrefatitle {{Understanding the Effects of Spacecraft Trajectories through Solar Coronal Mass Ejection Flux Ropes Using 3DCOREweb}} {{Understanding the Effects of Spacecraft Trajectories through Solar Coronal Mass Ejection Flux Ropes Using 3DCOREweb}}.{\BBCQ}
\newblock
\APACjournalVolNumPages{\apj}{973}{2}{150}.
\newblock
\begin{APACrefDOI} \doi{10.3847/1538-4357/ad660a} \end{APACrefDOI}
\PrintBackRefs{\CurrentBib}

\bibitem [\protect \citeauthoryear {%
{Ruffenach}%
\ \protect \BOthers {.}}{%
{Ruffenach}%
\ \protect \BOthers {.}}{%
{\protect \APACyear {2015}}%
}]{%
ruffenach2015statistical}
\APACinsertmetastar {%
ruffenach2015statistical}%
\begin{APACrefauthors}%
{Ruffenach}, A.%
, {Lavraud}, B.%
, {Farrugia}, C\BPBI J.%
, {D{\'e}moulin}, P.%
, {Dasso}, S.%
, {Owens}, M\BPBI J.%
\BDBL {}others%
\end{APACrefauthors}%
\unskip\
\newblock
\APACrefYearMonthDay{2015}{{\APACmonth{01}}}{}.
\newblock
{\BBOQ}\APACrefatitle {{Statistical study of magnetic cloud erosion by magnetic reconnection}} {{Statistical study of magnetic cloud erosion by magnetic reconnection}}.{\BBCQ}
\newblock
\APACjournalVolNumPages{\jgr (Space Physics)}{120}{1}{43-60}.
\newblock
\begin{APACrefDOI} \doi{10.1002/2014JA020628} \end{APACrefDOI}
\PrintBackRefs{\CurrentBib}

\bibitem [\protect \citeauthoryear {%
{Ruffenach}%
\ \protect \BOthers {.}}{%
{Ruffenach}%
\ \protect \BOthers {.}}{%
{\protect \APACyear {2012}}%
}]{%
ruffenach2012multispacecraft}
\APACinsertmetastar {%
ruffenach2012multispacecraft}%
\begin{APACrefauthors}%
{Ruffenach}, A.%
, {Lavraud}, B.%
, {Owens}, M\BPBI J.%
, {Sauvaud}, J\BPBI A.%
, {Savani}, N\BPBI P.%
, {Rouillard}, A\BPBI P.%
\BDBL {}others%
\end{APACrefauthors}%
\unskip\
\newblock
\APACrefYearMonthDay{2012}{{\APACmonth{09}}}{}.
\newblock
{\BBOQ}\APACrefatitle {{Multispacecraft observation of magnetic cloud erosion by magnetic reconnection during propagation}} {{Multispacecraft observation of magnetic cloud erosion by magnetic reconnection during propagation}}.{\BBCQ}
\newblock
\APACjournalVolNumPages{\jgr (Space Physics)}{117}{A9}{A09101}.
\newblock
\begin{APACrefDOI} \doi{10.1029/2012JA017624} \end{APACrefDOI}
\PrintBackRefs{\CurrentBib}

\bibitem [\protect \citeauthoryear {%
{Rust}%
}{%
{Rust}%
}{%
{\protect \APACyear {1994}}%
}]{%
rust1994spawning}
\APACinsertmetastar {%
rust1994spawning}%
\begin{APACrefauthors}%
{Rust}, D\BPBI M.%
\end{APACrefauthors}%
\unskip\
\newblock
\APACrefYearMonthDay{1994}{{\APACmonth{02}}}{}.
\newblock
{\BBOQ}\APACrefatitle {{Spawning and Shedding Helical Magnetic Fields in the Solar Atmosphere}} {{Spawning and Shedding Helical Magnetic Fields in the Solar Atmosphere}}.{\BBCQ}
\newblock
\APACjournalVolNumPages{\grl}{21}{4}{241-244}.
\newblock
\begin{APACrefDOI} \doi{10.1029/94GL00003} \end{APACrefDOI}
\PrintBackRefs{\CurrentBib}

\bibitem [\protect \citeauthoryear {%
Rüdisser%
\ \protect \BOthers {.}}{%
Rüdisser%
\ \protect \BOthers {.}}{%
{\protect \APACyear {2026}}%
}]{%
ruedisser2026towards}
\APACinsertmetastar {%
ruedisser2026towards}%
\begin{APACrefauthors}%
Rüdisser, H\BPBI T.%
, Davies, E\BPBI E.%
, Amerstorfer, U\BPBI V.%
, Möstl, C.%
, Weiler, E.%
, Weiss, A\BPBI J.%
\BDBL {}Nguyen, G.%
\end{APACrefauthors}%
\unskip\
\newblock
\APACrefYearMonthDay{2026}{}{}.
\newblock
\APACrefbtitle {Towards a Fully Automated Pipeline for Short-Term Forecasting of In Situ Coronal Mass Ejection Magnetic Field Structure.} {Towards a fully automated pipeline for short-term forecasting of in situ coronal mass ejection magnetic field structure.}
\newblock
\begin{APACrefURL} \url{https://arxiv.org/abs/2602.06926} \end{APACrefURL}
\PrintBackRefs{\CurrentBib}

\bibitem [\protect \citeauthoryear {%
{Salman}%
, {Winslow}%
\BCBL {}\ \BBA {} {Lugaz}%
}{%
{Salman}%
\ \protect \BOthers {.}}{%
{\protect \APACyear {2020}}%
}]{%
salman2020radial}
\APACinsertmetastar {%
salman2020radial}%
\begin{APACrefauthors}%
{Salman}, T\BPBI M.%
, {Winslow}, R\BPBI M.%
\BCBL {}\ \BBA {} {Lugaz}, N.%
\end{APACrefauthors}%
\unskip\
\newblock
\APACrefYearMonthDay{2020}{{\APACmonth{01}}}{}.
\newblock
{\BBOQ}\APACrefatitle {{Radial Evolution of Coronal Mass Ejections Between MESSENGER, Venus Express, STEREO, and L1: Catalog and Analysis}} {{Radial Evolution of Coronal Mass Ejections Between MESSENGER, Venus Express, STEREO, and L1: Catalog and Analysis}}.{\BBCQ}
\newblock
\APACjournalVolNumPages{\jgr (Space Physics)}{125}{1}{e27084}.
\newblock
\begin{APACrefDOI} \doi{10.1029/2019JA027084} \end{APACrefDOI}
\PrintBackRefs{\CurrentBib}

\bibitem [\protect \citeauthoryear {%
{Savani}%
, {Owens}%
, {Rouillard}%
, {Forsyth}%
\BCBL {}\ \BBA {} {Davies}%
}{%
{Savani}%
\ \protect \BOthers {.}}{%
{\protect \APACyear {2010}}%
}]{%
savani2010observational}
\APACinsertmetastar {%
savani2010observational}%
\begin{APACrefauthors}%
{Savani}, N\BPBI P.%
, {Owens}, M\BPBI J.%
, {Rouillard}, A\BPBI P.%
, {Forsyth}, R\BPBI J.%
\BCBL {}\ \BBA {} {Davies}, J\BPBI A.%
\end{APACrefauthors}%
\unskip\
\newblock
\APACrefYearMonthDay{2010}{{\APACmonth{05}}}{}.
\newblock
{\BBOQ}\APACrefatitle {{Observational Evidence of a Coronal Mass Ejection Distortion Directly Attributable to a Structured Solar Wind}} {{Observational Evidence of a Coronal Mass Ejection Distortion Directly Attributable to a Structured Solar Wind}}.{\BBCQ}
\newblock
\APACjournalVolNumPages{\apjl}{714}{1}{L128-L132}.
\newblock
\begin{APACrefDOI} \doi{10.1088/2041-8205/714/1/L128} \end{APACrefDOI}
\PrintBackRefs{\CurrentBib}

\bibitem [\protect \citeauthoryear {%
{Scherrer}%
\ \protect \BOthers {.}}{%
{Scherrer}%
\ \protect \BOthers {.}}{%
{\protect \APACyear {2012}}%
}]{%
Scherrer2012SoPh}
\APACinsertmetastar {%
Scherrer2012SoPh}%
\begin{APACrefauthors}%
{Scherrer}, P\BPBI H.%
, {Schou}, J.%
, {Bush}, R\BPBI I.%
, {Kosovichev}, A\BPBI G.%
, {Bogart}, R\BPBI S.%
, {Hoeksema}, J\BPBI T.%
\BDBL {}{Tomczyk}, S.%
\end{APACrefauthors}%
\unskip\
\newblock
\APACrefYearMonthDay{2012}{{\APACmonth{01}}}{}.
\newblock
{\BBOQ}\APACrefatitle {{The Helioseismic and Magnetic Imager (HMI) Investigation for the Solar Dynamics Observatory (SDO)}} {{The Helioseismic and Magnetic Imager (HMI) Investigation for the Solar Dynamics Observatory (SDO)}}.{\BBCQ}
\newblock
\APACjournalVolNumPages{\solphys}{275}{1-2}{207-227}.
\newblock
\begin{APACrefDOI} \doi{10.1007/s11207-011-9834-2} \end{APACrefDOI}
\PrintBackRefs{\CurrentBib}

\bibitem [\protect \citeauthoryear {%
{Scolini}%
\ \protect \BOthers {.}}{%
{Scolini}%
\ \protect \BOthers {.}}{%
{\protect \APACyear {2022}}%
}]{%
scolini2022complexity}
\APACinsertmetastar {%
scolini2022complexity}%
\begin{APACrefauthors}%
{Scolini}, C.%
, {Winslow}, R\BPBI M.%
, {Lugaz}, N.%
, {Salman}, T\BPBI M.%
, {Davies}, E\BPBI E.%
\BCBL {}\ \BBA {} {Galvin}, A\BPBI B.%
\end{APACrefauthors}%
\unskip\
\newblock
\APACrefYearMonthDay{2022}{{\APACmonth{03}}}{}.
\newblock
{\BBOQ}\APACrefatitle {{Causes and Consequences of Magnetic Complexity Changes within Interplanetary Coronal Mass Ejections: A Statistical Study}} {{Causes and Consequences of Magnetic Complexity Changes within Interplanetary Coronal Mass Ejections: A Statistical Study}}.{\BBCQ}
\newblock
\APACjournalVolNumPages{\apj}{927}{1}{102}.
\newblock
\begin{APACrefDOI} \doi{10.3847/1538-4357/ac3e60} \end{APACrefDOI}
\PrintBackRefs{\CurrentBib}

\bibitem [\protect \citeauthoryear {%
{Smith}%
\ \protect \BOthers {.}}{%
{Smith}%
\ \protect \BOthers {.}}{%
{\protect \APACyear {1998}}%
}]{%
smith1998ace}
\APACinsertmetastar {%
smith1998ace}%
\begin{APACrefauthors}%
{Smith}, C\BPBI W.%
, {L'Heureux}, J.%
, {Ness}, N\BPBI F.%
, {Acu{\~n}a}, M\BPBI H.%
, {Burlaga}, L\BPBI F.%
\BCBL {}\ \BBA {} {Scheifele}, J.%
\end{APACrefauthors}%
\unskip\
\newblock
\APACrefYearMonthDay{1998}{{\APACmonth{07}}}{}.
\newblock
{\BBOQ}\APACrefatitle {{The ACE Magnetic Fields Experiment}} {{The ACE Magnetic Fields Experiment}}.{\BBCQ}
\newblock
\APACjournalVolNumPages{\ssr}{86}{}{613-632}.
\newblock
\begin{APACrefDOI} \doi{10.1023/A:1005092216668} \end{APACrefDOI}
\PrintBackRefs{\CurrentBib}

\bibitem [\protect \citeauthoryear {%
{Solomon}%
\ \protect \BOthers {.}}{%
{Solomon}%
\ \protect \BOthers {.}}{%
{\protect \APACyear {2001}}%
}]{%
solomon2001messenger}
\APACinsertmetastar {%
solomon2001messenger}%
\begin{APACrefauthors}%
{Solomon}, S\BPBI C.%
, {McNutt}, R\BPBI L.%
, {Gold}, R\BPBI E.%
, {Acu{\~n}a}, M\BPBI H.%
, {Baker}, D\BPBI N.%
, {Boynton}, W\BPBI V.%
\BDBL {}{Zuber}, M\BPBI T.%
\end{APACrefauthors}%
\unskip\
\newblock
\APACrefYearMonthDay{2001}{{\APACmonth{12}}}{}.
\newblock
{\BBOQ}\APACrefatitle {{The MESSENGER mission to Mercury: scientific objectives and implementation}} {{The MESSENGER mission to Mercury: scientific objectives and implementation}}.{\BBCQ}
\newblock
\APACjournalVolNumPages{\planss}{49}{14-15}{1445-1465}.
\newblock
\begin{APACrefDOI} \doi{10.1016/S0032-0633(01)00085-X} \end{APACrefDOI}
\PrintBackRefs{\CurrentBib}

\bibitem [\protect \citeauthoryear {%
{Stone}%
\ \protect \BOthers {.}}{%
{Stone}%
\ \protect \BOthers {.}}{%
{\protect \APACyear {1998}}%
}]{%
stone1998advanced}
\APACinsertmetastar {%
stone1998advanced}%
\begin{APACrefauthors}%
{Stone}, E\BPBI C.%
, {Frandsen}, A\BPBI M.%
, {Mewaldt}, R\BPBI A.%
, {Christian}, E\BPBI R.%
, {Margolies}, D.%
, {Ormes}, J\BPBI F.%
\BCBL {}\ \BBA {} {Snow}, F.%
\end{APACrefauthors}%
\unskip\
\newblock
\APACrefYearMonthDay{1998}{{\APACmonth{07}}}{}.
\newblock
{\BBOQ}\APACrefatitle {{The Advanced Composition Explorer}} {{The Advanced Composition Explorer}}.{\BBCQ}
\newblock
\APACjournalVolNumPages{\ssr}{86}{}{1-22}.
\newblock
\begin{APACrefDOI} \doi{10.1023/A:1005082526237} \end{APACrefDOI}
\PrintBackRefs{\CurrentBib}

\bibitem [\protect \citeauthoryear {%
Sugiura%
}{%
Sugiura%
}{%
{\protect \APACyear {1964}}%
}]{%
sugiura1964}
\APACinsertmetastar {%
sugiura1964}%
\begin{APACrefauthors}%
Sugiura, M.%
\end{APACrefauthors}%
\unskip\
\newblock
\APACrefYearMonthDay{1964}{}{}.
\newblock
{\BBOQ}\APACrefatitle {Hourly values of equatorial Dst for the IGY} {Hourly values of equatorial dst for the igy}.{\BBCQ}
\newblock
\APACjournalVolNumPages{Annals of the International Geophysical Year}{}{}{}.
\PrintBackRefs{\CurrentBib}

\bibitem [\protect \citeauthoryear {%
{Svedhem}%
\ \protect \BOthers {.}}{%
{Svedhem}%
\ \protect \BOthers {.}}{%
{\protect \APACyear {2007}}%
}]{%
svedhem2007vex}
\APACinsertmetastar {%
svedhem2007vex}%
\begin{APACrefauthors}%
{Svedhem}, H.%
, {Titov}, D\BPBI V.%
, {McCoy}, D.%
, {Lebreton}, J\BHBI P.%
, {Barabash}, S.%
, {Bertaux}, J\BHBI L.%
\BDBL {}{Coradini}, M.%
\end{APACrefauthors}%
\unskip\
\newblock
\APACrefYearMonthDay{2007}{{\APACmonth{10}}}{}.
\newblock
{\BBOQ}\APACrefatitle {{Venus Express{\textemdash}The first European mission to Venus}} {{Venus Express{\textemdash}The first European mission to Venus}}.{\BBCQ}
\newblock
\APACjournalVolNumPages{\planss}{55}{12}{1636-1652}.
\newblock
\begin{APACrefDOI} \doi{10.1016/j.pss.2007.01.013} \end{APACrefDOI}
\PrintBackRefs{\CurrentBib}

\bibitem [\protect \citeauthoryear {%
Temerin%
\ \BBA {} Li%
}{%
Temerin%
\ \BBA {} Li%
}{%
{\protect \APACyear {2002}}%
}]{%
temerin2002newmodel}
\APACinsertmetastar {%
temerin2002newmodel}%
\begin{APACrefauthors}%
Temerin, M.%
\BCBT {}\ \BBA {} Li, X.%
\end{APACrefauthors}%
\unskip\
\newblock
\APACrefYearMonthDay{2002}{}{}.
\newblock
{\BBOQ}\APACrefatitle {A new model for the prediction of Dst on the basis of the solar wind} {A new model for the prediction of dst on the basis of the solar wind}.{\BBCQ}
\newblock
\APACjournalVolNumPages{Journal of Geophysical Research: Space Physics}{107}{A12}{SMP 31-1-SMP 31-8}.
\newblock
\begin{APACrefURL} \url{https://agupubs.onlinelibrary.wiley.com/doi/abs/10.1029/2001JA007532} \end{APACrefURL}
\newblock
\begin{APACrefDOI} \doi{https://doi.org/10.1029/2001JA007532} \end{APACrefDOI}
\PrintBackRefs{\CurrentBib}

\bibitem [\protect \citeauthoryear {%
{Temerin}%
\ \BBA {} {Li}%
}{%
{Temerin}%
\ \BBA {} {Li}%
}{%
{\protect \APACyear {2006}}%
}]{%
temerin2006magnetospheric}
\APACinsertmetastar {%
temerin2006magnetospheric}%
\begin{APACrefauthors}%
{Temerin}, M.%
\BCBT {}\ \BBA {} {Li}, X.%
\end{APACrefauthors}%
\unskip\
\newblock
\APACrefYearMonthDay{2006}{{\APACmonth{04}}}{}.
\newblock
{\BBOQ}\APACrefatitle {{Dst model for 1995-2002}} {{Dst model for 1995-2002}}.{\BBCQ}
\newblock
\APACjournalVolNumPages{\jgr (Space Physics)}{111}{A4}{A04221}.
\newblock
\begin{APACrefDOI} \doi{10.1029/2005JA011257} \end{APACrefDOI}
\PrintBackRefs{\CurrentBib}

\bibitem [\protect \citeauthoryear {%
{Temerin}%
\ \BBA {} {Li}%
}{%
{Temerin}%
\ \BBA {} {Li}%
}{%
{\protect \APACyear {2015}}%
}]{%
temerin2015underestimates}
\APACinsertmetastar {%
temerin2015underestimates}%
\begin{APACrefauthors}%
{Temerin}, M.%
\BCBT {}\ \BBA {} {Li}, X.%
\end{APACrefauthors}%
\unskip\
\newblock
\APACrefYearMonthDay{2015}{{\APACmonth{07}}}{}.
\newblock
{\BBOQ}\APACrefatitle {{The Dst index underestimates the solar cycle variation of geomagnetic activity}} {{The Dst index underestimates the solar cycle variation of geomagnetic activity}}.{\BBCQ}
\newblock
\APACjournalVolNumPages{Journal of Geophysical Research (Space Physics)}{120}{7}{5603-5607}.
\newblock
\begin{APACrefDOI} \doi{10.1002/2015JA021467} \end{APACrefDOI}
\PrintBackRefs{\CurrentBib}

\bibitem [\protect \citeauthoryear {%
{The pandas development team}%
}{%
{The pandas development team}%
}{%
{\protect \APACyear {2020}}%
}]{%
reback2020pandas}
\APACinsertmetastar {%
reback2020pandas}%
\begin{APACrefauthors}%
{The pandas development team}.%
\end{APACrefauthors}%
\unskip\
\newblock
\APACrefYearMonthDay{2020}{{\APACmonth{02}}}{}.
\newblock
\APACrefbtitle {pandas-dev/pandas: Pandas.} {pandas-dev/pandas: Pandas.}
\newblock
\APACaddressPublisher{}{Zenodo}.
\newblock
\begin{APACrefURL} \url{https://doi.org/10.5281/zenodo.3509134} \end{APACrefURL}
\newblock
\begin{APACrefDOI} \doi{10.5281/zenodo.3509134} \end{APACrefDOI}
\PrintBackRefs{\CurrentBib}

\bibitem [\protect \citeauthoryear {%
{{\v{C}}alogovi{\'c}}%
\ \protect \BOthers {.}}{%
{{\v{C}}alogovi{\'c}}%
\ \protect \BOthers {.}}{%
{\protect \APACyear {2021}}%
}]{%
calogovic2021dbem}
\APACinsertmetastar {%
calogovic2021dbem}%
\begin{APACrefauthors}%
{{\v{C}}alogovi{\'c}}, J.%
, {Dumbovi{\'c}}, M.%
, Sudar, D.%
, {Vr{\v{s}}nak}, B.%
, {Martini{\'c}}, K.%
, Temmer, M.%
\BCBL {}\ \BBA {} Veronig, A\BPBI M.%
\end{APACrefauthors}%
\unskip\
\newblock
\APACrefYearMonthDay{2021}{{\APACmonth{07}}}{}.
\newblock
{\BBOQ}\APACrefatitle {Probabilistic Drag-Based Ensemble Model (DBEM) Evaluation for Heliospheric Propagation of CMEs} {Probabilistic drag-based ensemble model (dbem) evaluation for heliospheric propagation of cmes}.{\BBCQ}
\newblock
\APACjournalVolNumPages{Solar Physics}{296}{7}{}.
\newblock
\begin{APACrefURL} \url{http://dx.doi.org/10.1007/s11207-021-01859-5} \end{APACrefURL}
\newblock
\begin{APACrefDOI} \doi{10.1007/s11207-021-01859-5} \end{APACrefDOI}
\PrintBackRefs{\CurrentBib}

\bibitem [\protect \citeauthoryear {%
{Verbeke}%
\ \protect \BOthers {.}}{%
{Verbeke}%
\ \protect \BOthers {.}}{%
{\protect \APACyear {2023}}%
}]{%
verbeke2023quantifying}
\APACinsertmetastar {%
verbeke2023quantifying}%
\begin{APACrefauthors}%
{Verbeke}, C.%
, {Mays}, M\BPBI L.%
, {Kay}, C.%
, {Riley}, P.%
, {Palmerio}, E.%
, {Dumbovi{\'c}}, M.%
\BDBL {}{Hinterreiter}, J.%
\end{APACrefauthors}%
\unskip\
\newblock
\APACrefYearMonthDay{2023}{{\APACmonth{12}}}{}.
\newblock
{\BBOQ}\APACrefatitle {{Quantifying errors in 3D CME parameters derived from synthetic data using white-light reconstruction techniques}} {{Quantifying errors in 3D CME parameters derived from synthetic data using white-light reconstruction techniques}}.{\BBCQ}
\newblock
\APACjournalVolNumPages{Advances in Space Research}{72}{12}{5243-5262}.
\newblock
\begin{APACrefDOI} \doi{10.1016/j.asr.2022.08.056} \end{APACrefDOI}
\PrintBackRefs{\CurrentBib}

\bibitem [\protect \citeauthoryear {%
Virtanen%
\ \protect \BOthers {.}}{%
Virtanen%
\ \protect \BOthers {.}}{%
{\protect \APACyear {2020}}%
}]{%
scipy2020}
\APACinsertmetastar {%
scipy2020}%
\begin{APACrefauthors}%
Virtanen, P.%
, Gommers, R.%
, Oliphant, T\BPBI E.%
, Haberland, M.%
, Reddy, T.%
, Cournapeau, D.%
\BDBL {}{SciPy 1.0 Contributors}%
\end{APACrefauthors}%
\unskip\
\newblock
\APACrefYearMonthDay{2020}{}{}.
\newblock
{\BBOQ}\APACrefatitle {{{SciPy} 1.0: Fundamental Algorithms for Scientific Computing in Python}} {{{SciPy} 1.0: Fundamental Algorithms for Scientific Computing in Python}}.{\BBCQ}
\newblock
\APACjournalVolNumPages{Nature Methods}{17}{}{261--272}.
\newblock
\begin{APACrefDOI} \doi{10.1038/s41592-019-0686-2} \end{APACrefDOI}
\PrintBackRefs{\CurrentBib}

\bibitem [\protect \citeauthoryear {%
{Vourlidas}%
, {Patsourakos}%
\BCBL {}\ \BBA {} {Savani}%
}{%
{Vourlidas}%
\ \protect \BOthers {.}}{%
{\protect \APACyear {2019}}%
}]{%
vourlidas2019predicting}
\APACinsertmetastar {%
vourlidas2019predicting}%
\begin{APACrefauthors}%
{Vourlidas}, A.%
, {Patsourakos}, S.%
\BCBL {}\ \BBA {} {Savani}, N\BPBI P.%
\end{APACrefauthors}%
\unskip\
\newblock
\APACrefYearMonthDay{2019}{{\APACmonth{07}}}{}.
\newblock
{\BBOQ}\APACrefatitle {{Predicting the geoeffective properties of coronal mass ejections: current status, open issues and path forward}} {{Predicting the geoeffective properties of coronal mass ejections: current status, open issues and path forward}}.{\BBCQ}
\newblock
\APACjournalVolNumPages{Philosophical Transactions of the Royal Society of London Series A}{377}{2148}{20180096}.
\newblock
\begin{APACrefDOI} \doi{10.1098/rsta.2018.0096} \end{APACrefDOI}
\PrintBackRefs{\CurrentBib}

\bibitem [\protect \citeauthoryear {%
{Vr{\v{s}}nak}%
\ \protect \BOthers {.}}{%
{Vr{\v{s}}nak}%
\ \protect \BOthers {.}}{%
{\protect \APACyear {2013}}%
}]{%
vrsnak2013propagation}
\APACinsertmetastar {%
vrsnak2013propagation}%
\begin{APACrefauthors}%
{Vr{\v{s}}nak}, B.%
, {{\v{Z}}ic}, T.%
, {Vrbanec}, D.%
, {Temmer}, M.%
, {Rollett}, T.%
, {M{\"o}stl}, C.%
\BDBL {}{Shanmugaraju}, A.%
\end{APACrefauthors}%
\unskip\
\newblock
\APACrefYearMonthDay{2013}{Jul}{}.
\newblock
{\BBOQ}\APACrefatitle {{Propagation of Interplanetary Coronal Mass Ejections: The Drag-Based Model}} {{Propagation of Interplanetary Coronal Mass Ejections: The Drag-Based Model}}.{\BBCQ}
\newblock
\APACjournalVolNumPages{\solphys}{285}{1-2}{295-315}.
\newblock
\begin{APACrefDOI} \doi{10.1007/s11207-012-0035-4} \end{APACrefDOI}
\PrintBackRefs{\CurrentBib}

\bibitem [\protect \citeauthoryear {%
Wanliss%
\ \BBA {} Showalter%
}{%
Wanliss%
\ \BBA {} Showalter%
}{%
{\protect \APACyear {2006}}%
}]{%
wanliss2006symh}
\APACinsertmetastar {%
wanliss2006symh}%
\begin{APACrefauthors}%
Wanliss, J\BPBI A.%
\BCBT {}\ \BBA {} Showalter, K\BPBI M.%
\end{APACrefauthors}%
\unskip\
\newblock
\APACrefYearMonthDay{2006}{}{}.
\newblock
{\BBOQ}\APACrefatitle {High-resolution global storm index: Dst versus SYM-H} {High-resolution global storm index: Dst versus sym-h}.{\BBCQ}
\newblock
\APACjournalVolNumPages{Journal of Geophysical Research: Space Physics}{111}{A2}{}.
\newblock
\begin{APACrefURL} \url{https://agupubs.onlinelibrary.wiley.com/doi/abs/10.1029/2005JA011034} \end{APACrefURL}
\newblock
\begin{APACrefDOI} \doi{https://doi.org/10.1029/2005JA011034} \end{APACrefDOI}
\PrintBackRefs{\CurrentBib}

\bibitem [\protect \citeauthoryear {%
{Weiler}%
\ \protect \BOthers {.}}{%
{Weiler}%
\ \protect \BOthers {.}}{%
{\protect \APACyear {2025}}%
}]{%
weiler2025superstorm}
\APACinsertmetastar {%
weiler2025superstorm}%
\begin{APACrefauthors}%
{Weiler}, E.%
, {M{\"o}stl}, C.%
, {Davies}, E\BPBI E.%
, {Veronig}, A\BPBI M.%
, {Amerstorfer}, U\BPBI V.%
, {Amerstorfer}, T.%
\BDBL {}{Reiss}, M.%
\end{APACrefauthors}%
\unskip\
\newblock
\APACrefYearMonthDay{2025}{{\APACmonth{03}}}{}.
\newblock
{\BBOQ}\APACrefatitle {{First Observations of a Geomagnetic Superstorm With a Sub-L1 Monitor}} {{First Observations of a Geomagnetic Superstorm With a Sub-L1 Monitor}}.{\BBCQ}
\newblock
\APACjournalVolNumPages{Space Weather}{23}{3}{2024SW004260}.
\newblock
\begin{APACrefDOI} \doi{10.1029/2024SW004260} \end{APACrefDOI}
\PrintBackRefs{\CurrentBib}

\bibitem [\protect \citeauthoryear {%
A\BPBI J.~{Weiss}%
\ \protect \BOthers {.}}{%
A\BPBI J.~{Weiss}%
\ \protect \BOthers {.}}{%
{\protect \APACyear {2021}}%
}]{%
weiss2021triple}
\APACinsertmetastar {%
weiss2021triple}%
\begin{APACrefauthors}%
{Weiss}, A\BPBI J.%
, {M{\"o}stl}, C.%
, {Davies}, E\BPBI E.%
, {Amerstorfer}, T.%
, {Bauer}, M.%
, {Hinterreiter}, J.%
\BDBL {}{Baumjohann}, W.%
\end{APACrefauthors}%
\unskip\
\newblock
\APACrefYearMonthDay{2021}{{\APACmonth{12}}}{}.
\newblock
{\BBOQ}\APACrefatitle {{Multi-point analysis of coronal mass ejection flux ropes using combined data from Solar Orbiter, BepiColombo, and Wind}} {{Multi-point analysis of coronal mass ejection flux ropes using combined data from Solar Orbiter, BepiColombo, and Wind}}.{\BBCQ}
\newblock
\APACjournalVolNumPages{\aap}{656}{}{A13}.
\newblock
\begin{APACrefDOI} \doi{10.1051/0004-6361/202140919} \end{APACrefDOI}
\PrintBackRefs{\CurrentBib}

\bibitem [\protect \citeauthoryear {%
J.~{Weiss} A%
\ \protect \BOthers {.}}{%
J.~{Weiss} A%
\ \protect \BOthers {.}}{%
{\protect \APACyear {2021}}%
}]{%
weiss2021analysis}
\APACinsertmetastar {%
weiss2021analysis}%
\begin{APACrefauthors}%
{Weiss}, J., A%
, {M{\"o}stl}, C.%
, {Amerstorfer}, T.%
, {Bailey}, R\BPBI L.%
, {Reiss}, M\BPBI A.%
, {Hinterreiter}, J.%
\BDBL {}{Bauer}, M.%
\end{APACrefauthors}%
\unskip\
\newblock
\APACrefYearMonthDay{2021}{{\APACmonth{01}}}{}.
\newblock
{\BBOQ}\APACrefatitle {{Analysis of Coronal Mass Ejection Flux Rope Signatures Using 3DCORE and Approximate Bayesian Computation}} {{Analysis of Coronal Mass Ejection Flux Rope Signatures Using 3DCORE and Approximate Bayesian Computation}}.{\BBCQ}
\newblock
\APACjournalVolNumPages{\apjs}{252}{1}{9}.
\newblock
\begin{APACrefDOI} \doi{10.3847/1538-4365/abc9bd} \end{APACrefDOI}
\PrintBackRefs{\CurrentBib}

\bibitem [\protect \citeauthoryear {%
{W}es {M}c{K}inney%
}{%
{W}es {M}c{K}inney%
}{%
{\protect \APACyear {2010}}%
}]{%
mckinney2010pandas}
\APACinsertmetastar {%
mckinney2010pandas}%
\begin{APACrefauthors}%
{W}es {M}c{K}inney.%
\end{APACrefauthors}%
\unskip\
\newblock
\APACrefYearMonthDay{2010}{}{}.
\newblock
{\BBOQ}\APACrefatitle {{D}ata {S}tructures for {S}tatistical {C}omputing in {P}ython} {{D}ata {S}tructures for {S}tatistical {C}omputing in {P}ython}.{\BBCQ}
\newblock
\BIn{} {S}t\'efan van~der {W}alt\ \BBA {} {J}arrod {M}illman\ (\BEDS), \APACrefbtitle {{P}roceedings of the 9th {P}ython in {S}cience {C}onference} {{P}roceedings of the 9th {P}ython in {S}cience {C}onference}\ (\BPG~56 - 61).
\newblock
\begin{APACrefDOI} \doi{10.25080/Majora-92bf1922-00a} \end{APACrefDOI}
\PrintBackRefs{\CurrentBib}

\bibitem [\protect \citeauthoryear {%
{Winslow}%
\ \protect \BOthers {.}}{%
{Winslow}%
\ \protect \BOthers {.}}{%
{\protect \APACyear {2015}}%
}]{%
winslow2015interplanetary}
\APACinsertmetastar {%
winslow2015interplanetary}%
\begin{APACrefauthors}%
{Winslow}, R\BPBI M.%
, {Lugaz}, N.%
, {Philpott}, L\BPBI C.%
, {Schwadron}, N\BPBI A.%
, {Farrugia}, C\BPBI J.%
, {Anderson}, B\BPBI J.%
\BCBL {}\ \BBA {} {Smith}, C\BPBI W.%
\end{APACrefauthors}%
\unskip\
\newblock
\APACrefYearMonthDay{2015}{{\APACmonth{08}}}{}.
\newblock
{\BBOQ}\APACrefatitle {{Interplanetary coronal mass ejections from MESSENGER orbital observations at Mercury}} {{Interplanetary coronal mass ejections from MESSENGER orbital observations at Mercury}}.{\BBCQ}
\newblock
\APACjournalVolNumPages{\jgr (Space Physics)}{120}{8}{6101-6118}.
\newblock
\begin{APACrefDOI} \doi{10.1002/2015JA021200} \end{APACrefDOI}
\PrintBackRefs{\CurrentBib}

\bibitem [\protect \citeauthoryear {%
{Wold}%
\ \protect \BOthers {.}}{%
{Wold}%
\ \protect \BOthers {.}}{%
{\protect \APACyear {2018}}%
}]{%
wold2018verification}
\APACinsertmetastar {%
wold2018verification}%
\begin{APACrefauthors}%
{Wold}, A\BPBI M.%
, {Mays}, M\BPBI L.%
, {Taktakishvili}, A.%
, {Jian}, L\BPBI K.%
, {Odstrcil}, D.%
\BCBL {}\ \BBA {} {MacNeice}, P.%
\end{APACrefauthors}%
\unskip\
\newblock
\APACrefYearMonthDay{2018}{{\APACmonth{03}}}{}.
\newblock
{\BBOQ}\APACrefatitle {{Verification of real-time WSA-ENLIL+Cone simulations of CME arrival-time at the CCMC from 2010 to 2016}} {{Verification of real-time WSA-ENLIL+Cone simulations of CME arrival-time at the CCMC from 2010 to 2016}}.{\BBCQ}
\newblock
\APACjournalVolNumPages{Journal of Space Weather and Space Climate}{8}{}{A17}.
\newblock
\begin{APACrefDOI} \doi{10.1051/swsc/2018005} \end{APACrefDOI}
\PrintBackRefs{\CurrentBib}

\bibitem [\protect \citeauthoryear {%
{Xie}%
, {Ofman}%
\BCBL {}\ \BBA {} {Lawrence}%
}{%
{Xie}%
\ \protect \BOthers {.}}{%
{\protect \APACyear {2004}}%
}]{%
xie2004cone}
\APACinsertmetastar {%
xie2004cone}%
\begin{APACrefauthors}%
{Xie}, H.%
, {Ofman}, L.%
\BCBL {}\ \BBA {} {Lawrence}, G.%
\end{APACrefauthors}%
\unskip\
\newblock
\APACrefYearMonthDay{2004}{{\APACmonth{03}}}{}.
\newblock
{\BBOQ}\APACrefatitle {{Cone model for halo CMEs: Application to space weather forecasting}} {{Cone model for halo CMEs: Application to space weather forecasting}}.{\BBCQ}
\newblock
\APACjournalVolNumPages{Journal of Geophysical Research (Space Physics)}{109}{A3}{A03109}.
\newblock
\begin{APACrefDOI} \doi{10.1029/2003JA010226} \end{APACrefDOI}
\PrintBackRefs{\CurrentBib}

\bibitem [\protect \citeauthoryear {%
{Zhao}%
, {Plunkett}%
\BCBL {}\ \BBA {} {Liu}%
}{%
{Zhao}%
\ \protect \BOthers {.}}{%
{\protect \APACyear {2002}}%
}]{%
zhao2002determination}
\APACinsertmetastar {%
zhao2002determination}%
\begin{APACrefauthors}%
{Zhao}, X\BPBI P.%
, {Plunkett}, S\BPBI P.%
\BCBL {}\ \BBA {} {Liu}, W.%
\end{APACrefauthors}%
\unskip\
\newblock
\APACrefYearMonthDay{2002}{{\APACmonth{08}}}{}.
\newblock
{\BBOQ}\APACrefatitle {{Determination of geometrical and kinematical properties of halo coronal mass ejections using the cone model}} {{Determination of geometrical and kinematical properties of halo coronal mass ejections using the cone model}}.{\BBCQ}
\newblock
\APACjournalVolNumPages{Journal of Geophysical Research (Space Physics)}{107}{A8}{1223}.
\newblock
\begin{APACrefDOI} \doi{10.1029/2001JA009143} \end{APACrefDOI}
\PrintBackRefs{\CurrentBib}

\bibitem [\protect \citeauthoryear {%
{Zhou}%
\ \protect \BOthers {.}}{%
{Zhou}%
\ \protect \BOthers {.}}{%
{\protect \APACyear {2020}}%
}]{%
zhou2020relationship}
\APACinsertmetastar {%
zhou2020relationship}%
\begin{APACrefauthors}%
{Zhou}, Z.%
, {Liu}, R.%
, {Cheng}, X.%
, {Jiang}, C.%
, {Wang}, Y.%
, {Liu}, L.%
\BCBL {}\ \BBA {} {Cui}, J.%
\end{APACrefauthors}%
\unskip\
\newblock
\APACrefYearMonthDay{2020}{{\APACmonth{03}}}{}.
\newblock
{\BBOQ}\APACrefatitle {{The Relationship between Chirality, Sense of Rotation, and Hemispheric Preference of Solar Eruptive Filaments}} {{The Relationship between Chirality, Sense of Rotation, and Hemispheric Preference of Solar Eruptive Filaments}}.{\BBCQ}
\newblock
\APACjournalVolNumPages{\apj}{891}{2}{180}.
\newblock
\begin{APACrefDOI} \doi{10.3847/1538-4357/ab7666} \end{APACrefDOI}
\PrintBackRefs{\CurrentBib}

\bibitem [\protect \citeauthoryear {%
{Zhuang}%
\ \protect \BOthers {.}}{%
{Zhuang}%
\ \protect \BOthers {.}}{%
{\protect \APACyear {2024}}%
}]{%
zhuang2024combining}
\APACinsertmetastar {%
zhuang2024combining}%
\begin{APACrefauthors}%
{Zhuang}, B.%
, {Lugaz}, N.%
, {Al-Haddad}, N.%
, {Scolini}, C.%
, {Farrugia}, C\BPBI J.%
, {Regnault}, F.%
\BDBL {}{Galvin}, A\BPBI B.%
\end{APACrefauthors}%
\unskip\
\newblock
\APACrefYearMonthDay{2024}{{\APACmonth{02}}}{}.
\newblock
{\BBOQ}\APACrefatitle {{Combining STEREO heliospheric imagers and Solar Orbiter to investigate the evolution of the 2022 March 10 CME}} {{Combining STEREO heliospheric imagers and Solar Orbiter to investigate the evolution of the 2022 March 10 CME}}.{\BBCQ}
\newblock
\APACjournalVolNumPages{\aap}{682}{}{A107}.
\newblock
\begin{APACrefDOI} \doi{10.1051/0004-6361/202347561} \end{APACrefDOI}
\PrintBackRefs{\CurrentBib}

\bibitem [\protect \citeauthoryear {%
{Zurbuchen}%
\ \BBA {} {Richardson}%
}{%
{Zurbuchen}%
\ \BBA {} {Richardson}%
}{%
{\protect \APACyear {2006}}%
}]{%
zurbuchen2006situ}
\APACinsertmetastar {%
zurbuchen2006situ}%
\begin{APACrefauthors}%
{Zurbuchen}, T\BPBI H.%
\BCBT {}\ \BBA {} {Richardson}, I\BPBI G.%
\end{APACrefauthors}%
\unskip\
\newblock
\APACrefYearMonthDay{2006}{{\APACmonth{03}}}{}.
\newblock
{\BBOQ}\APACrefatitle {{In-Situ Solar Wind and Magnetic Field Signatures of Interplanetary Coronal Mass Ejections}} {{In-Situ Solar Wind and Magnetic Field Signatures of Interplanetary Coronal Mass Ejections}}.{\BBCQ}
\newblock
\APACjournalVolNumPages{\ssr}{123}{1-3}{31-43}.
\newblock
\begin{APACrefDOI} \doi{10.1007/s11214-006-9010-4} \end{APACrefDOI}
\PrintBackRefs{\CurrentBib}

\bibitem [\protect \citeauthoryear {%
{Zwickl}%
\ \protect \BOthers {.}}{%
{Zwickl}%
\ \protect \BOthers {.}}{%
{\protect \APACyear {1998}}%
}]{%
zwickl1998rtswace}
\APACinsertmetastar {%
zwickl1998rtswace}%
\begin{APACrefauthors}%
{Zwickl}, R\BPBI D.%
, {Doggett}, K\BPBI A.%
, {Sahm}, S.%
, {Barrett}, W\BPBI P.%
, {Grubb}, R\BPBI N.%
, {Detman}, T\BPBI R.%
\BDBL {}{Maruyama}, T.%
\end{APACrefauthors}%
\unskip\
\newblock
\APACrefYearMonthDay{1998}{{\APACmonth{07}}}{}.
\newblock
{\BBOQ}\APACrefatitle {{The NOAA Real-Time Solar-Wind (RTSW) System using ACE Data}} {{The NOAA Real-Time Solar-Wind (RTSW) System using ACE Data}}.{\BBCQ}
\newblock
\APACjournalVolNumPages{\ssr}{86}{}{633-648}.
\newblock
\begin{APACrefDOI} \doi{10.1023/A:1005044300738} \end{APACrefDOI}
\PrintBackRefs{\CurrentBib}

\end{thebibliography}


\end{document}